\documentclass[letterpaper,12pt]{article}
\pdfoutput=1
 
\usepackage[english]{babel}
\usepackage[latin1]{inputenc}
\usepackage{graphicx}
\usepackage{indentfirst}
\usepackage{fancyhdr}
\usepackage{bbold}
\usepackage{cite} %package for using Bibtex
\usepackage{url}
\usepackage{color}
\usepackage[usenames,dvipsnames,svgnames,table]{xcolor}
\usepackage{xspace}
\usepackage{framed}
\usepackage{cite} %package for using Bibtex
\usepackage[font=small,labelfont=bf]{caption}
\usepackage{setspace}
\usepackage[bottom=2.7cm]{geometry}
\usepackage{amssymb}
\usepackage{amsmath}
\usepackage{latexsym}
\usepackage{amsthm}
\usepackage{amsfonts}
\usepackage{amssymb}
\definecolor{LNKcolor}{rgb}{0.00,0.10,0.45}
\usepackage[hyperindex=false,pdftex,bookmarks=true,bookmarksopen=true,pdftitle={2d-4d-spectrum generator},pdfauthor={Longhi},colorlinks=true,linkcolor=LNKcolor,citecolor=LNKcolor,urlcolor=LNKcolor,filecolor=LNKcolor]{hyperref}
\newcommand{\beq}{\begin{eqnarray}}
\newcommand{\eeq}{\end{eqnarray}}
\newcommand{\ba}{\beq\begin{aligned}}
\newcommand{\ea}{\end{aligned}\eeq}

\definecolor{shadecolor}{rgb}{0.85,0.85,0.85}
\def\CX {{\cal X}}
\def\CY {{\cal Y}}
\def\P {{\cal P}}
\def\N {{\cal N}}
\def\K {{\cal K}}

\def\CP {{\mathbb{CP}}}

\def\e {{\mathrm{e}}}
\def\d {{\mathrm{d}}}
\def\II {\cite{GMN2}\xspace}
\def\IV {\cite{GMN4}\xspace}
\def\ie {{\emph{i.e.}}\xspace}

\begin{document}
\begin{flushright}
\date \\
\normalsize
\end{flushright}

\vspace{0.5cm}

\begin{center}
{\LARGE {THE BPS SPECTRUM GENERATOR\\ \vspace{10pt} IN 2D-4D SYSTEMS}} \\

\vspace{3.6 cm} {Pietro Longhi\footnote{\href{mailto:longhi@physics.rutgers.edu}{longhi@physics.rutgers.edu}}}\\

\vspace{8pt} {\it { NHETC and Department of Physics \& Astronomy, \\ Rutgers University, Piscataway, NJ 08855-0849, USA }}

\vspace{2.5cm}

%{\bf Abstract}
\end{center}

\begin{quotation}
\noindent 
\linespread{2.1}
We apply the techniques provided by the recent works Gaiotto, Moore and Neitzke, to derive the most general spectrum generating functions for coupled 2d-4d $A_1$ theories of class ${\cal S}$, in presence of surface and line defects. As an application of the result, some well-known BPS spectra are reproduced. Our results apply to a large class of coupled 2d-4d systems, the corresponding spectrum generating functions can be easily derived from our general expressions.
\end{quotation}
\newpage

%\vspace{1cm}
%\noindent\HRule
%\vspace{.5pt}
%\tableofcontents
%\vspace{.5pt}
%\noindent\HRule

\linespread{1}
\section{Introduction}
In recent works \cite{GMN1,GMN2,GMN3,GMN4,GMN5} of Gaiotto, Moore and Neitzke, a framework capable of probing the BPS spectrum of a large class of theories, specifically the ``$A_1$ theories of class S'', has been developed. As the name suggests, these are $4d$ $\N=2$ gauge theories obtained from compactifications of the six-dimensional $(2,0)$ theory on a punctured Riemann surface $C$. Such theories are well known to exhibit BPS spectra \cite{1994NuPhB.426...19S,1994NuPhB.431..484S,1997NuPhB.490217F,Shapere:fk,Ferrari:1996uq,Bilal:1996fk,Alim:2011zr,1997NuPhB.501...53F}: at generic points on the Coulomb branch $\mathcal{B}$ the gauge symmetry is broken down to an abelian subgroup, and BPS states carry electric-magnetic charges $\gamma\in\Gamma$, a sublattice of $H_1(\Sigma,\mathbb{Z})$, where $\Sigma$ is the corresponding Seiberg-Witten curve. One of the key ingredients of the GMN framework is that, at a generic $u\in\mathcal{B}$, the BPS degeneracies $\Omega(\gamma)$ are encoded into an ordered product of Poisson transformations $\mathbf{S}=\prod \K_\gamma^{\Omega(\gamma)}$, acting on certain Fock-Goncharov coordinates $\CX_\gamma$ on the moduli space of BPS field configurations $\mathcal{M}$. The structure behind this statement is quite rich, as we partially review below: it relies crucially on the connection to the 6d setting \cite{Brane-Construction,Klemm:1996kx}, which is exploited by establishing a correspondence between such theories and a class of Hitchin systems (see also\cite{Donagi:1996oq,CK1,CK2}), the data of Hitchin system allows then to make contact with the Fock-Goncharov construction of coordinates on $\mathcal{M}$.

A nice feature of this approach is that one needs not use all the sophisticated machinery employed by the authors, in order to investigate the BPS spectra of such theories. Indeed, a few basic rules outlined in the reference allow to derive the \emph{spectrum generating functions} $S_i$, namely the action of $\mathbf{S}$ on a set of ``basis'' coordinates $\CX_{\gamma_i}$. The full BPS spectrum is then encoded\footnote{The spectrum in only known implicitly, since it is a nontrivial task to extract the BPS degeneracies from the set of $S_i$.} in the set of spectrum generating functions $S_i$, and can be read off by considering suitable filtrations of the $\CX_\gamma$ algebra.\\

This class of theories can be enriched by considering both line- and surface defects \cite{Drukker:2009vn,Drukker:2010ys,Gukov1,Gukov2,Surface-ops,GukovWitten,Kapustin,GMN3,GMN4} ``coupled'' to the 4d theory.
% , these extensions arise naturally from the 6d viewpoint
Particularly interesting is a certain class of surface defects, which preserve four of the eight supersymmetries of the 4d theory, and support $\N=(2,2)$ 2d field theories with finite sets of massive vacua. These 2d theories are interesting companions to the 4d ones, because they carry well-understood BPS spectra of their own \cite{CV1,CV2,CV3}, the 2d BPS states are solitons interpolating between different vacua. As discussed in \cite{GMN4}, the subspace of BPS states is greatly enhanced if one considers ``coupled'' 2d-4d systems: for each couple of vacua of the surface defect, there can now be infinitely many solitons interpolating between them, carrying gauge and flavor charges of the 4d bulk theory. Another new feature of the 2d-4d setting are 2d solitons sitting in a single vacuum, carrying four-dimensional gauge and flavor charge.

The study of this new type of spectrum can be carried out by methods that generalize those of \cite{GMN2}, this approach was exploited in great detail in \cite{GMN4}, we review some relevant aspects of this construction below. Once again, the structure involved is rather rich, and requires some effort to be understood and applied.\\

This brings us to the main motivation for this paper. A natural question to ask is whether there exists an analog of the spectrum generating functions of the pure 4d case, that applies to the 2d-4d setting. More precisely, upon introducing suitably generalized FG coordinates $\CY_{\gamma_{ij}}$ \cite{GMN4}, the 2d-4d BPS degeneracies $\omega,\mu$ are encoded into a product of Poisson morphisms acting on these new functions\footnote{here $\mathcal{K}_\gamma,\mathcal{S}_{\gamma_{ij}}$ correspond respectively to morphisms induced by 4d and 2d states, and $:\ :$ denotes an ordering by phases of central charges of the product. }
\ba
      &\CY'_{\gamma_{ij}}\,=\,\mathbf{S}\,\CY_{\gamma_{ij}} \\
      &{\bf S}\,=\,:\prod_\gamma \K_\gamma^\omega\,\prod_{\gamma_{ij}} \mathcal{S}_{\gamma_{ij}}^\mu:
\ea
Detailed formulae for the transformation $\bf S$ are given in equations (\ref{eq:AD_WC}), (\ref{eq:AD_WC_bis}), (\ref{eq:SG-AD-large}), (\ref{eq:SG-AD-N2}) and (\ref{eq:SG-CP1}) below, and in the equations right above them.
It would be nice to find a general formula for the transformed $\CY'_{\gamma_{ij}}$, expressed only in terms of the original $\CY_{\gamma_{ij}}$. This would allow to recover the spectrum generator $\mathbf{S}$ and therefore the BPS degeneracies, without having to employ the full machinery of GMN.

In this paper we investigate the existence of such expressions, our main result is a set of formulae for the $\CY'_{\gamma_{ij}}$, expressed in terms of the original coordinates. The expressions we derive encode the action of the most general $\mathbf{S}$ on the $\CY_{\gamma_{ij}}$, and they can be easily applied to any specific situation. Just like in the pure 4d case, once adapted to the specific situation under study, our formulae encode the 2d-4d degeneracies. These formulae provide a systematic approach to the study of 2d-4d spectra: as the examples below show, it is fairly easy to recover the simplest finite spectra from our general expressions, while more complicated theories will probably require some sort of algorithm to extract the degeneracies.\\

The outline of the paper is the following. Section \ref{sec:formal_work} begins with a quick review of certain aspects of the GMN construction that are relevant for this paper, we then move on to analyzing how small flat sections at regular singularities of the Hitchin system behave under an ``omnipop'', this will allow us to express the omnipop transformation for FG coordinates associated to the solitonic charges: these expressions are the analog of the  4d spectrum generating functions for 2d-4d coupled systems. We then extend our analysis to include irregular singularities and line defects, and derive the building blocks for the most general 2d-4d spectrum generating functions for this type of theories.

In section \ref{sec:examples} we apply our results to derive some well-known spectra: we will work in detail through the simplest Argyres-Douglas theories coupled to one or two surface defects, as well as through an example of a $\mathbb{CP}^1$ $\sigma$ model. As we show below, obtaining the generating functions specific to each theory, when starting from our general formulae, is achieved in just a few steps.

\vfill

\section{Spectrum generating functions in 2d-4d systems}\label{sec:formal_work}

\subsection{A short account of the GMN construction}
Let us begin by recalling some fundamental ingredients from papers \cite{GMN2,GMN3,GMN4}. At generic $u$ on the Coulomb branch $\mathcal{B}$ of $\N=2$ four-dimensional SYM, there is a family of WKB triangulations $T_{WKB}^\vartheta$ of $C$, defined as the isotopy class of the flow
\ba
    \langle\partial_t,\lambda\rangle\in e^{i\vartheta}\mathbb{R}^\times
\ea
where $\lambda$ is the Seiberg-Witten differential. These triangulations are piecewise independent of $\vartheta$ and the BPS spectrum manifests itself through jumps of  $T_{WKB}^\vartheta$ at values of $\vartheta$ coinciding with $Arg\,Z_{\gamma_{BPS}}$. The Hitchin system arises by considering a different kind of compactification, namely taking the the 6d theory on a circle, which leads to 5d super Yang-Mills, then considering the ``4d BPS instantons'' on the space-like directions, finally reducing the corresponding self-duality equations on $C$. The M-theory engineering of these theories provides the data that is necessary to specify the corresponding Hitchin system, in particular the behavior of the Higgs field $\varphi$ and of the connection $A$ at singular points, in correspondence of M5-brane intersections \cite{2012JHEP...08..034G,2012arXiv1203.2930C}. The solutions of the Hitchin system that satisfy the boundary conditions, modulo gauge, are parameterized by a hyperkahler manifold $\mathcal{M}$ which is also the moduli space of flat $sl(2,\mathbb{C})$ connections defined by
\ba
    \mathcal{A}=\frac{R}{\zeta}\varphi + A+ R\zeta\bar\varphi,\qquad\zeta\in\mathbb{C}^\times.
\ea
At each singularity there is a monodromy matrix $M_i$ associated with this connection, which depends on the boundary data as well as on $\zeta, R$. Each of the $M_{i}$ has two eigen-sections $s_i,\, \tilde s_i$, with respective eigenvalues $\mu_i,\,\mu_i^{-1}$. The \emph{small flat section} at a singularity is defined to be the $\mathcal{A}$-flat eigen-section $s_{i}$ , chosen between the two such that its norm decreases when evaluated along WKB lines that flow into the singularity. The choice of small flat section at each singularity is a ``decoration'' of $T_{WKB}$.

Each edge of the triangulation corresponds to a homology cycle of the spectral curve of the Hitchin system. Let $\Gamma$ denote the lattice generated by such cycles, to each $\gamma\in\Gamma$ one associates a corresponding Fock-Goncharov coordinate $\CX_{\gamma}$ defined by
\ba
	\CX_{\gamma}=-\frac{(s_{i}\wedge s_{j})(s_{k}\wedge s_{\ell})}{(s_{j}\wedge s_{k})(s_{\ell}\wedge s_{i})}
\ea
where $i,j,k,\ell$ denote the four singularities -ordered counterclockwise- at the corners of a quadrilateral with $E_{\gamma}$ stretching between singularities $i,k$. Here the cycles $\gamma$ belong to $\Gamma$, the homology sublattice of gauge charges. The jumps of $T_{WKB}^\vartheta$ are quantitatively described by Poisson transformations $\K_{\gamma'}$ acting on the $\CX_\gamma$, there are two main types of jumps: one due to  a BPS hypermultiplet and one due to a vectormultiplet. Naively, keeping track of the jumps as $\vartheta$ varies is one way to recover the BPS spectrum, however typical spectra are infinite and in practice cannot be obtained with this method. The BPS spectrum divides evenly into \emph{particles} and their CPT conjugates, by adopting an appropriate definition of particle , the central charges can be taken to lie within a half of the complex plane; therefore varying $\vartheta$ over an angle of $\pi$ captures all the states of interest. The full BPS spectrum is then encoded into the transformations $\CX_{\gamma_{i}}^{\vartheta}\mapsto\CX_{\gamma_{i}}^{\vartheta+\pi}$  ($\gamma_{i}$ runs over a basis of $\Gamma$), these determine\footnote{The Fock-Goncharov coordinates obey the multiplication rules 
\ba
\CX_{\gamma}\CX_{\gamma'}=\CX_{\gamma+\gamma'}
\ea
thus, the transformations of \emph{basis coordinates} (those associated to a basis of $\Gamma$) determine that of any $\CX_{\gamma}$.} an operator $\mathbf{S}$ having a \emph{unique} factorization 
\ba
	{\bf S}=\prod_\gamma{\K_\gamma^{\Omega(\gamma,u)}}
\ea
where $\Omega(\gamma,u)$ are the BPS degeneracies for states of charge $\gamma$.

At this point, this method might appear quite inconvenient, because of the difficulties involved in constructing explicitly Fock-Goncharov coordinates. As a matter of fact, however, there are simple expressions for the $\CX_{\gamma_{i}}^{\vartheta+\pi}$ in terms of the initial FG coordinates. In \cite{GMN2} a fairly easy recipe for writing them down was provided, which applies to all $A_{1}$ theories of class $\mathcal{S}$. For later convenience we briefly recall the key idea behind it. The crucial step is to notice that $T_{WKB}^{\vartheta+\pi}$ has the same topology as $T_{WKB}^{\vartheta}$, but inverted flow direction, the overall effect of this is to switch the definition of small flat section at each singularity, this amounts to a change in the \emph{decoration} of $T_{WKB}$ also known as the \emph{omnipop}. One can then write down  the \emph{spectrum generating functions} $S_i$, defined by
\ba
	\CX_{\gamma_{i}}^{\vartheta+\pi}={\bf S}\,\CX^{\vartheta}_{\gamma_i}\,=\, S_i\,\CX^{\vartheta}_{\gamma_i}
\ea
by noting that, in the ratio $\CX_{\gamma_{i}}^{\vartheta+\pi}/\CX_{\gamma_{i}}^{\vartheta}$, terms containing the new small flat sections cancel out. As a side note, we stress that the spectrum generating functions only implicitly encode the BPS spectrum, while the factorization of the spectrum generator $\mathbf{S}$ into $\K$ operators ultimately encodes the spectrum in an explicit fashion.

The aim of this paper is to extend this technique to the case of coupled $2d-4d$ systems. In fact, just like gauge charges of the 4d IR theory have a nice geometric interpretation as representatives of $H_1(\Sigma,\mathbb{Z})$, correspondingly charges of $2d$ solitons are described by introducing a set of $\Gamma$-torsors $\Gamma_{ij}$, $i,j\in{\cal V}$ the set of vacua of the defect.\footnote{Physically, the rationale is that an element $\gamma_{ij}$ of one of these torsors represents a $2d$ soliton state carrying some $4d$ gauge charge.} An element of $\Gamma_{ij}$ corresponds then to a representative of the relative homology class of oriented open paths on $\Sigma$, running from $z_{i}$ to $z_{j}$, two of the lifts of $z\in C$ to $\Sigma$, where $z$ is the position of the defect. This interpretation, together with the natural notion of composition of oriented paths, determines whether two charges can be ``added together''. The $2d-4d$ BPS spectrum is studied by introducing an enlarged set of Fock-Goncharov variables $\CY_{a}$\footnote{The $\CY$ obey a \emph{twisted} multiplication rule 
\ba
\CY_{a} \CY_{b} = \left\{  \begin{array}{lr} \sigma(a,b)\CY_{a+b} \qquad & \text{if $a+b$ is defined} \\ 0 & \text{otherwise} \end{array}\right.
\ea
the definition of the twisting function $\sigma(a,b)$ can be found in \S 7 of \cite{GMN4}}, where $a$ is any charge belonging to $\Gamma,\Gamma_{ij}$.\\
The construction of the $\CY$ is analogous to that of the $\CX$ for the pure gauge charges, while we review it below for the other types of charges. Two distinct sets of degeneracies are employed to describe the full 2d-4d spectrum: the $\omega:\Gamma\times\coprod_{i,j}\Gamma_{ij}\to\mathbb{Z}$, satisfying $\omega(\gamma,a+b)=\omega(\gamma,a)+\omega(\gamma,b)$ and the $\mu:\Gamma_{ij}\to \mathbb{Z}$ defined for each $\gamma_{ij},\ i\neq j$. The picture is closely analogous to the 4d one: the spectrum manifests itself through an ordered product of transformations acting on the $\CY$, this time however there are two different types of transformations.
\ba
      {\bf S}=:\prod_\gamma \K_\gamma^\omega\,\prod_{\gamma_{ij}} \mathcal{S}_{\gamma_{ij}}^\mu:
\ea
where the $:\,\,:$ indicate that the product is ordered according to the phases of the central charges of the BPS states involved. 

\subsection{The conjugate section at a singularity}
For later convenience, we now determine explicitly the ``large'' flat section $\tilde s_P$ at a singular point $P\in C$, written in terms of $s_a$ and of the $\CX_a$ in the star-shaped neighborhood of $P$.

\begin{figure}[h!]
\begin{center}
\leavevmode
\includegraphics[width=0.43 \textwidth]{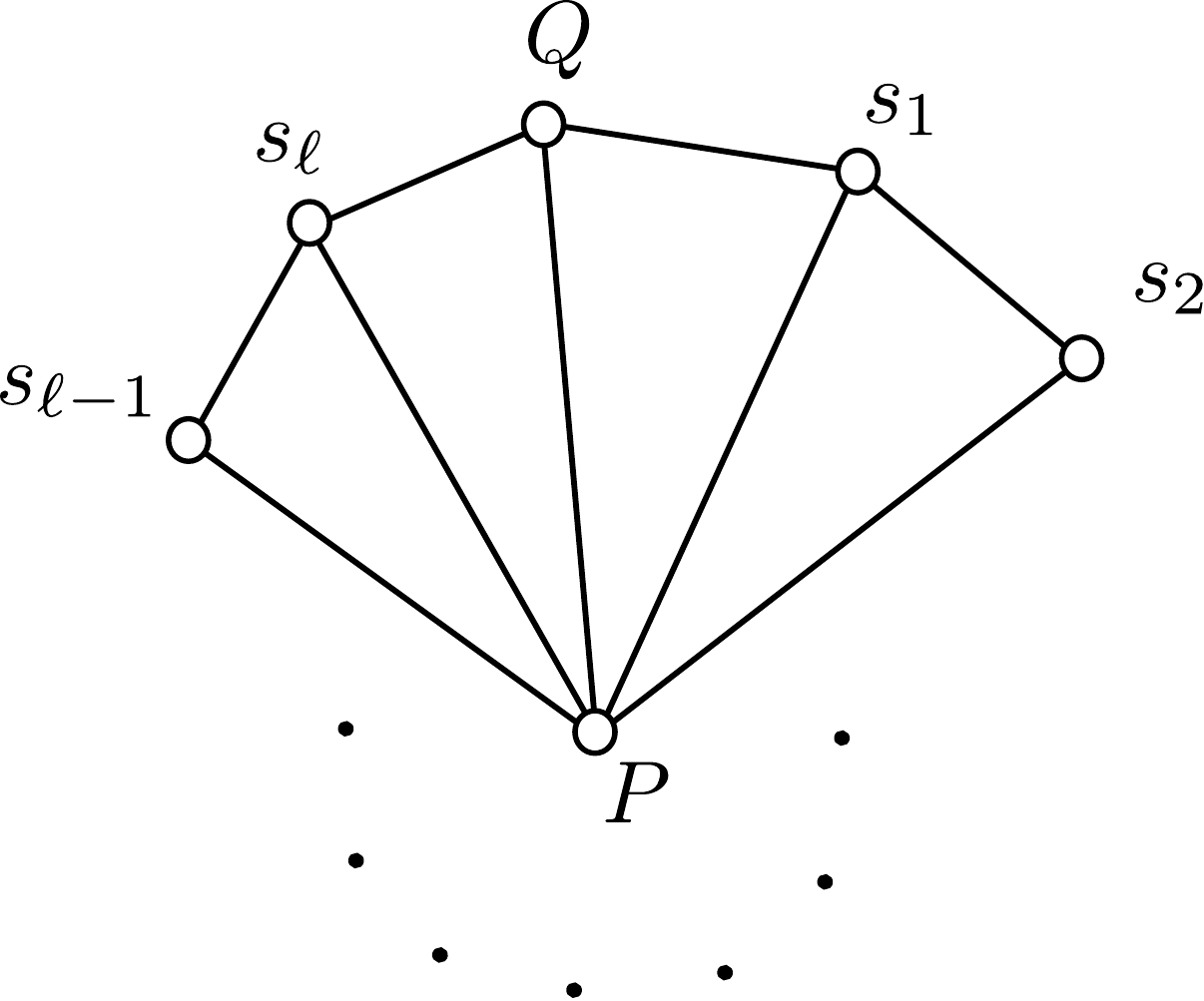}
\end{center}
\caption{The triangulation of $C$ around a generic regular singularity.}\label{fig:regular_singularity}
\end{figure}

Let us begin with the case, shown in fig.\ref{fig:regular_singularity}, in which $P$ is a regular singular point, and every point in its neighborhood is a regular singular point. Define 
\ba
  \Sigma(P;Q\to Q)=1+\CX_{P,\ell}+\CX_{P,\ell}\CX_{P,\ell-1}+\cdots+(\CX_{P,\ell}\cdots\CX_{P,1})
\ea 
this can be expressed \II in terms of flat sections as
\ba
  \Sigma(P; Q \to Q) =   (1- \mu_P^{2}) \frac{ (s_P \wedge s_\ell)(s_Q \wedge \tilde s_P)}{(s_Q \wedge s_\ell)(s_P \wedge \tilde s_P)}. \label{eq:SigmaInSmallFlatSections}
\ea
solving for $\tilde s_P$ we have\footnote{To lighten notation, we use $(s_a,s_b)\equiv(s_a\wedge s_b)$ from here on.}
\ba
	\tilde s_P =  \xi_P \left[   ( \Sigma(P; Q\to Q)  (s_Q,s_\ell)   s_P  -  (1-\mu_P^{2}) (s_P, s_{\ell})   s_Q \right]\label{eq:s_conj}
\ea
where $\xi_P$ is a constant depending on the normalization convention we choose. Both $\Sigma(P;Q\to Q)$ and $\mu_{P}$ have explicit expressions in terms of the $\CX_{a}$, so this is the form of $\tilde s_{P}$ we were after. This has a straightforward extension to the case in which any of $P$ and its neighbors are irregular punctures, we deal with it below.

\subsection{The omnipop for solitonic FG coordinates}\label{sec:soliton_pop}
We focus on the \emph{shortest} representatives of each relative homology class. The spectrum generator for more general ones can be obtained by employing the twisted product law of the $\CY$: writing $\CY_{\gamma^{0}_{ij}+\gamma}$ as a twisted product of coordinates corresponding to the ``simplest'' solitonic charges $\CY_{\gamma^0_{ij}}$ together with purely gauge $\CY_\gamma$. 

\begin{figure}[h!]
\begin{center}
\leavevmode
\includegraphics[width=0.45\textwidth]{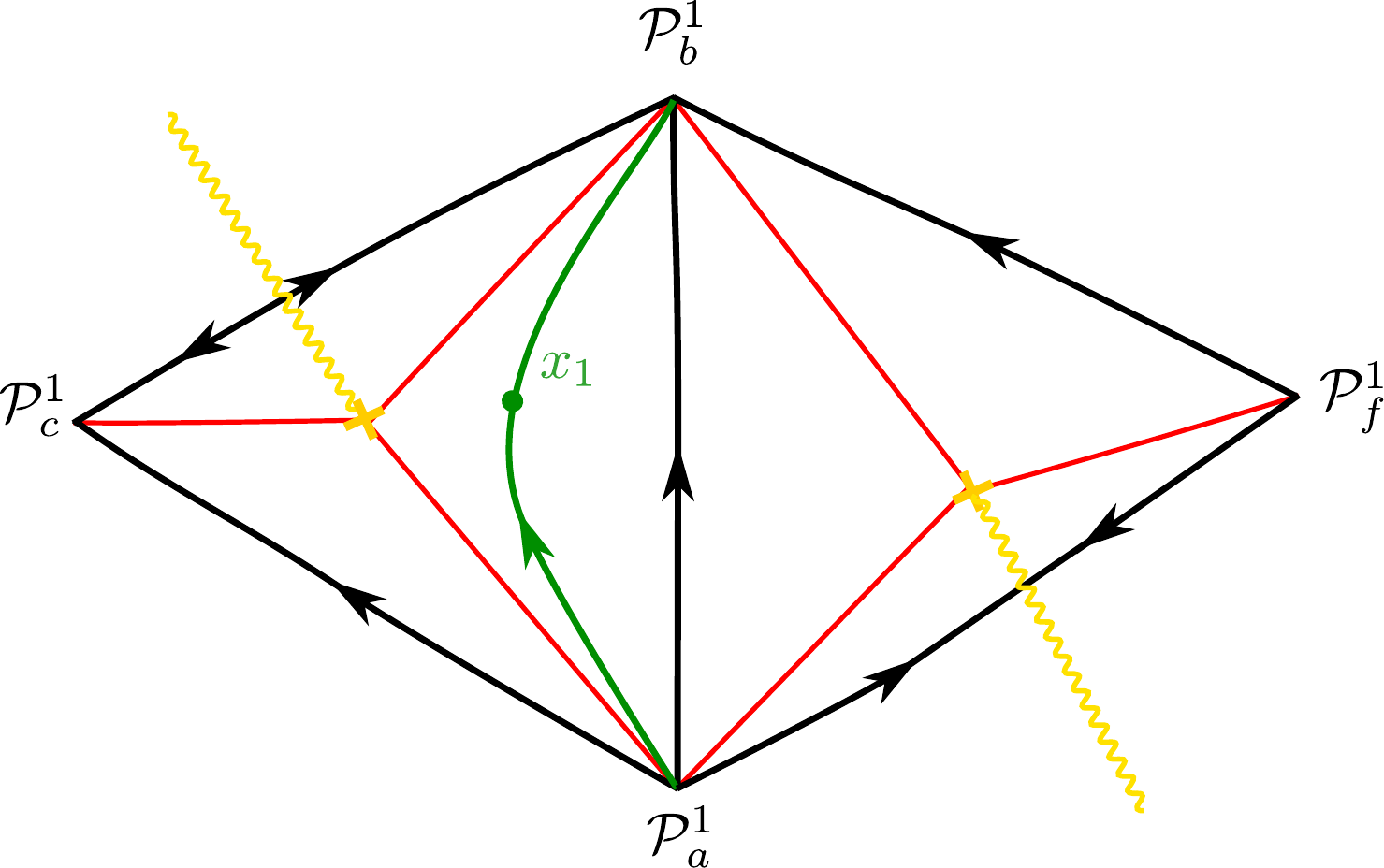}\hspace{0.08\textwidth}\includegraphics[width=0.45\textwidth]{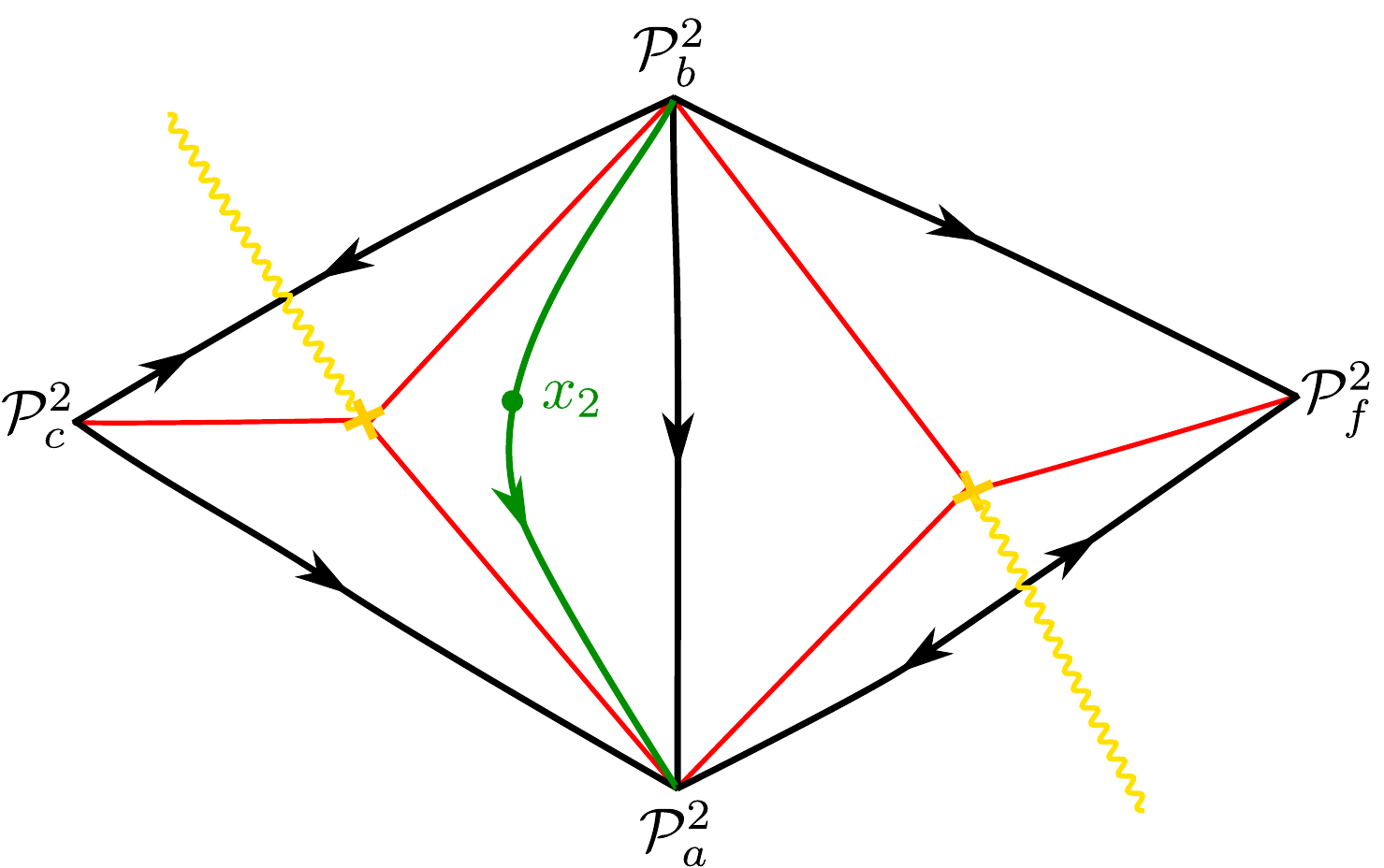}
\end{center}
\caption{A quadrilateral containing $z$, at an angle $\vartheta$: sheet 1 on the left, sheet 2 on the right. The path $\gamma_{12}$ runs from $x_1$ (the lift of $z$ in sheet 1) straight to the turning point inside triangle $abc$ on sheet 1, then back from the turning point to $x_2$ on sheet 2.}\label{fig:scan}
\end{figure}

Any WKB triangulation of $C$ carries a corresponding decomposition into quadrilateral \emph{cells} $C_{ab}$ bounded by four \emph{separating} WKB lines \II, i.e. with vertices consisting of two turning points and two singularities $a$ and $b$, subject to the constraint that there aren't any singularities, nor turning points within the cell. As a first example, let us consider a single surface defect located at $z\in C_{ab}$, within a quadrilateral which vertices are all regular punctures, our goal is to compute the omnipop for the $\CY$ corresponding to a BPS $ij$-soliton. This type of situation was examined in \S7.5.2 of \IV, $T^\vartheta_{\text{WKB}}$ is shown in  Fig.\ref{fig:scan}. 

According to eq.(7.36) of \cite{GMN4}
\ba
	\CY_{\gamma_{12}}=\frac{(s_a,s_c)}{(s_b,s_a)(s_b,s_c)} s_b(z)\otimes s_b(z)
\ea

We briefly review the rules leading to this result.
\subsubsection{Defining the $\CY_{\gamma_{ij}}$}
There are two equivalent definitions of the $\CY$, we quickly review how these coordinates are obtained and set our conventions for the rest of this work. For clarity, we report the pictures of the two sheets of $\Sigma$ in a neighborhood of cell $C_{ab}$, indicating the direction (as given by the sign of $\langle\lambda,\partial_t\rangle\e^{-i\vartheta}$) of WKB lines on each sheet. We use $\P_{\alpha}^{1,2}$ for the lifts of singular points on sheets $1,2$ respectively. Adopting the conventions of \IV: $\gamma_{12}$ is the simplest path from $z$ to the turning point in triangle $abc$, when lifted, it flows from sheet 1 to sheet 2. \\

\noindent {\bf First method}

We begin by applying the methods described in section 7.5.2 of \IV. We first identify a homotopy equivalent to $\gamma_{12}$, made from edges of the WKB triangulation, on $\Sigma$: we choose the path
\beq
  x_{1} \to \P_b^1 \to \P_c^2 \to \P_a^2 \to \P_b^2 \to x_{2} \label{eq:A}
\eeq
we must pass through $\P_b^1$ at the beginning because of the direction of  the WKB line through $x_{1}$ and similarly for $\P_b^2$ at the end because of the direction of the WKB line through $x_{2}$.
We have the equivalence
\beq
  \gamma_{12} \sim \hat E_{b,x_{1}} - \hat E_{b,c} + \hat E_{c,a} - \hat E_{a,b} + \hat E_{b,x_{2}} \label{eq:B}
\eeq
where the signs are dictated by comparison of the direction of the path of our choice with that of the WKB edges employed. The $\hat E$ are oriented lifts of the edges on $C$ and they are understood to be taken on the sheet on which we are working, that is actually specified by eq. (\ref{eq:A}). Equation $(7.27)$ of \cite{GMN4} defines 
\ba
	 \CX_{n,b,b}^\vartheta(z,z) &=\prod_{\alpha,\beta}{(s_{\alpha},s_{\beta})^{n_{\alpha\beta}}\ s_b^\vartheta(z)\otimes s_b^\vartheta(z) }
\ea
where $n_{\alpha\beta}$ is a matrix such that $\gamma_{12}=\sum n_{\alpha\beta}\hat E_{\alpha\beta}$, therefore in our case (\ref{eq:B}) yields
\ba
  \CX_{n,b,b}^\vartheta(z,z) = s_b(z)\otimes s_b(z)\ \frac{(s_c,s_a)}{(s_a,s_b)(s_b,s_c)}\label{eq:C}.
\ea
There is a sign for passing from the untwisted $\CX$ to the twisted $\CY$, which is positive according to the rules outlined in appendix F.1 of \cite{GMN4}. So (\ref{eq:C}) gives $\CY_{\gamma_{12}}$.\\

\noindent {\bf Second method}

As discussed in appendix F in \IV, the proper definition of the $\CY_{a}$ is slightly more involved, we now recall it and then use it to re-derive the result (\ref{eq:C}). This method has two advantages: signs are fixed unambiguously, and the procedure is somewhat faster. We will use this second type of construction throughout the rest of this work. 

As we mentioned above, to each WKB triangulation corresponds a cell decomposition into quadrilaterals. Considering the union of the edges from $T_{WKB}$ and those from the cell decomposition yields a finer decomposition into ``sectors''. For example, in fig.\ref{fig:scan} the sector containing $z$ is the triangle whose vertices are $a,b$ and the branch point on the left, and whose edges are the the generic path (in black) from $a$ to $b$, together with the two separating WKB paths running between the branch point and $a,b$ respectively. \\
More generally, denote by $S$ the sector containing the surface defect. Also, let $z_{i},z_{j}$ be two of the lifts of $z$. Consider a path in $C$ from $z$ to the turning point on the boundary of $S$, and denote by $\gamma_{ij,S}$ the odd sum of its lifts, namely an oriented open path in $\Sigma$ running from $z_{i}$ to $z_{j}$. Let $a$ be the vertex of $T_{WKB}$ reached by flowing along a lifted WKB path from $z_{i}$, and let $(abc)$ be the vertices of the triangle in counterclockwise order, then define $s_{i,S}:=s_{a}\,(s_{b},s_{c})$ and similarly define $s_{j,S}$.
Finally, $\CY_{\gamma_{ij},S}$ is defined to be the fiber endomorphism of the rank-2 bundle over the point $z$ that maps 
\ba
s_{i,S}	\mapsto 0\, ,\qquad s_{j,S}\mapsto \nu_{i,S} s_{i,S}
\ea
with $\nu_{i,S}=+1$ (respectively $-1$) if the lifted generic WKB path through $z_{i}$ runs counterclockwise (clockwise) around the triangle. Letting $\gamma_{ii,0}$ be the element of $\Gamma_{ii}$ corresponding to $0\in\Gamma$, define $\CY_{\gamma_{ii,0}}$ to be the fiber endomorphism
\ba
s_{i}\mapsto s_{i}\, ,\qquad s_{j}\mapsto 0.
\ea
Variables $\CY_{\gamma_{ij}+\gamma}$ and $\CY_{\gamma_{ii,0}+\gamma},\gamma\in\Gamma$, corresponding to more general elements of the $\Gamma$-torsors are obtained via the twisted multiplication laws.
In our specific case, we have
\ba
	s_{1,S}=s_b (s_c,s_a) \qquad s_{2,S}=s_a (s_b,s_c),
\ea
and the fiber endomorphisms 
\ba
	&\CY_{\gamma_{ii}}=\frac{s_{j,S}(z)\otimes s_{i,S}(z)}{(s_{i,S},s_{j,S})} \\
	&\CY_{\gamma_{ij}}=\nu_{i,S}\frac{s_{i,S}(z)\otimes s_{i,S}(z)}{(s_{j,S},s_{i,S})} \label{eq:fiber_endomorphisms}
\ea
where $\nu_1=1=-\nu_2$. Therefore, we have explicitly
\ba
	\CY_{\gamma_{11},0}=\frac{s_{a}(z)\otimes s_{b}(z)}{(s_{b},s_{a})} \qquad & \CY_{\gamma_{22},0}=\frac{s_{b}(z)\otimes s_{a}(z)}{(s_{a},s_{b})}\\
	\CY_{\gamma_{12}}=s_{b}(z)\otimes s_{b}(z) \, \frac{(s_{c},s_{a})}{(s_{a},s_{b})(s_{b},s_{c})} \qquad & \CY_{\gamma_{21}}=s_{a}(z)\otimes s_{a}(z) \, \frac{(s_{b},s_{c})}{(s_{a},s_{b})(s_{c},s_{a})}
\ea
Although the definition of Fock-Goncharov coordinates might appear somewhat unmotivated, this in fact generalizes the $\CX_{\gamma},\gamma\in\Gamma$  in agreement both with the twisted multiplication laws of the $\CY_{a}$, and with the morphisms induced by crossing ${\cal S}$ or ${\cal K}$ walls, details can be found in appendix F of \cite{GMN4}.

\subsubsection{The case of regular punctures}\label{sec:regular}
We now set about deriving the expression for $\tilde \CY_{\gamma_{12}}:=\CY^{\vartheta+\pi}_{\gamma_{12}}$ in terms of the $\CY_a:=\CY_a^{\vartheta}$. To begin with, recall that sending $\vartheta\to\vartheta+\pi$ inverts the direction of the WKB flow, as well as switching decorations at the punctures. The inversion of the WKB flow has different effects on gauge and solitonic charges.  Sending $\vartheta\to\vartheta+\pi$ yields $\gamma^{\vartheta+\pi}=-\gamma^\vartheta$ since, for 4d gauge charges, the orientation of cycles $\gamma$ is defined by the intersection with WKB lines. In contrast, for soliton charges a path $\gamma_{ij}$ is specified to go from sheet $i$ to sheet $j$, within a certain relative homology class, thus its orientation will remain unchanged under an omnipop. As a consequence, the procedure for obtaining the spectrum generator will be slightly different from the one for the pure 4d case. More precisely, for gauge charges one can derive $\mathbf{S}$ by evaluating the transformation \II
\ba
	\CX_{\gamma}^{\vartheta+\pi}=\CX^\vartheta_{\gamma}\,\cdot\,\big( \CX_{E}^{T_{\text{WKB}(\vartheta,\lambda^2)}}\,\tilde\CX_{E}^{T_{\text{WKB}(\vartheta,\lambda^2)}} \big)^{-1}.
\ea 
This result relies on the fact that, under the omnipop, $$\CX_{\gamma^\vartheta_E} \mapsto \CX_{\gamma^{\vartheta+\pi}_E}=\CX_{-\gamma^{\vartheta}_E}=\CX_{\gamma^\vartheta_E}^{-1}$$ 
As we mentioned, this is not the case with solitonic charges: after sending $\vartheta\mapsto\vartheta+\pi$ a charge $\gamma_{ij}$ still runs from sheet $i$ to sheet $j$. Since we can no longer employ this trick, we will instead directly inspect $\tilde\CY_a$. 

\begin{figure}[h!]
\begin{center}
\leavevmode
\includegraphics[width=0.45\textwidth]{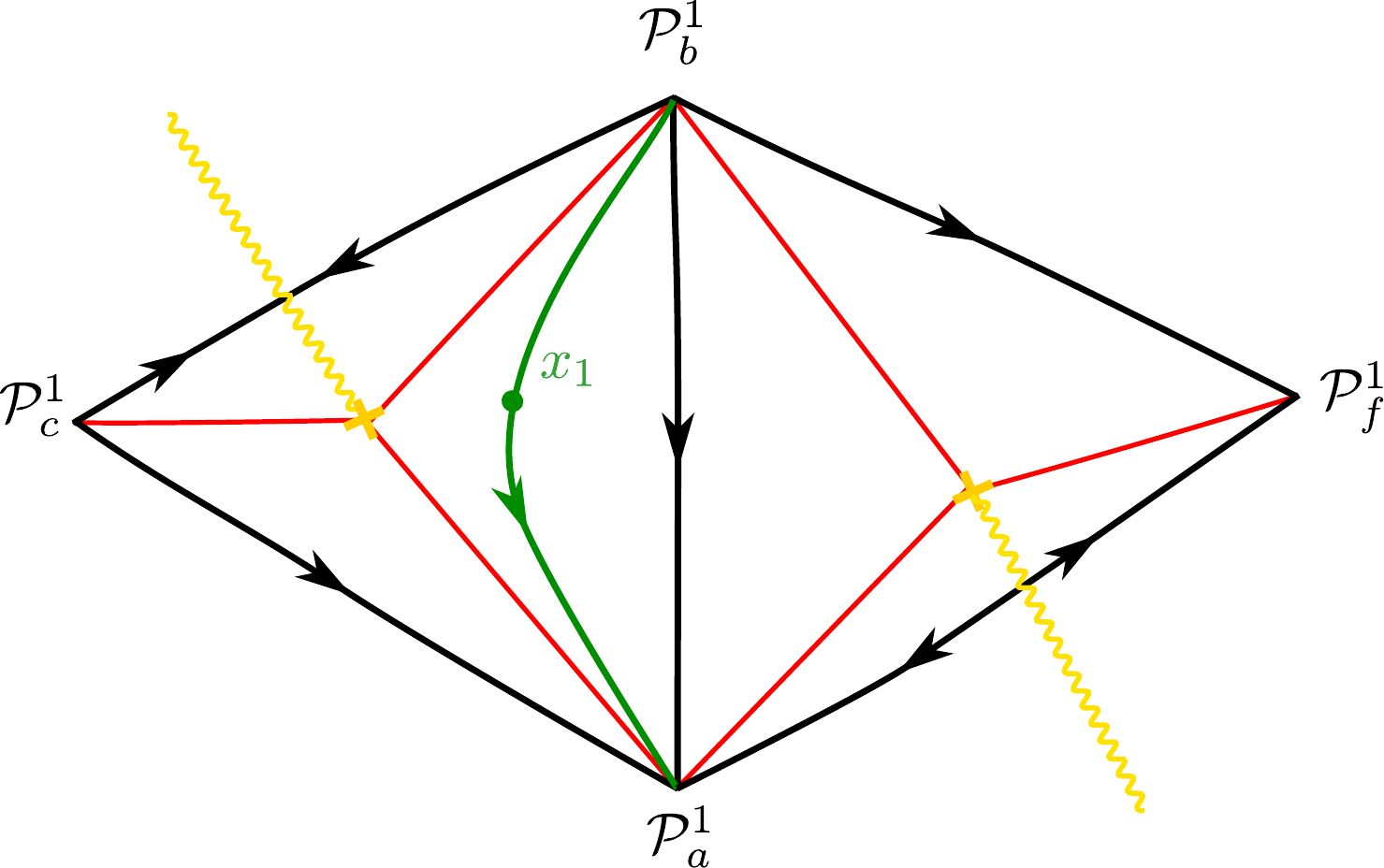}\hspace{0.08\textwidth}\includegraphics[width=0.45\textwidth]{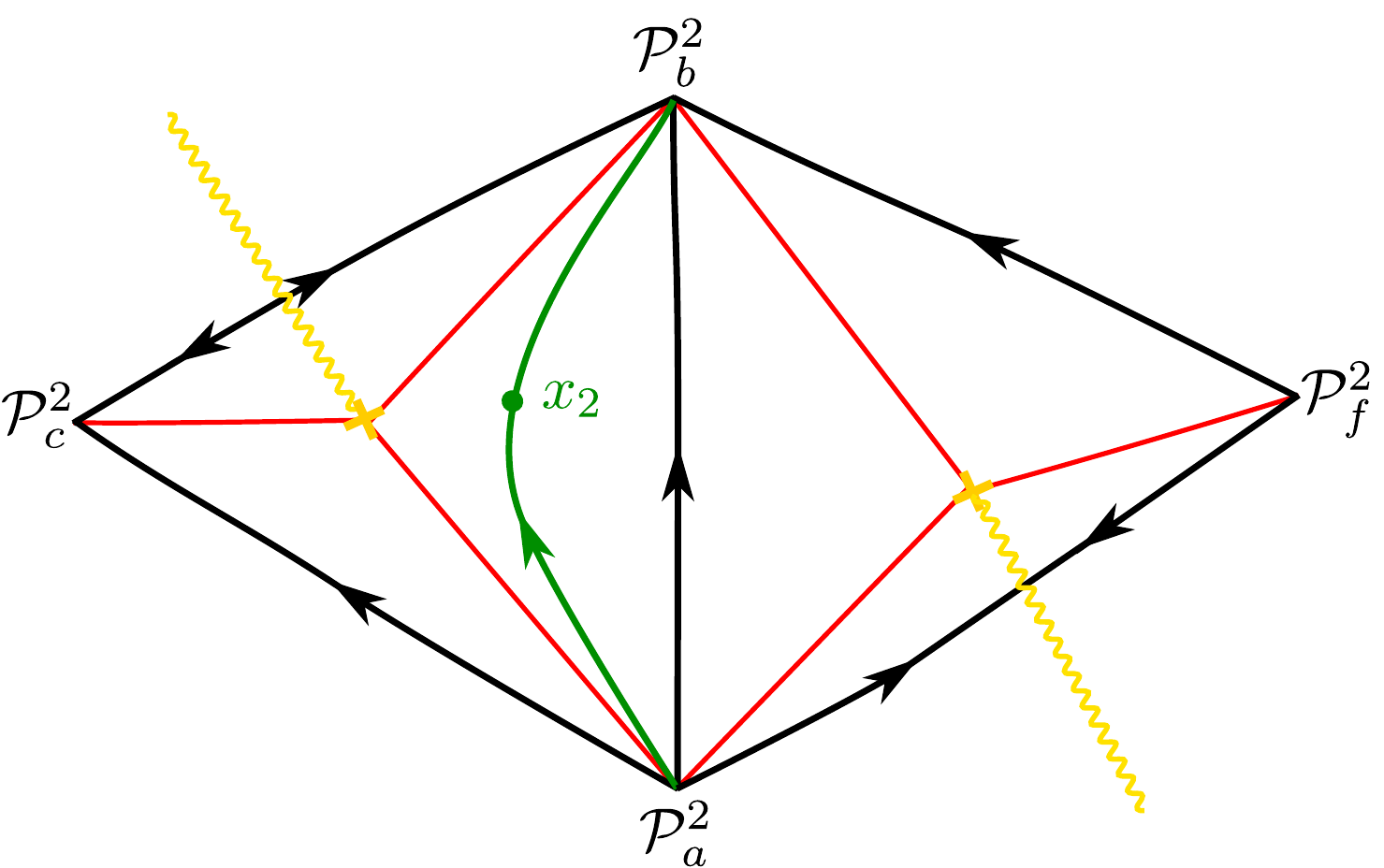}
\end{center}
\caption{A quadrilateral containing $z$, at an angle $\vartheta+\pi$: sheet 1 on the left, sheet 2 on the right. The path $\gamma_{12}$ runs from $x_1$ straight to the turning point in triangle $abc$, on sheet 1, then back from the turning point to $x_2$ on sheet 2.}\label{fig:scan1}
\end{figure}

\begin{figure}[h!]
\begin{center}
\leavevmode
\includegraphics[width=0.60\textwidth]{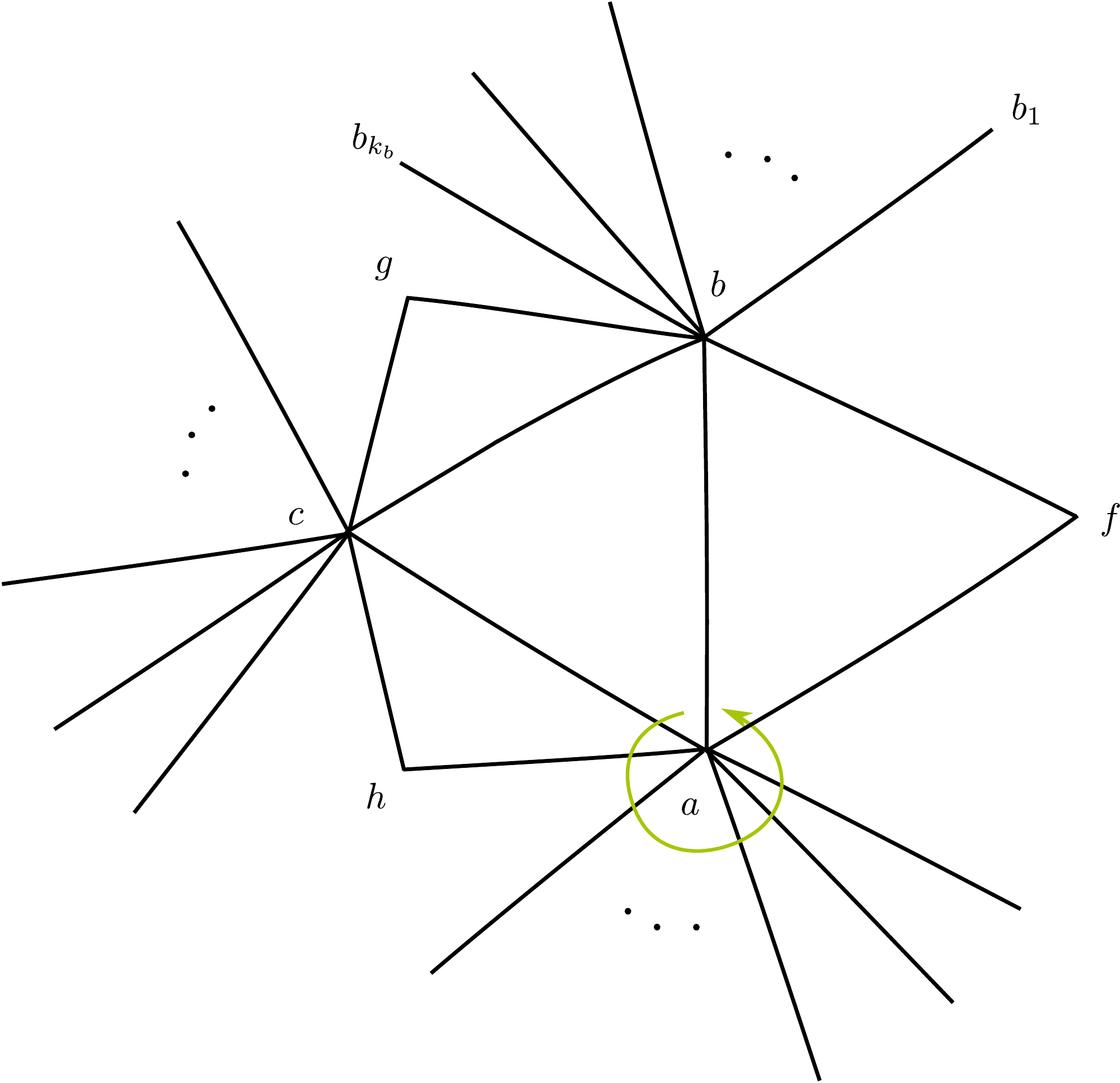}
\end{center}
\caption{Punctures $a,b,c$ are regular, indicated is the path for computing $\Sigma(a;b \to b)$}\label{fig:scan2}
\end{figure}

We begin by applying again the rules in (\ref{eq:fiber_endomorphisms}) to write down $\CY^{\vartheta+\pi}_{\gamma_{12}}$
in terms of the new sections: we now have
\ba
	\tilde s_1(z)=\tilde s_a(z) (\tilde s_b,\tilde s_c) \qquad \tilde s_2(z)=\tilde s_b(z) (\tilde s_c,\tilde s_a) \qquad \tilde\nu_1=-\tilde\nu_2=-
\ea
therefore, applying (\ref{eq:fiber_endomorphisms}) gives
\ba
	\tilde\CY_{\gamma_{12}} = \tilde s_a(z) \otimes \tilde s_a(z)\ \frac{(\tilde s_b,\tilde s_c)}{(\tilde s_a,\tilde s_b)(\tilde s_c,\tilde s_a)} \label{eq:CX_pop}
\ea
The next step is finding an expression for $\tilde\CY_{\gamma_{12}}$ in terms of $\CY_a$. In order to do so, let us consider a neighborhood of triangle $abc$, as shown in Fig.\ref{fig:scan2}, we can employ eq.(\ref{eq:s_conj}) to get the conjugate flat sections:
\ba
	& P \to a, \quad Q \to b, \quad \ell \to c \\
	& \tilde s_a\, = \, \xi_a \left[\Sigma_{a}^{b \to b} (b,c) s_a + (1-\mu_a^{2}) (c,a) s_b \right] \label{eq:apop}
\ea
where we understand the shorthands $\Sigma_{\alpha}^{\beta\to\beta}:=\Sigma(\alpha;\beta\to\beta)$ and $(\alpha,\beta):=(s_\alpha,s_\beta)$. Similarly, we have the other sections by cyclic permutation of the indices
\ba
	& P \to b, \quad Q \to c, \quad \ell \to a \\
	& \tilde s_b\, = \, \xi_b \left[\Sigma_{b}^{c \to c} (c,a) s_b + (1-\mu_b^{2}) (a,b) s_c \right] \\
	& \\
	& P \to c, \quad Q \to a, \quad \ell \to b \\
	& \tilde s_c\, = \, \xi_c \left[\Sigma_{c}^{a \to a} (a,b) s_c + (1-\mu_c^{2}) (b,c) s_a \right] \label{eq:bcpop}
\ea

We first compute
\ba
	\tilde s_a(z) \otimes \tilde s_a(z) &= \xi_a^2\left\{ [\Sigma_{a}^{b \to b} (b,c)]^2\, s_a\otimes s_a + (1-\mu_a^{2})^2 (a,c)^2 s_b\otimes s_b \right.\\
	& \left. + \Sigma_{a}^{b \to b}(1-\mu_a^{2}) (b,c)(c,a) [s_a\otimes s_b + s_b\otimes s_a] \right\},
\ea
from now on, we'll drop the normalization factors $\xi_\alpha$, since they cancel out in eq.(\ref{eq:CX_pop}). In order to compute the other piece of eq.(\ref{eq:CX_pop}) we first evaluate
\ba
	(\tilde s_a, \tilde s_b) &= \left[\Sigma_{a}^{b \to b} \Sigma_{b}^{c \to c} - \Sigma_{a}^{b \to b} (1-\mu_b^{2}) + (1-\mu_a^{2})(1-\mu_b^{2}) \right]\\
	&\times (a,b)(b,c)(c,a), \label{eq:bi}
\ea
together with cyclic permutations of the three indices. 
For later convenience, we define the quantities
\ba
	\Xi(X,Y;x,y)&:=[XY - X(1-y^{2}) + (1-x^{2})(1-y^{2})] \\
	\omega_{a,b,c}&:=(s_a,s_b)(s_b,s_c)(s_c,s_a), \label{eq:Xi_def}
\ea
for each triple of vertices $a,b,c$ of a triangle, labeled counter-clockwise.\\
Equations (\ref{eq:bi}) can be summarized as
\ba
	(\tilde s_a, \tilde s_b)&=\Xi(\Sigma_a^{b \to b},\Sigma_b^{c \to c};\mu_a,\mu_b)\,\omega_{a,b,c} \label{eq:bi_compact}
\ea

In order to avoid confusion below, let us stress here that some care is needed, when using this notation: the $\Sigma$'s appearing into $\Xi$ must be those related to the triangle $abc$, thus e.g. for fig.\ref{fig:line_b}, one \emph{cannot} write $(\tilde s_c, \tilde s_b)=\Xi(\Sigma_c^{b \to b},\Sigma_b^{c \to c};\mu_c,\mu_b)\,\omega_{a,b,c}$, because the clockwise labeling would impose to work in triangle $c,b,g$. One can, of course, use $(\tilde s_c, \tilde s_b)=-(\tilde s_b, \tilde s_c)$ and work on triangle $a,b,c$ instead.

Notice that, since $\omega_{a,b,c}$ is antisymmetric under odd permutations of the indices, in particular $a,b$, then we must have $\Xi(A,B;x,y)=\Xi(B,A;y,x)$ which is \emph{not} trivial from the definition of $\Xi$, but must be justified by studying the properties of the $\Sigma$'s.

From (\ref{eq:bi}) follows
\ba
	\frac{(\tilde s_b,\tilde s_c)}{(\tilde s_a,\tilde s_b)(\tilde s_c,\tilde s_a)} & = %
	\frac{1}{\omega_{a,b,c}}\ \Xi(\Sigma_b^{c \to c},\Sigma_c^{a \to a};\mu_b,\mu_c) \\
	& \times \left[\Xi(\Sigma_a^{b \to b},\Sigma_b^{c \to c};\mu_a,\mu_b)\,\Xi(\Sigma_c^{a \to a},\Sigma_a^{b \to b};\mu_c,\mu_a) \right]^{-1}
\ea
The factors $\Xi$ have well defined expressions in terms of $\CY_\gamma$: see eq.(11.9) of \II. Eventually, we come to the explicit expression for $\tilde\CY_{\gamma_{12}}$
\ba
	\tilde\CY_{\gamma_{12}} &= \frac{1}{\omega_{a,b,c}}\ \Xi(\Sigma_b^{c \to c},\Sigma_c^{a \to a};\mu_b,\mu_c) \\
	& \times \left[\Xi(\Sigma_a^{b \to b},\Sigma_b^{c \to c};\mu_a,\mu_b)\,\Xi(\Sigma_c^{a \to a},\Sigma_a^{b \to b};\mu_c,\mu_a) \right]^{-1}\\
	& \times \left\{ [\Sigma_{a}^{b \to b} (b,c)]^2\, s_a\otimes s_a + (1-\mu_a^{2})^2 (a,c)^2 s_b\otimes s_b \right.\\
	& \left. + \Sigma_{a}^{b \to b} (1-\mu_a^{2}) (b,c)(c,a) [s_a\otimes s_b + s_b\otimes s_a] \right\} \\
	& = \Xi(\Sigma_b^{c \to c},\Sigma_c^{a \to a};\mu_b,\mu_c) \\
	& \times \left[\Xi(\Sigma_a^{b \to b},\Sigma_b^{c \to c};\mu_a,\mu_b)\,\Xi(\Sigma_c^{a \to a},\Sigma_a^{b \to b};\mu_c,\mu_a) \right]^{-1}\\
	& \times \left\{(\Sigma_{a}^{b \to b})^2 \CY_{\gamma_{21}} + (1-\mu_a^{2})^2 \CY_{\gamma_{12}} + \Sigma_{a}^{b \to b}(1-\mu_a^{2}) \left[ \CY_{\gamma_{22}=0} - \CY_{\gamma_{11}=0}\right] \right\} \label{eq:CY_pop_reg}
\ea
Similar, tedious but straightforward, calculations give
\ba
	\tilde\CY_{\gamma_{11=0}}&=\frac{\tilde s_b(z) \otimes \tilde s_a(z)}{(\tilde s_a,\tilde s_b)} \\
	&=\left[\Xi(\Sigma_a^{b \to b},\Sigma_b^{c \to c};\mu_a,\mu_b)\right]^{-1}\\
	&\times \left\{\Sigma_{a}^{b \to b}\Sigma_{b}^{c \to c}\CY_{\gamma_{22}=0} + \Sigma_b^{c \to c}(1-\mu_a^{2}) \CY_{\gamma_{12}} \right.\\
	&\left.- \Sigma_{a}^{b \to b}(1-\mu_b^{2}) \left[ \CY_{\gamma_{22}=0} + \CY_{\gamma_{21}}\right] - (1-\mu_a^{2})(1-\mu_b^{2}) \left[ \CY_{\gamma_{12}} - \CY_{\gamma_{11}=0}\right]\right\} \label{eq:CY_pop_reg_bis}\\
	\tilde\CY_{\gamma_{22=0}}&=\frac{\tilde s_a(z) \otimes \tilde s_b(z)}{(\tilde s_b,\tilde s_a)} \\
	&= -\left[\Xi(\Sigma_a^{b \to b},\Sigma_b^{c \to c};\mu_a,\mu_b)\right]^{-1}\\
	&\times \left\{-\Sigma_{a}^{b \to b}\Sigma_{b}^{c \to c}\CY_{\gamma_{11}=0} + \Sigma_b^{c \to c}(1-\mu_a^{2}) \CY_{\gamma_{12}} \right.\\
	&\left.- \Sigma_{a}^{b \to b}(1-\mu_b^{2}) \left[\CY_{\gamma_{21}} - \CY_{\gamma_{11}=0}\right] - (1-\mu_a^{2})(1-\mu_b^{2}) \left[ \CY_{\gamma_{12}} + \CY_{\gamma_{22}=0}\right]\right\}
	& \\
	\tilde\CY_{\gamma_{21}}&=\tilde s_b(z) \otimes \tilde s_b(z)\frac{(\tilde s_c,\tilde s_a)}{(\tilde s_a,\tilde s_b)(\tilde s_b,\tilde s_c)}  \\
	& = \Xi(\Sigma_c^{a \to a},\Sigma_a^{b \to b};\mu_c,\mu_a) \\
	& \times \left[\Xi(\Sigma_a^{b \to b},\Sigma_b^{c \to c};\mu_a,\mu_b)\,\Xi(\Sigma_b^{c \to c},\Sigma_c^{a \to a};\mu_b,\mu_c) \right]^{-1}\\
	& \times \left\{(\Sigma_{b}^{c \to c})^2 \CY_{\gamma_{12}} - \Sigma_b^{c \to c} (1-\mu_b^{2}) \left[ 2 \CY_{\gamma_{12}} + \CY_{\gamma_{22}=0} - \CY_{\gamma_{11}=0} \right] \right. \\
	& \left. + (1-\mu_b^{2})^2 \left[ \CY_{\gamma_{12}} + \CY_{\gamma_{21}} + \CY_{\gamma_{22}=0} - \CY_{\gamma_{11}=0}\right] \right\} 	\nonumber
\ea

\subsubsection{Extension to irregular punctures: one irregular puncture}\label{sec:irregular_single}
Suppose $b,c$ are regular punctures and $a$ is now irregular. A pop acts in two combined ways on the decorated triangulation at the irregular puncture, as explained in \S8 of \II:
\begin{itemize}
 \item on the decoration -- a pop can be regarded as the action by 1 on the $\mathbb{Z}$ torsor of decorations at the irregular puncture: labeling vertices $\dots Q_j,Q_{j+1}\dots$ clockwise with corresponding decorations $\dots s_n,s_{n+1}\dots$, then after sending $\vartheta \to \vartheta+\pi$ the decorations associated to vertices will be $\dots s_{n-1},s_{n}\dots$.
 \item on the vertices -- a pop acts as a cyclic permutation of the vertices associated with an irregular puncture, this is easy to see e.g. in AD theories, where the irregular puncture is at infinity: here we can actually follow the evolution of the triangulation as $\vartheta\mapsto\vartheta+\pi$, we see the WKB rays rotating counterclockwise by an angle $2\pi/(N+2)$
\end{itemize}
The overall effect is a combination of these two, they don't add up, rather, they describe the same behavior.

\begin{figure}[bth]
\begin{center}
\leavevmode
\includegraphics[width=0.35\textwidth]{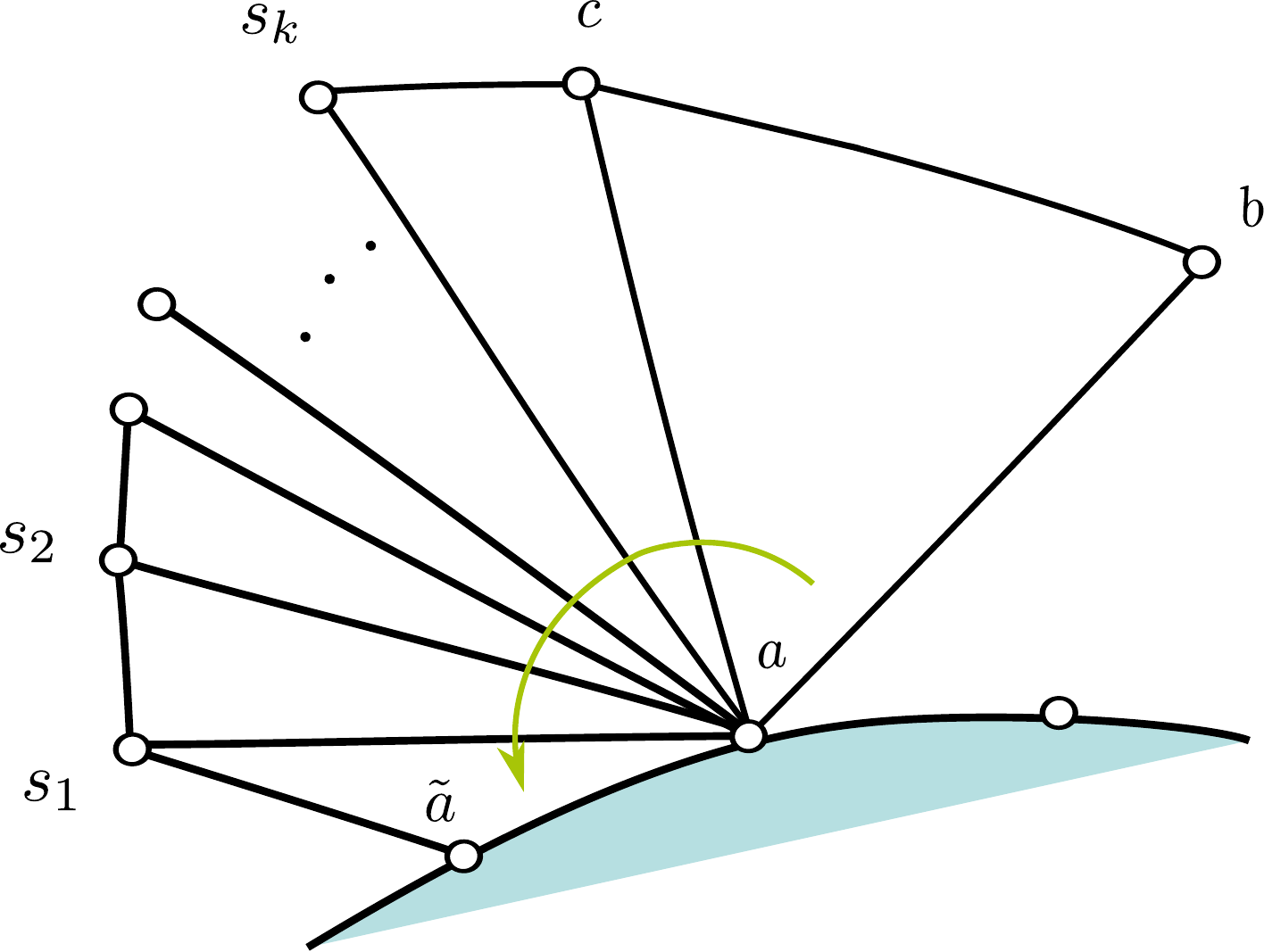}\hspace{.08\textwidth}\includegraphics[width=0.35\textwidth]{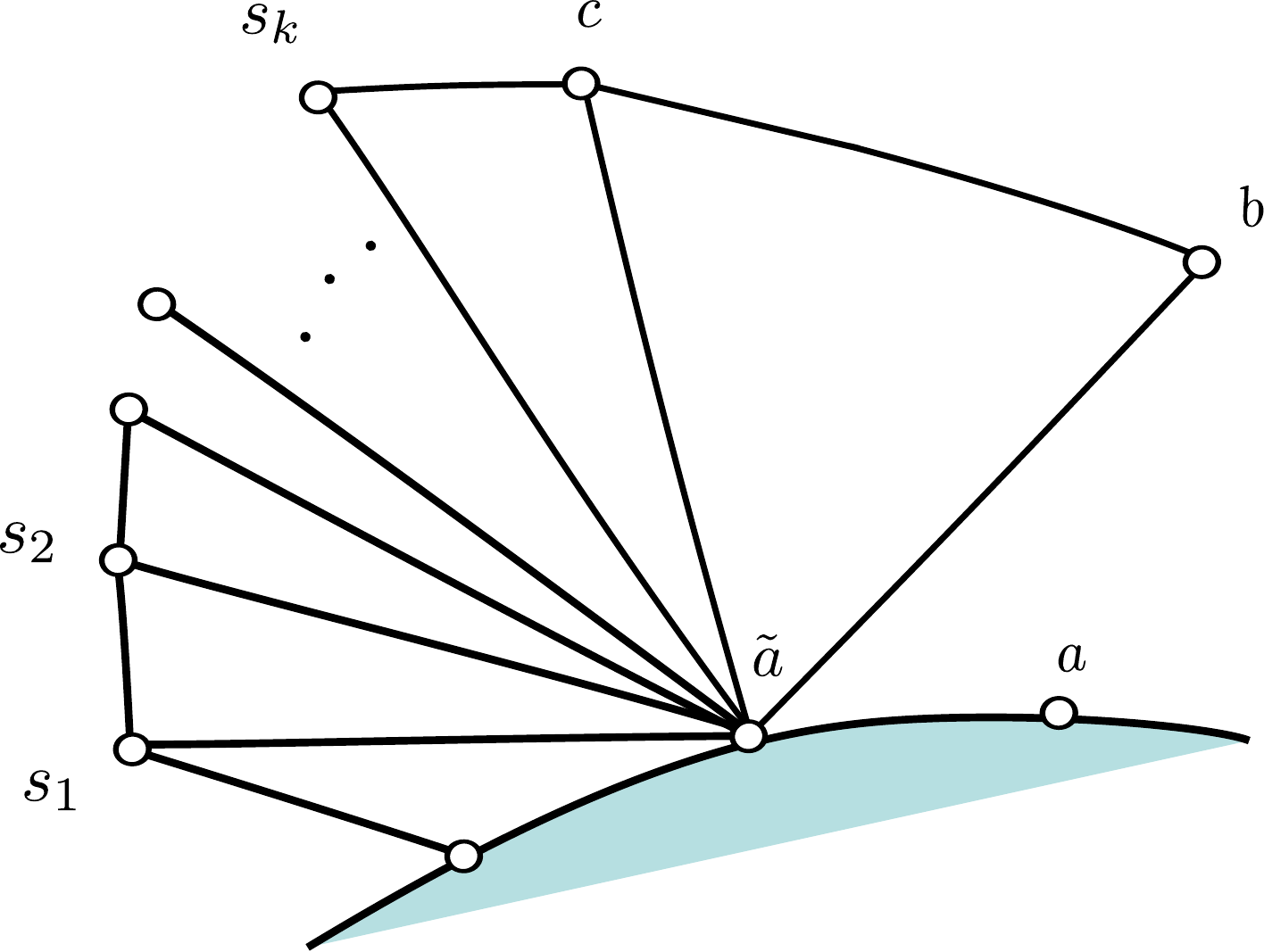}
\end{center}
\caption{The defect's parameter $z$ sits in cell $C_{ab}$, within a triangle whose vertex $a$ is an irregular puncture. Indicated in yellow is the path for parallel transport, employed in constructing $\Sigma_a^{b\to\tilde a}$}\label{fig:irr_a}
\end{figure}

Therefore, referring to fig.\ref{fig:irr_a}, equation (\ref{eq:CX_pop}) still holds. The rules for expressing $\tilde s_b, \tilde s_c$ don't change, while we have $\tilde s_a=s_{\tilde a}$.

The suitable generalization of (\ref{eq:SigmaInSmallFlatSections}) is (see \II)
\ba
	\Sigma_a^{b \to \tilde a}=\frac{(s_{\tilde a},s_b)(s_a,s_c)}{(s_b,s_c)(s_{\tilde a},s_a)} \label{eq:irr_identity}
\ea
from which we derive\footnote{In the case of N=1 AD theory, this result together with identity (\ref{eq:section_identity}) gives the expected $\tilde s_a=s_c$.}
\ba
	s_{\tilde a}=\xi_{\tilde a}[\Sigma_a^{b\to\tilde a}\,(b,c)\, s_a\,+\,(c,a)\,s_b ] \label{eq:irr_pop}
\ea
Notice that this corresponds to making the replacements $\Sigma_a^{b\to b}\to \Sigma_a^{b\to\tilde a},\ \mu^{2}_a\to 0$ into eq (\ref{eq:apop}). Therefore, we can immediately write down the outer products of sections at popped vertices, in a way that is analogous to (\ref{eq:bi_compact})
\ba
	(\tilde s_a,\tilde s_b)&=\Xi(\Sigma_a^{b\to\tilde a},\Sigma_b^{c\to c};0,\mu_b)\,\omega_{a,b,c}\\
	(\tilde s_c,\tilde s_a)&=\Xi(\Sigma_c^{a\to a},\Sigma_a^{b \to \tilde a};\mu_c,0)\,\omega_{a,b,c}\\
\ea
similarly, computing again $s_{\tilde a}\otimes s_{\tilde a}$ just involves making the above-mentioned replacements.
Eventually, we have the new expression for $\tilde \CY_{\gamma_{12}}$
\ba
	\tilde\CY_{\gamma_{12}} &= \Xi(\Sigma_b^{c \to c},\Sigma_c^{a \to a};\mu_b,\mu_c) \\
	& \times \left[\Xi(\Sigma_a^{b \to \tilde a},\Sigma_b^{c \to c};0,\mu_b)\,\Xi(\Sigma_c^{a \to a},\Sigma_a^{b \to \tilde a};\mu_c,0) \right]^{-1}\\
	& \times \left\{(\Sigma_{a}^{b \to \tilde a})^2 \CY_{\gamma_{21}} + \CY_{\gamma_{12}} + \Sigma_{a}^{b \to \tilde a} \left[ \CY_{\gamma_{22}=0} - \CY_{\gamma_{11}=0}\right] \right\}
\ea
Similar expressions for $\tilde\CY_{\gamma_{21}},\, \tilde\CY_{\gamma_{11}=0},\, \tilde\CY_{\gamma_{22}=0}$ are obtained after applying the proper substitutions.

\subsubsection{Extension to irregular punctures: general case}\label{sec:irregular_general}
\begin{figure}[h!]
\begin{center}
\leavevmode
\includegraphics[width=0.60\textwidth]{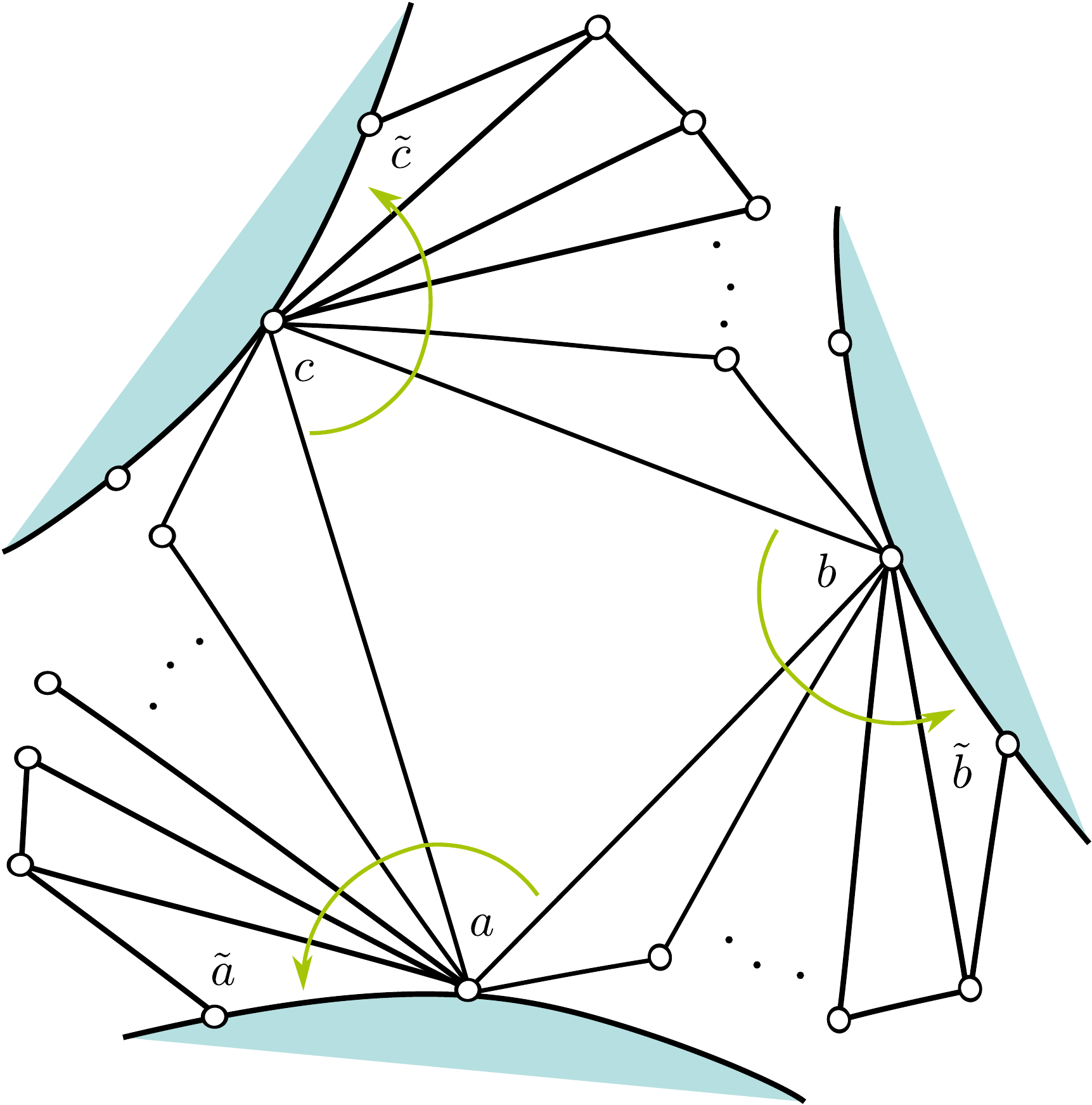}
\end{center}
\caption{The defect's parameter $z$ sits in cell $C_{ab}$, within a triangle whose vertices are all irregular punctures}\label{fig:irr_abc}
\end{figure}
It is straightforward to extend the above reasoning to the case in which any combination of $a,b,c$ are regular or irregular. If a vertex $\alpha$ is irregular, one must apply the corresponding  replacements: $\Sigma_\alpha^{\beta\to\beta}\to \Sigma_\alpha^{\beta\to\tilde \beta}$ and $\mu^{2}_\alpha\to 0$, into expression (\ref{eq:CY_pop_reg}), where $\alpha,\beta,\gamma$ is an ordered triple valued in $\{a,b,c\}$.

\subsection{Including line defects}\label{sec:including_line_defects}
We now want to derive the spectrum generator in presence of line defects. The simplest case of a single line defect will be considered. 

\subsubsection{Defects in the same cell, different sectors}\label{sec:same_cell}
\begin{figure}[h]
\begin{center}
\leavevmode
\includegraphics[width=0.55\textwidth]{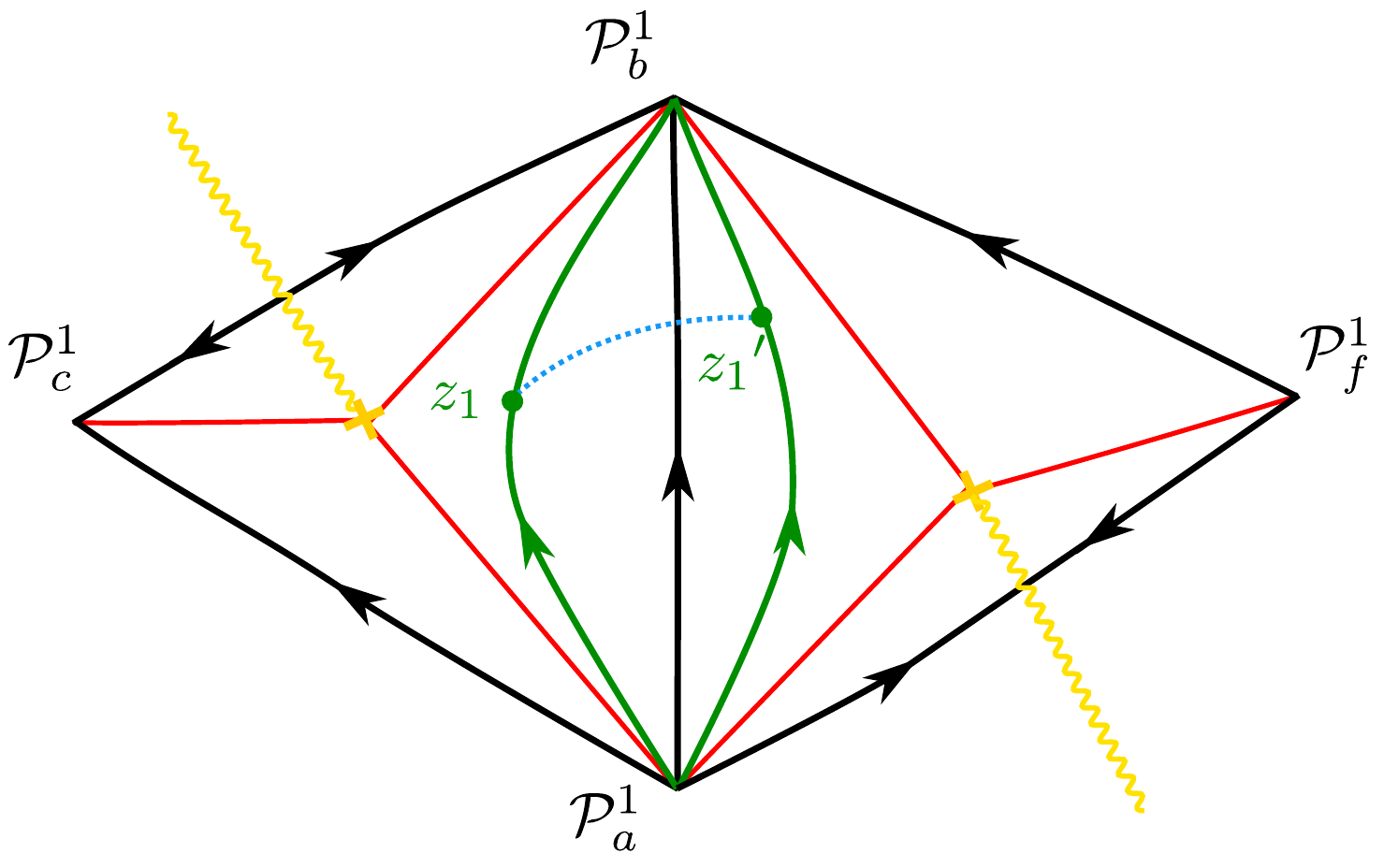}
\end{center}
\caption{Sheet 1, at angle $\vartheta$}\label{fig:line_a}
\end{figure}
Let's start with the situation of fig.\ref{fig:line_a}: both $z,z'$ lie within cell $C_{ab}$. We first assume all punctures to be regular, and later generalize.

In analogy with what we have seen so far, we define the WKB coordinates corresponding to framed BPS states as follows. Let $S,S'$ be the sectors\footnote{a sector is a subset of $C$ bounded by a WKB edge and two separatrices} in which $z,z'$ lie respectively. On sheet $i$ we define $s_{i,S}(z)=s_1(z)\,(s_2,s_3)$ ($s_{i',S'}(z')=s_1(z')\,(s_2,s_3)$) where $s_1(z)$ ($s_1(z')$) is the small flat section of the vertex into which the WKB line through $z$ ($z'$) flows, evaluated at $z$ ($z'$). The triple $123$ indicates a counterclockwise labeling of the vertices of the triangles containing $z,\, z'$. 

We define the bundle morphisms (understanding $i=i'\neq j=j'$)
\ba
	&\CY_{ii'}:\left\{\begin{array}{l} s_i(z) \to s_{i'}(z') \\ s_j \to 0 \end{array}\right.\quad = \frac{s_j(z) \otimes s_{i'}(z')}{(s_{i'},s_j)} \\
	&\CY_{ij'}:\left\{\begin{array}{l} s_i(z) \to 0 \\ s_j(z) \to \nu_i\,s_{i'}(z')  \end{array}\right.\quad = \nu_i\,\frac{s_i(z) \otimes s_{i'}(z')}{(s_j,s_{i'})} \label{eq:line_morphisms}
\ea
in close analogy to the reasoning in appendix F of \IV; we omitted the subscripts indicating the cells, since such information is specified by whether we evaluate at $z$ or $z'$. The sign $\nu_{i,S}$ is positive if the WKB line through $s_i(z)$ flows counterclockwise within the triangle containing $z$, negative otherwise: it really depends on the sector, not just on the cell, of $z$.

A remark is in order here: this definition is not explicitly stated in \IV, but it follows naturally by extending what is defined in appendix F of that reference for the fiber endomorphisms (see eq (\ref{eq:fiber_endomorphisms})), along with enforcing normalization invariance. However, in eq. (8.11) of \IV, in order to get $\CY_{-+'}$, the second line of (\ref{eq:line_morphisms}) must be changed to 
\ba
    \CY_{ij'} = \nu_j\,\frac{s_j(z) \otimes s_{j'}(z')}{(s_i,s_{j'})} \label{eq:change}
\ea
leaving the first line unchanged. Eventually, all the spectra we will derive in the rest of this work will match with those of cases analyzed in \IV, provided we make the proper identifications.

Therefore in our case we identify
\ba
	s_1(z)=s_b(z)\,(s_c,s_a) \qquad & s_{1'}(z')=s_b(z')\,(s_a,s_f) \\
	s_2(z)=s_a(z)\,(s_b,s_c) \qquad & s_{2'}(z')=s_a(z')\,(s_f,s_b) \label{eq:AAA}
\ea
together with the signs $\nu_1=-\nu_{1'}=-\nu_2=\nu_{2'}=+$.

Consider the two ``simplest paths'' (not crossing any separatrices) between $z$ and $z'$, lying entirely on sheets 1 and 2, to them associate respectively: $\CY^\vartheta_{11'}\equiv\CY_{11'},\, \CY^\vartheta_{22'}\equiv\CY_{22'}$; similarly, the ``simplest'' paths from sheet 1 to 2, and vice versa, are $\CY^\vartheta_{12'}\equiv\CY_{12'}$, $\CY^\vartheta_{21'}\equiv\CY_{21'}$. They read
\ba
	\CY_{11'}=\frac{s_a(z)\otimes s_b(z')}{(s_b,s_a)} \qquad & \CY_{22'}=\frac{s_b(z)\otimes s_a(z')}{(s_a,s_b)}, \\
	\CY_{12'}=s_b(z)\otimes s_b(z')\,\frac{(s_c,s_a)}{(s_a,s_b)(s_b,s_c)} \qquad &	\CY_{21'}=s_a(z)\otimes s_a(z')\,\frac{(s_b,s_c)}{(s_a,s_b)(s_c,s_a)}.
\ea
After sending $\vartheta\to\vartheta+\pi$, we have the new quantities
\ba
	\tilde s_1(z)=\tilde s_a(z)\,(\tilde s_b,\tilde s_c) \qquad & \tilde s_{1'}(z')=\tilde s_a(z')\,(\tilde s_f,\tilde s_b)  \\
	\tilde s_2(z)=\tilde s_b(z)\,(\tilde s_c,\tilde s_a) \qquad & \tilde s_{2'}(z')=\tilde s_b(z')\,(\tilde s_a,\tilde s_f) \label{eq:AAB}
\ea
together with the new signs $\tilde \nu_1=-\tilde \nu_{1'}=-\tilde \nu_2=\tilde \nu_{2'}=-$.

Applying again definitions (\ref{eq:line_morphisms}), the new coordinates read
\ba
  \CY_{11'}^{\vartheta+\pi}&=:\tilde\CY_{11'}=\frac{\tilde s_b(z)\otimes \tilde s_a(z')}{(\tilde s_a,\tilde s_b)},\\ 
  \CY_{22'}^{\vartheta+\pi}&=:\tilde\CY_{22'}=\frac{\tilde s_a(z)\otimes \tilde s_b(z')}{(\tilde s_b,\tilde s_a)}\\
  \CY_{12'}^{\vartheta+\pi}&=:\tilde\CY_{12'}=\tilde s_a(z)\otimes \tilde s_a(z')\,\frac{(\tilde s_b,\tilde s_c)}{(\tilde s_a,\tilde s_b)(\tilde s_c,\tilde s_a)},\\ 
  \CY_{21'}^{\vartheta+\pi}&=:\tilde\CY_{21'}=\tilde s_b(z)\otimes \tilde s_b(z')\,\frac{(\tilde s_c,\tilde s_a)}{(\tilde s_a,\tilde s_b)(\tilde s_b,\tilde s_c)}.
\ea
Employing eq.s (\ref{eq:apop})(\ref{eq:bcpop}) and (\ref{eq:bi_compact}) we obtain
\ba
  \tilde\CY_{11'} &= \left[ \Xi(\Sigma_{a}^{b \to b},\Sigma_{b}^{c \to c}; \mu_a,\mu_b )\right]^{-1}\\
  &\times \Big\{ \Sigma_{a}^{b \to b} \Sigma_{b}^{c \to c}\, \CY_{22'}^{\vartheta} \, + \,\Sigma_b^{c\to c}(1-\mu_a^{2}) \, \CY_{12'}^\vartheta  \\
  &+ (1-\mu_a^{2})(1-\mu_b^{2}) \,\frac{s_c(z) \otimes s_b(z')}{(s_b,s_c)}\,+\, \Sigma_a^{b\to b} (1-\mu_b^{2})\,\frac{s_c(z) \otimes s_a(z')}{(s_c,s_a)} \Big\}
\ea
In order to fix the second row, \ie $\ $to eliminate $s_c(z)$, we employ the identity 
\ba
  s_a(z)\,(s_b,s_c) \,+\, s_b(z)\,(s_c,s_a) \,+\, s_c(z)\,(s_a,s_b)\,=\,0 \label{eq:section_identity}
\ea
and get
\ba
  	\tilde\CY_{11'}&=\left[ \Xi(\Sigma_{a}^{b \to b},\Sigma_{b}^{c \to c}; \mu_a,\mu_b )\right]^{-1}\\
	&\times \left\{ \Sigma_{a}^{b \to b} \Sigma_{b}^{c \to c}\, \CY_{22'} \, + \,\Sigma_b^{c\to c}(1-\mu_a^{2}) \, \CY_{12'} \right. \\
  	&- (1-\mu_a^{2})(1-\mu_b^{2}) \,\left[\CY_{12'} -  \CY_{11'} \right]\\
	&-\left. \Sigma_a^{b\to b} (1-\mu_b^{2})\,\left[ \CY_{21'} + \CY_{22'} \right] \right\} \label{eq:CY_pop_line_reg}
\ea
Similar, tedious but straightforward, calculations show that
\ba
	\tilde\CY_{22'}&=-\left[ \Xi(\Sigma_{a}^{b \to b},\Sigma_{b}^{c \to c}; \mu_a,\mu_b )\right]^{-1}\\
	&\times \left\{ -\Sigma_{a}^{b \to b} \Sigma_{b}^{c \to c}\, \CY_{11'} \, + \,\Sigma_b^{c\to c}(1-\mu_a^{2}) \, \CY_{12'} \right. \\
  	&- (1-\mu_a^{2})(1-\mu_b^{2}) \,\left[\CY_{12'} +  \CY_{22'} \right]\\
	&-\left. \Sigma_a^{b\to b} (1-\mu_b^{2})\,\left[ \CY_{21'} - \CY_{11'} \right] \right\} \label{eq:CY_pop_line_reg_bis}
\ea
\ba
	\tilde\CY_{12'}& = \Xi(\Sigma_b^{c \to c},\Sigma_c^{a \to a};\mu_b,\mu_c) \\
	& \times \left[\Xi(\Sigma_a^{b \to b},\Sigma_b^{c \to c};\mu_a,\mu_b)\,\Xi(\Sigma_c^{a \to a},\Sigma_a^{b \to b};\mu_c,\mu_a) \right]^{-1}\\
	& \times \left\{(\Sigma_{a}^{b \to b})^2 \CY_{21'} + (1-\mu_a^{2})^2 \CY_{12'} + \Sigma_{a}^{b \to b}(1-\mu_a^{2}) \left[ \CY_{22'} - \CY_{11'}\right] \right\} \\
	& \\
	\tilde\CY_{21'}& = \Xi(\Sigma_c^{a \to a},\Sigma_a^{b \to b};\mu_c,\mu_a) \\
	& \times \left[\Xi(\Sigma_a^{b \to b},\Sigma_b^{c \to c};\mu_a,\mu_b)\,\Xi(\Sigma_b^{c \to c},\Sigma_c^{a \to a};\mu_b,\mu_c) \right]^{-1}\\
	& \times \left\{(\Sigma_{b}^{c \to c})^2 \CY_{12'} - \Sigma_b^{c \to c} (1-\mu_b^{2}) \left[ 2 \CY_{12'} + \CY_{22'} - \CY_{11'} \right] \right. \\
	& \left. + (1-\mu_b^{2})^2 \left[ \CY_{12'} + \CY_{21'} + \CY_{22'} - \CY_{11'}\right] \right\}  \nonumber
\ea
\vspace{6pt}\\

\noindent{\bf Extension to irregular punctures} 

If one or more of the vertices $a,b,c,f$ belong to an \emph{irregular puncture}, we obtain the new spectrum generator by using the analogue of eq.(\ref{eq:irr_pop}) in place of (\ref{eq:apop}), (\ref{eq:bcpop}), ultimately this amounts to performing the substitutions mentioned in subsection \ref{sec:irregular_general}.

\subsubsection{Defects in same sector} \label{sec:same_sector}
Suppose $z,z'$ are in the same sector of cell $C_{ab}$, mark a point $z''$ in the \emph{other} sector of cell $C_{ab}$: then use the twisted product law of the $\CY_a$, to glue together paths $z\to z''$ and $z'' \to z'$.

For future convenience, we point out a few features here. Looking back at last section, one might notice that the only changes come from substituting $s_{1'}=s_b(z')(s_c,s_a),\,s_{2'}=s_a(z')(s_b,s_c)$ in eq.(\ref{eq:AAA}) and $\tilde s_{1'}=\tilde s_a(z')(\tilde s_b,\tilde s_c),\,\tilde s_{2'}=\tilde s_b(z')(\tilde s_c,\tilde s_a)$ in eq.(\ref{eq:AAB}), together with flipping the signs $\nu_{i'},\,\tilde\nu_{i'}$.

The point is that none of what we change comes into play in the expressions of the $\CY, \, \tilde\CY$ ($s_f$ never appears, indeed, due to cancellations). Henceforth if we move $z'$ across the $ab$ edge nothing changes, as long as it keeps lying in the same cell.

\subsubsection{Defects in neighboring cells}\label{sec:line_defect_neighboring_cells}
\begin{figure}[h]
\begin{center}
\leavevmode
\includegraphics[width=0.50\textwidth]{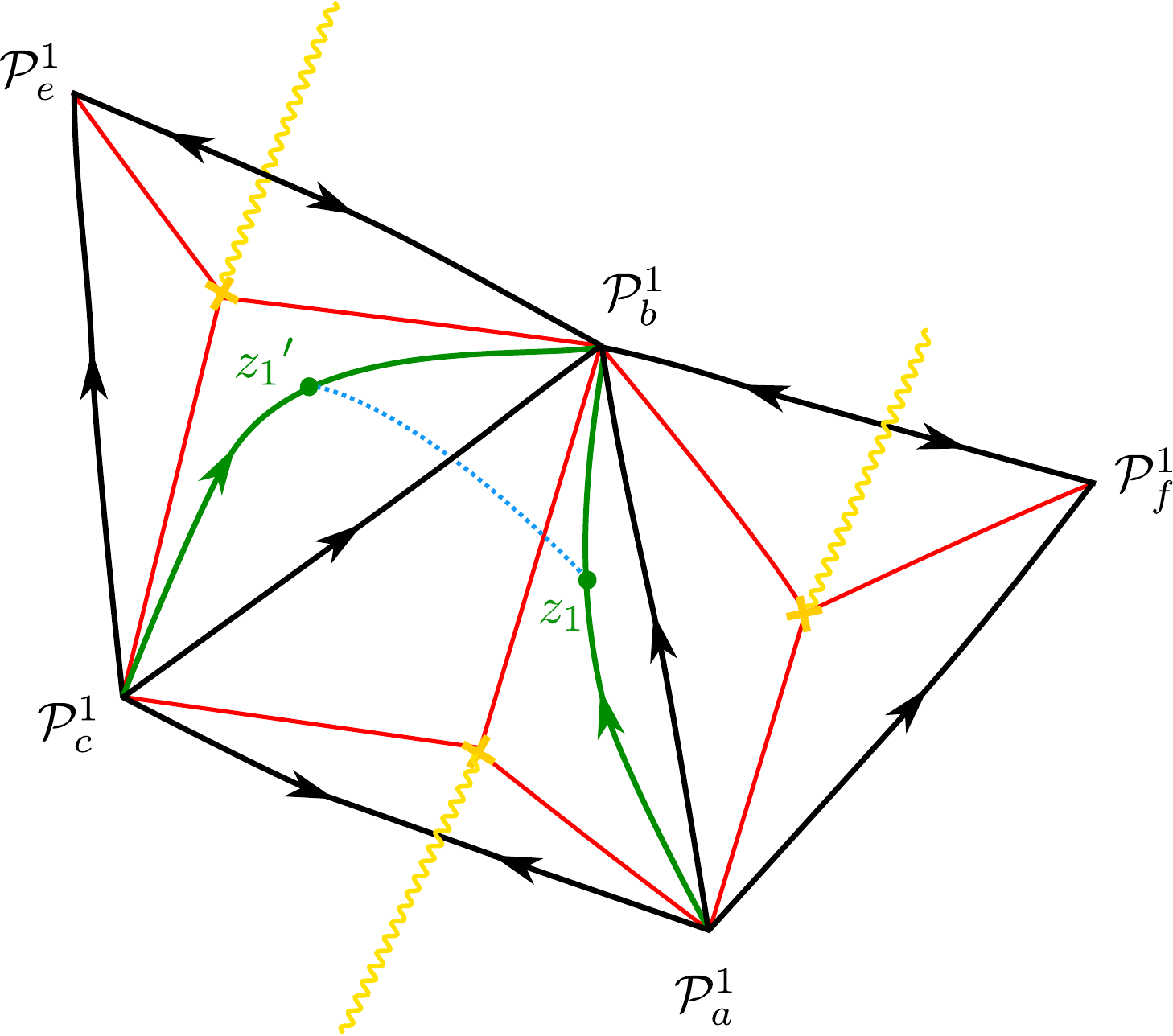}
\end{center}
\caption{}\label{fig:line_b}
\end{figure}
Referring to fig.\ref{fig:line_b} we consider the case where $z,z'$ are in neighboring cells with the choice of sectors shown in the picture. Other choices of sectors can be achieved by using the twisted multiplication rule of the $\CY$ to attach to $z$ or $z'$ the paths described in the previous subsection.

Following the rules sketched above, we have now
\ba
	s_1(z)=s_b(z)\,(s_c,s_a) \qquad & s_{1'}(z')=s_b(z')\,(s_e,s_c) \\
	s_2(z)=s_a(z)\,(s_b,s_c) \qquad & s_{2'}(z')=s_c(z')\,(s_b,s_e) \\
	\nu_1=\nu_{1'}=-\nu_2=&-\nu_{2'}=+
\ea
yielding
\ba
	\CY_{11'}^\vartheta&=\frac{s_{2}(z)\otimes s_{1'}(z')}{(s_{1'},s_{2})}=\frac{s_a(z)\otimes s_b(z')}{(s_b,s_a)}, \\ 
	\CY_{22'}^\vartheta&=\frac{s_{1}(z)\otimes s_{2'}(z')}{(s_{2'},s_{1})}=\frac{s_b(z)\otimes s_c(z')}{(s_c,s_b)}, \\
	\CY_{12'}^\vartheta&=\nu_1\frac{s_{1}(z)\otimes s_{1'}(z')}{(s_{2},s_{1'})}=s_b(z)\otimes s_b(z')\,\frac{(s_c,s_a)}{(s_a,s_b)(s_b,s_c)}, \\
 	\CY_{21'}^\vartheta&=\nu_2\frac{s_{2}(z)\otimes s_{2'}(z')}{(s_{1},s_{2'})}=-\frac{s_a(z)\otimes s_c(z')}{(s_c,s_a)}.
\ea
After sending $\vartheta\to\vartheta+\pi$ we have
\ba
	\tilde s_1(z)=\tilde s_a(z)\,(\tilde s_b,\tilde s_c) \qquad & \tilde s_{1'}(z')=\tilde s_c(z')\,(\tilde s_b,\tilde s_e) \\
	\tilde s_2(z)=\tilde s_b(z)\,(\tilde s_c,\tilde s_a) \qquad & \tilde s_{2'}(z')=\tilde s_b(z')\,(\tilde s_e,\tilde s_c) \\
	\tilde\nu_1=\tilde\nu_{1'}=-\tilde\nu_2=&-\tilde\nu_{2'}=-
\ea
therefore, applying the definitions gives
\ba
	\tilde\CY_{11'}=\frac{\tilde s_b(z)\otimes \tilde s_c(z')}{(\tilde s_c,\tilde s_b)} \qquad & \tilde\CY_{22'}=\frac{\tilde s_a(z)\otimes \tilde s_b(z')}{(\tilde s_b,\tilde s_a)} \\
	\tilde\CY_{12'}=-\frac{\tilde s_a(z)\otimes \tilde s_c(z')}{(\tilde s_c,\tilde s_a)} \qquad & \tilde\CY_{21'}=\tilde s_b(z)\otimes \tilde s_b(z')\,\frac{(\tilde s_c,\tilde s_a)}{(\tilde s_a,\tilde s_b)(\tilde s_b,\tilde s_c)}.
\ea
We start working on $\CY^{\vartheta+\pi}_{11'}\equiv\tilde\CY_{11'}$, at the denominator we have $(\tilde s_c,\tilde s_b)$, in employing (\ref{eq:bi_compact})\footnote{As we stressed in defining $\Xi$, it would be a mistake to write e.g. $$(\tilde s_c, \tilde s_b)=\Xi(\Sigma_c^{b \to b}, \Sigma_b^{c\to c};\mu_c,\mu_b)\omega_{a,b,c}.$$ There are two possibilities here: $$(\tilde s_c, \tilde s_b)=-(\tilde s_b, \tilde s_c)=-\Xi(\Sigma_b^{c \to c}, \Sigma_c^{a\to a};\mu_b,\mu_c)=+\Xi(\Sigma_{c}^{b \to b},\Sigma_{b}^{e \to e};\mu_c,\mu_b)\omega_{c,b,e}$$} we choose to work within triangle $abc$. Thus we refer to eq.s (\ref{eq:bcpop}), and have
\ba
	\tilde\CY_{11'} &= [-\Xi(\Sigma_b^{c\to c}, \Sigma_c^{a \to a};\mu_b,\mu_c)\omega_{a,b,c}]^{-1} \\
	& \times\left\{\Sigma_{b}^{c \to c}\Sigma_{c}^{a \to a} \,(c,a)(a,b) \, s_c(z)\otimes s_c(z') \right. \\
	& + \Sigma_b^{c \to c} \,(c,a)(b,c) (1-\mu_c^{2})\, s_b(z)\otimes s_a(z') \\
	& + \Sigma_c^{a \to a} \,(a,b)^2 (1-\mu_b^{2})\, s_c(z)\otimes s_c(z') \\
	& \left. + (1-\mu_b^{2})(1-\mu_c^{2})\,(a,b)(b,c)\, s_c(z)\otimes s_c(z') \right\}
\ea
in order to have proper asymptotics, one must use the analogue of (\ref{eq:section_identity}) to substitute both $s_a(z')$ and $s_c(z)$, notice that since we are ``working in triangle $abc$'' it is convenient to apply such identities among these vertices, in order to benefit from proper cancellations. The resulting expression reads
\ba
	\tilde\CY_{11'} &=[\Xi(\Sigma_b^{c\to c}, \Sigma_c^{a \to a};\mu_b,\mu_c)]^{-1} \\
	& \times \left\{\Sigma_{b}^{c \to c}\Sigma_{c}^{a \to a} \CY_{22'} + \Sigma_b^{c \to c}\, (1-\mu_c^{2}) (\CY_{12'}-\CY_{22'})\right. \\
	& - (\Sigma_c^{a \to a} +\mu_c^{2} -1) (1-\mu_b^{2}) (\CY_{21'} + \CY_{22'}) \\
	& \left. + (1-\mu_b^{2})(1-\mu_c^{2}) (\CY_{11'}-\CY_{12'})\right\} \label{eq:CY_pop_framed_neighbor}
\ea

A similar procedure leads to
\ba
	\tilde\CY_{22'} &=[\Xi(\Sigma_a^{b\to b}, \Sigma_b^{c \to c};\mu_a,\mu_b)]^{-1} \, \left\{\Sigma_a^{b\to b}\Sigma_{b}^{c \to c} \CY_{11'} + \Sigma_a^{b \to b}\, (1-\mu_b^{2}) \CY_{21'}\right. \\
	& \left. - \Sigma_b^{c \to c} (1-\mu_a^{2}) \CY_{12'} + (1-\mu_a^{2})(1-\mu_b^{2}) \CY_{22'} \right\} \\
	& \\
	\tilde\CY_{12'} &=[\Xi(\Sigma_c^{a\to a}, \Sigma_a^{b \to b};\mu_c,\mu_a)]^{-1} \, \left\{\Sigma_a^{b\to b}\Sigma_{c}^{a \to a} \CY_{21'} + \Sigma_c^{a \to a}\, (1-\mu_a^{2}) \CY_{22'}\right. \\
	& \left. - \Sigma_a^{b \to b} (1-\mu_c^{2}) (\CY_{11'} + \CY_{21'}) + (1-\mu_a^{2})(1-\mu_c^{2}) (\CY_{12'} - \CY_{22'}) \right\} \\
	& \\\tilde\CY_{21'} &=\frac{\Xi(\Sigma_c^{a\to a}, \Sigma_a^{b \to b};\mu_c,\mu_a)}{\Xi(\Sigma_a^{b\to b}, \Sigma_b^{c \to c};\mu_a,\mu_b)\Xi(\Sigma_b^{c\to c}, \Sigma_c^{a \to a};\mu_b,\mu_c)} \\
	& \times \left\{(\Sigma_{b}^{c \to c})^2 \CY_{12'} - \Sigma_b^{c \to c}\, (1-\mu_b^{2}) (\CY_{22'}+\CY_{12'}+\CY_{11'})\right. \\
	& \left. + (1-\mu_b^{2})^2 (\CY_{21'}+ \CY_{22'})\right\} \label{eq:CY_pop_framed_neighbor_bis}
\ea

\subsubsection{Variation on the case of neighboring cells}\label{sec:variation}
Here we briefly repeat the above calculation, but the path from $z$ to $z'$ now goes clockwise around the turning point. For simplicity we take the case in which both $z,z'$ belong to the same triangle. This will be useful when considering $N=1$ AD theory with a line defect.
\begin{figure}[!htb]
 \centering
 \includegraphics[width=0.25\textwidth]{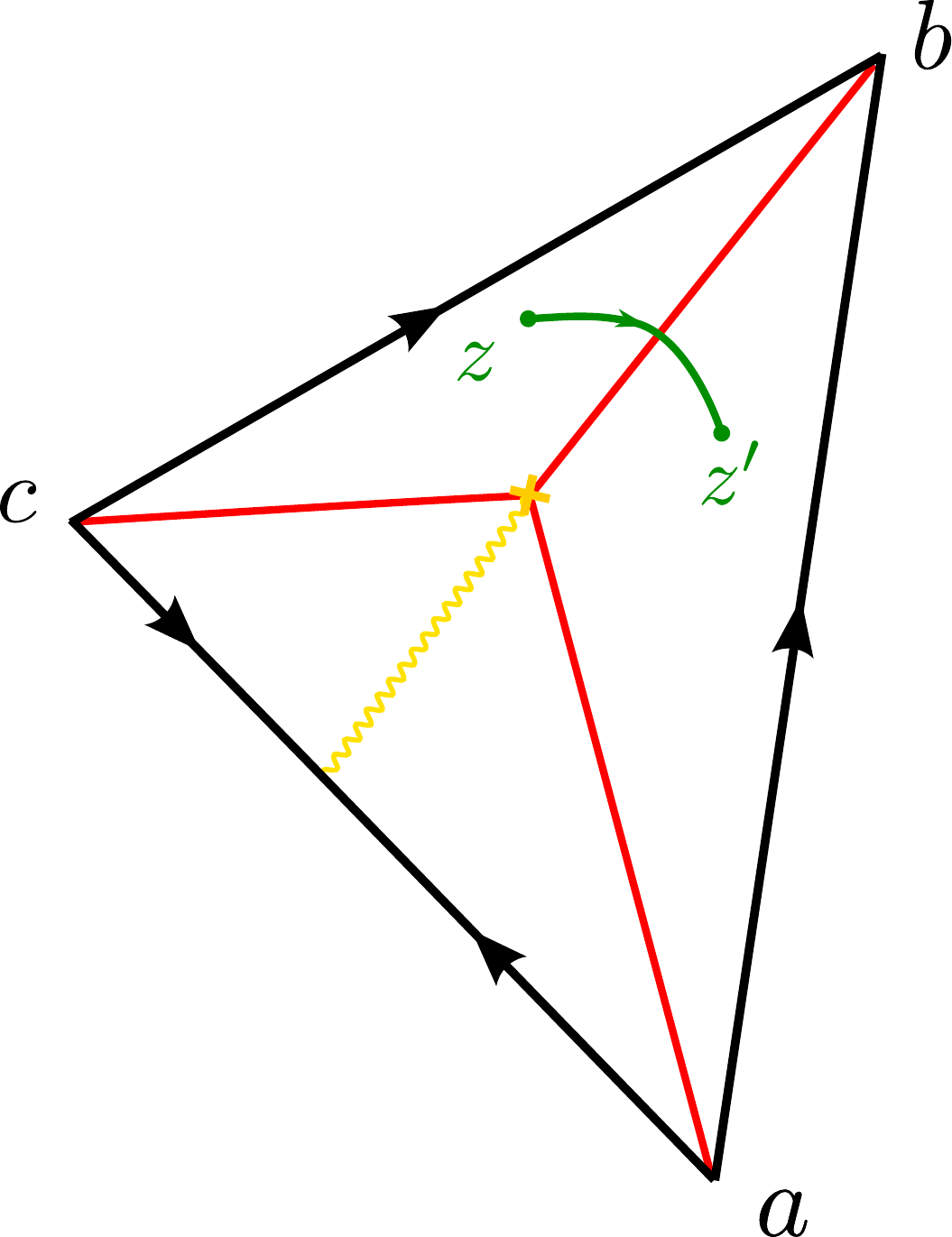}
 \caption{}
 \label{fig:line_defect_neigboring_clockwise}
\end{figure}

We refer to fig.\ref{fig:line_defect_neigboring_clockwise}.
\ba
	s_1(z)=s_b(z)\,(s_c,s_a) \qquad & s_2(z)=s_c(z)\,(s_a,s_b) \\
	s_{1'}(z')=s_b(z')\,(s_c,s_a) \qquad & s_{2'}(z')=s_a(z')\,(s_b,s_c) \\
	\nu_1 = -\nu_2 &= -
\ea
From the definitions (\ref{eq:line_morphisms}) follow
\ba
	\CY_{11'} &= \frac{s_2(z) \otimes s_{1'}(z')}{(s_{1'},s_2)} = \frac{s_c(z) \otimes s_b(z')}{(s_b,s_c)} \\
	\CY_{22'} &= \frac{s_1(z) \otimes s_{2'}(z')}{(s_{2'},s_1)} = \frac{s_b(z) \otimes s_a(z')}{(s_a,s_b)} \\
	\CY_{12'} &= \nu_1\,\frac{s_1(z) \otimes s_{1'}(z')}{(s_{2},s_{1'})} = s_b(z) \otimes s_b(z') \frac{(s_c,s_a)}{(s_a,s_b)(s_b,s_c)} \\
	\CY_{21'} &= \nu_2\,\frac{s_2(z) \otimes s_{2'}(z')}{(s_{1},s_{2'})} = \frac{s_c(z) \otimes s_a(z')}{(s_a,s_c)} 
\ea
After sending $\vartheta\to\vartheta+\pi$ we have instead
\ba
	\tilde s_1(z)=\tilde s_c(z)\,(\tilde s_a,\tilde s_b) \qquad & \tilde s_2(z)=\tilde s_b(z)\,(\tilde s_c,\tilde s_a) \\
	\tilde s_{1'}(z')=\tilde s_a(z')\,(\tilde s_b,\tilde s_c) \qquad & \tilde s_{2'}(z')=\tilde s_b(z')\,(\tilde s_c,\tilde s_a) \\
	\tilde \nu_1 = -\tilde \nu_2 = +
\ea
From the definitions (\ref{eq:line_morphisms}), and employing eq.s (\ref{eq:apop}), (\ref{eq:bcpop}), (\ref{eq:bi_compact}) follow
\ba
	\tilde\CY_{11'} &= [\Xi(\Sigma_{a}^{b \to b}, \Sigma_{b}^{c \to c};\mu_a,\mu_b)]^{-1}\,\left\{ \Sigma_{a}^{b \to b} \Sigma_{b}^{c \to c} \CY_{22'} + \Sigma_{b}^{c \to c} (1-\mu_a^{2}) \CY_{12'} \right. \\
	& \left. -\Sigma_{a}^{b \to b}(1-\mu_b^{2}) \CY_{21'} + (1-\mu_a^{2})(1-\mu_b^{2}) \CY_{11'} \right\} \\
	& \\
	\tilde\CY_{22'} &= [\Xi(\Sigma_{b}^{c \to c}, \Sigma_{c}^{a \to a};\mu_b,\mu_c)]^{-1}\,\left\{ \Sigma_{b}^{c \to c} \Sigma_{c}^{a \to a} \CY_{11'} - \Sigma_{c}^{a \to a} (1-\mu_b^{2}) [\CY_{11'}-\CY_{21'}] \right. \\
	& \left. -\Sigma_{b}^{c \to c}(1-\mu_c^{2}) [\CY_{12'}+\CY_{11'}] + (1-\mu_b^{2})(1-\mu_c^{2}) [\CY_{11'} + \CY_{22'} + 
	\CY_{12'} - \CY_{21'}] \right\} \\
	& \\
	\tilde\CY_{12'} &= [\Xi(\Sigma_{c}^{a \to a}, \Sigma_{a}^{b \to b};\mu_c,\mu_a)]^{-1}\,\left\{ \Sigma_{c}^{a \to a} \Sigma_{a}^{b \to b} \CY_{21'} - \Sigma_{c}^{a \to a} (1-\mu_a^{2}) \CY_{11'} \right. \\
	& \left. + \Sigma_{a}^{b \to b}(1-\mu_c^{2}) [\CY_{22'}-\CY_{21'}] + (1-\mu_c^{2})(1-\mu_a^{2}) [\CY_{12'} + \CY_{11'}] \right\} \\
	& \\
	\tilde\CY_{21'} &= \Xi(\Sigma_{c}^{a \to a}, \Sigma_{a}^{b \to a};\mu_c,\mu_a)[\Xi(\Sigma_{b}^{c \to c}, \Sigma_{c}^{a \to a};\mu_b,\mu_c)\Xi(\Sigma_{a}^{b \to b}, \Sigma_{b}^{c \to c};\mu_a,\mu_b)]^{-1} \\
	& \times \left\{ (\Sigma_{b}^{c \to c})^2 \CY_{12'} + (1-\mu_b^{2})^2 [\CY_{21'}-\CY_{11'}]  \right. \\
	& \left. + \Sigma_{b}^{c \to c} (1-\mu_b^{2}) [\CY_{11'} - \CY_{22'} - \CY_{12'}] \right\} \label{eq:CY_pop_framed_variation}
\ea
\vspace{.5cm}

\section{Extracting the spectrum using ${\bf S}$}\label{sec:examples}
We now provide some examples of how to apply the results of \S\ref{sec:formal_work}. In the cases reviewed below, using ${\bf S}$ to extrapolate the spectrum is quick and easy. As we will see, it essentially amounts to \emph{matching pictures}, which is achieved by matching vertex labels and sheet labels. According to this picture, we expect to have wall crossing for solitonic charges whenever labels jump: this occurs when $z$ crosses a separating line or a branch cut\footnote{A branch cut is not physically meaningful, so the reader may be puzzled by its relevance in jumps of the spectrum. In this context if $z$ crosses the cut, this amounts to a deck transformation, or to exchanging the lifts of $z$ to $\Sigma$. In other words, when $z$ crosses a cut we switch $\gamma_{ij}\leftrightarrow\gamma_{ji}$ which explains why we see a ``jump'' in the spectrum: we are really just switching notation, accordingly the jump must reflect such exchange of charges.}, in agreement with the results of \IV.

\subsection{$N=1$ AD theory without line defects}
\begin{figure}[h!]
\begin{center}
\leavevmode
\includegraphics[width=0.25\textwidth]{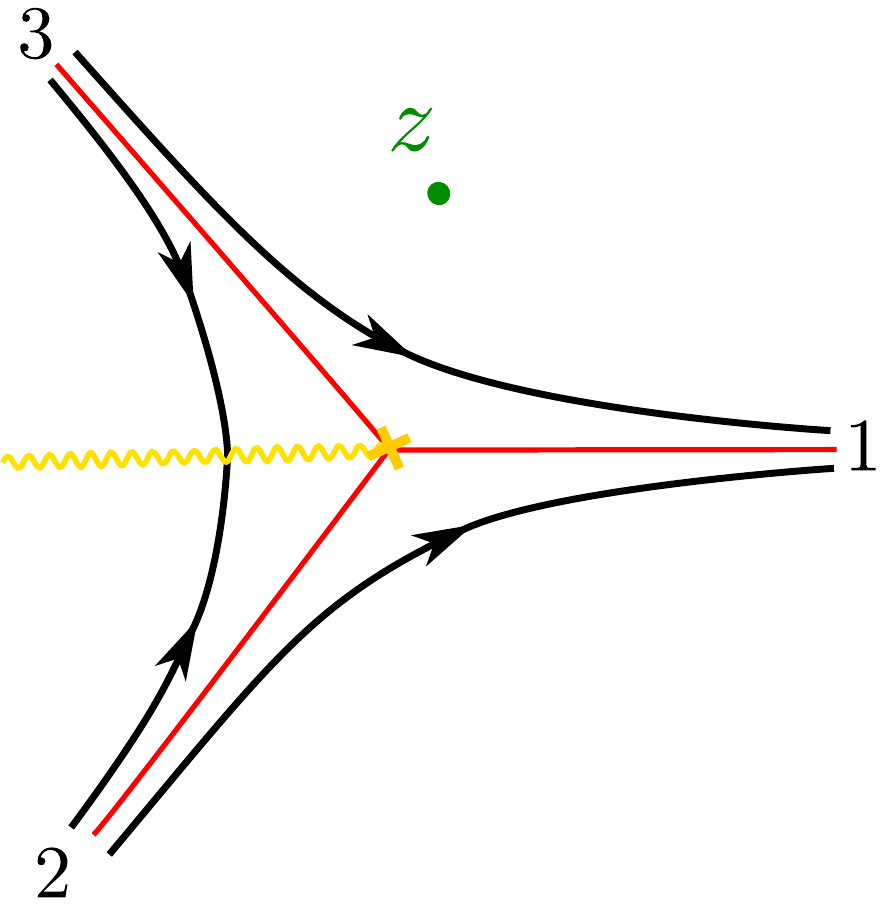}%
\hspace{0.15\textwidth}%
\includegraphics[width=0.25\textwidth]{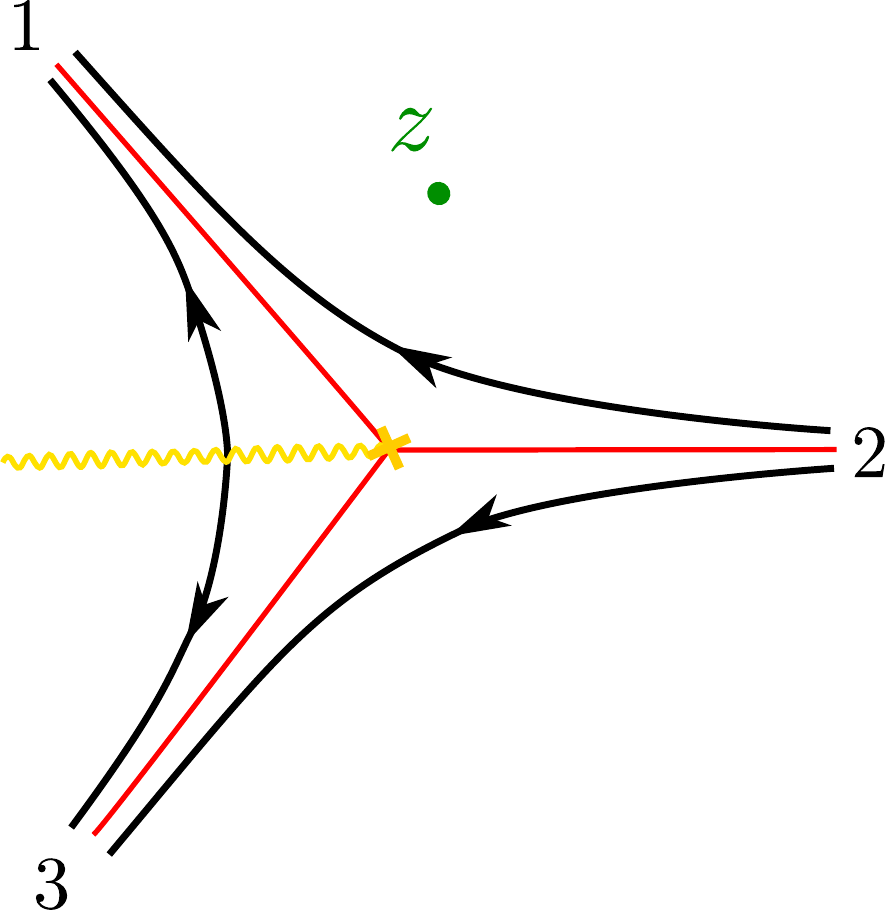}
\end{center}
\caption{The $+$ sheet, on the left at angle $\vartheta=0$, on the right at angle $\vartheta=\pi$.}\label{fig:AD_1}
\end{figure}
For the $N=1$ Argyres-Douglas theory we have $\phi_2(z)=-z\,\d z^2$, on $C=\CP^1$, meaning that there is one irregular singularity at infinity, a turning point in $z=0$, and $\Sigma$ is a two-sheeted cover of $C$. We consider the theory in presence of a single surface defect sitting at $z$, with $Arg(z)\in[0,2\pi/3]$, as in fig.\ref{fig:AD_1}. Notice that vertices are labeled clockwise (counter-clockwise as seen from a neighborhood of the irregular puncture at infinity).

We now want to apply the results obtained in section \ref{sec:soliton_pop} to derive the spectrum. Compare figures \ref{fig:scan} and \ref{fig:AD_1}: matching the flow of WKB lines through $z$ in both pictures entails the identifications for the vertices $a \leftrightarrow 1,\ b \leftrightarrow 3,\ c \leftrightarrow 2 $ together with the identification of sheets 
\ba
\text{sheet }2 \leftrightarrow \text{sheet }+
\ea 
and similarly for sheets $1$ and $-$.

Indeed, \emph{identifying the sheets properly is all we need}: this entails
\ba
    & \CY_{11}\leftrightarrow\CY_{--}, \qquad \CY_{12}\leftrightarrow\CY_{-+}, \qquad \dots \\
    & \tilde \CY_{11}\leftrightarrow \tilde \CY_{--}, \qquad \tilde \CY_{12}\leftrightarrow\tilde \CY_{-+}, \qquad \dots
\ea
thus, employing expressions (\ref{eq:CY_pop_reg}) and (\ref{eq:CY_pop_reg_bis}), and making the replacements explained in \S\S \ref{sec:irregular_single}, \ref{sec:irregular_general} should give the transformation generated by ${\bf S}$. As a matter of fact, in this particularly simple example, we have only degenerate edges, thus we replace all $\mu_\alpha^{2}=0$, and all $\Sigma=\Xi=1$.

With these rules, eq.s (\ref{eq:CY_pop_reg}) and (\ref{eq:CY_pop_reg_bis}) give the transformation 
\ba
      \left\{\begin{array}{l}
      \tilde \CY_{12} = \CY_{12} + \CY_{21} + \CY_{22} - \CY_{11} \\
      \tilde \CY_{21} = \CY_{21} \\
      \tilde \CY_{11} = \CY_{11}-\CY_{21} \\
      \tilde \CY_{22} = \CY_{22} + \CY_{21}
      \end{array} \right.\leftrightarrow %
      \left\{\begin{array}{l}
	\tilde\CY_{-+}=\CY_{-+}+\CY_{+-}+\CY_{++}-\CY_{--}\\
	\tilde\CY_{+-}=\CY_{+-}\\
	\tilde\CY_{--}=\CY_{--}-\CY_{+-}\\
	\tilde\CY_{++} = \CY_{++}+\CY_{+-}       
      \end{array}\right.
\ea
This corresponds to the transformation
\ba
	{\cal S}_{{+-}}^\mu: \,\CY_{++} \mapsto (1-\mu_{+-} \CY_{+-})\,\CY_{++}\,(1+\mu_{+-} \CY_{+-}) \label{eq:AD_WC}
\ea
with $\mu_{+-}=+1$, and similarly for the other $\CY$'s. This means that the BPS spectrum contains only one soliton $\gamma_{+-}$: ${\bf S}={\cal S}_{{+-}}^\mu$ (together with its antiparticle), consequently there can't be any marginal stability wall, nor wall crossing.\\

\subsection{$N=1$ AD theory: framed wall crossing}
\subsubsection{Small angular separation}
\begin{figure}[h!]
\begin{center}
\leavevmode
\includegraphics[width=0.25\textwidth]{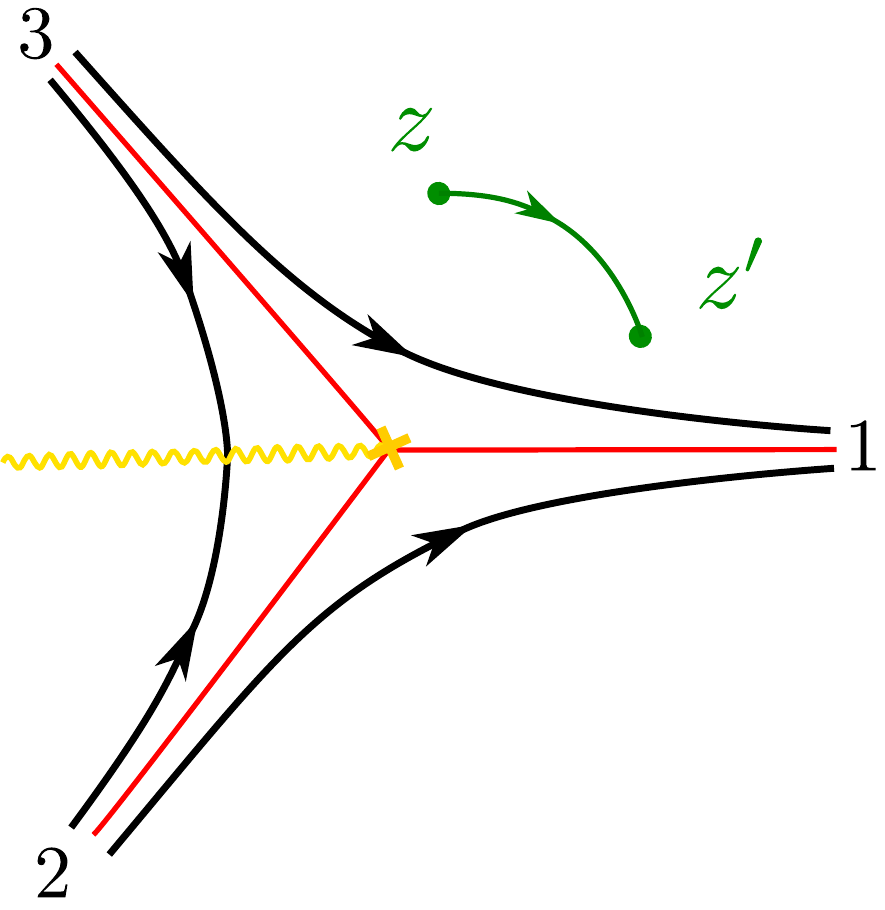}%
\hspace{0.15\textwidth}%
\includegraphics[width=0.25\textwidth]{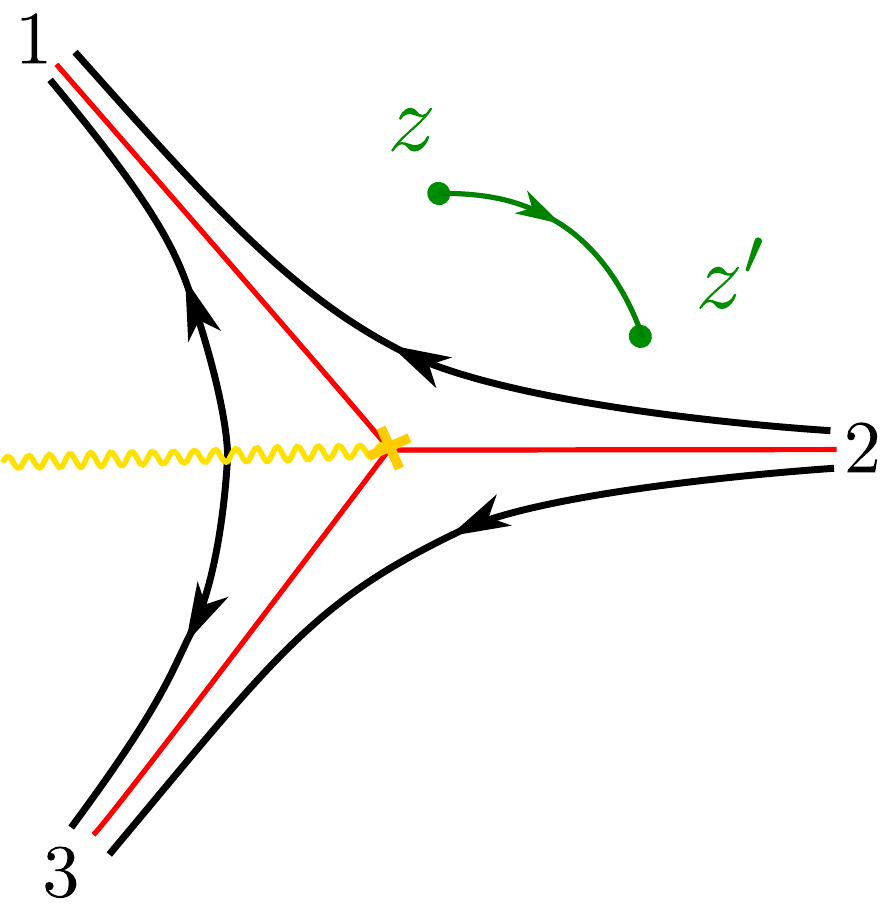}
\end{center}
\caption{The $+$ sheet, on the left at angle $\vartheta=0$, on the right at angle $\vartheta=\pi$.}\label{fig:AD_1_framed}
\end{figure}

We now consider what happens in presence of a line defect \cite{GMN3,Kapustin}, namely an interface between two surface defects \cite{GMN4}. Let the two surface defects sit at points $z,z'$ on $C$, we consider $\CY^{\vartheta=0}_{++'}$, associated to the path shown in fig.\ref{fig:AD_1_framed}.

We can then apply the machinery of sections \ref{sec:same_cell} and \ref{sec:same_sector} and quickly obtain the spectrum generator.

Comparing figures \ref{fig:line_a} and \ref{fig:AD_1_framed} we immediately see that to adapt the general discussion to our case we must identify 
\ba
    \text{sheet} 2 \, \leftrightarrow \, \text{sheet} +
\ea
and accordingly identify sheets $1$ and $-$.
Having to deal with an irregular puncture, we must make the replacements mentioned at the end of section \ref{sec:same_cell}. Since all edges involved are boundary edges we just have to replace all $\mu^{2}=0$, $\Sigma=1=\Xi$ into equations (\ref{eq:CY_pop_line_reg}), (\ref{eq:CY_pop_line_reg_bis}).

Doing so, gives immediately the \emph{omnipop} transformation
\ba
      \left\{\begin{array}{l}
      \tilde \CY_{12'} = \CY_{12'} + \CY_{21'} + \CY_{22'} - \CY_{11'} \\
      \tilde \CY_{21'} = \CY_{21'} \\
      \tilde \CY_{11'} = \CY_{11'}-\CY_{21'} \\
      \tilde \CY_{22'} = \CY_{22'} + \CY_{21'}
      \end{array} \right.\leftrightarrow %
      \left\{\begin{array}{l}
	\tilde\CY_{-+'}=\CY_{-+'}+\CY_{+-'}+\CY_{++'}-\CY_{--'}\\
	\tilde\CY_{+-'}=\CY_{+-'}\\
	\tilde\CY_{--'}=\CY_{--'}-\CY_{+-'}\\
	\tilde\CY_{++'} = \CY_{++'}+\CY_{+-'}       
      \end{array}\right.
\ea
All these transformations can be traced back to the action of 
\ba
{\bf S}={\cal S}_{+'-'} {\cal S}_{+-}, \label{eq:AD_WC_bis}
\ea
with $\mu(+'-')=\mu(+-)=+1$, as one could naively expect from the discussion of the previous section: the spectrum contains two BPS solitons, one for each surface defect, and each of them undergoes exactly the same wall crossing as that of eq. (\ref{eq:AD_WC}), since both defects will get crossed by the same WKB ray as in that case.

Notice that, in order to match our result with the example analyzed in \IV section 8.1.1, one needs to apply the label exchange $ij \leftrightarrow ji$ as described in remark (\ref{eq:change}). After doing this, one recovers the transformations of eq.s (8.9), (8.10) in the reference.

% \textcolor{Blue}{%\footnotesize
As a check, we give a full derivation of the spectrum generator, this will also be useful in analyzing line defects below. According to the rules sketched above
\ba
	s_+(z)=s_1(z)\,(s_3,s_2) \qquad & s_{+'}(z')=s_1(z')\,(s_3,s_2) \\
	s_-(z)=s_3(z)\,(s_2,s_1) \qquad & s_{-'}(z')=s_3(z')\,(s_2,s_1) \\
	\nu_+=-\nu_-= - = \nu_{+'}&=-\nu_{-'}
\ea
therefore, from (\ref{eq:line_morphisms}) we have
\ba
	\CY_{++'}^\vartheta =\frac{s_3(z)\otimes s_1(z')}{(s_1,s_3)} \qquad & \CY_{--'}^\vartheta =\frac{s_1(z)\otimes s_3(z')}{(s_3,s_1)} \\
	\CY_{+-'}^\vartheta=s_1(z)\otimes s_1(z')\,\frac{(s_3,s_2)}{(s_1,s_3)(s_2,s_1)} \qquad & \CY_{-+'}^\vartheta=s_3(z)\otimes s_3(z')\,\frac{(s_2,s_1)}{(s_1,s_3)(s_3,s_2)},
\ea
setting $\vartheta=\pi$ gives the situation of fig.\ref{fig:AD_1_framed}.
\ba
	\tilde s_+(z)= s_1(z)\,( s_3, s_2) \qquad & \tilde s_{+'}(z')= s_1(z')\,( s_3, s_2) \\
	\tilde s_-(z)= s_2(z)\,( s_1, s_3) \qquad & \tilde s_{-'}(z')= s_2(z')\,( s_1, s_3) \\
	\tilde \nu_+=-\tilde \nu_-= + = \tilde \nu_{+'}&=-\tilde \nu_{-'}
\ea
Therefore, rules (\ref{eq:line_morphisms}) now give
\ba
	\tilde \CY_{++'}= \frac{ s_2(z)\otimes  s_1(z')}{( s_1, s_2)}\qquad & %
	\tilde \CY_{--'}= \frac{ s_1(z)\otimes  s_2(z')}{( s_2, s_1)}\\
	\tilde \CY_{+-'}=  s_1(z)\otimes  s_1(z') \frac{( s_3, s_2)}{( s_2, s_1)( s_1, s_3)}\qquad & %
	\tilde \CY_{-+'}=  s_2(z)\otimes  s_2(z') \frac{( s_1, s_3)}{( s_3, s_2)( s_2, s_1)} .
\ea
We can now employ eq. (\ref{eq:section_identity}), and get e.g.
\ba
	\tilde \CY_{++'}=\CY_{++'}\,+\,\CY_{+-'} \label{eq:AD_1_framed_pops}
\ea
similarly, repeating the machinery one finds
\ba
	&\tilde\CY_{--'}=\CY_{--'}-\CY_{+-'}\\
	&\tilde\CY_{-+'}=\CY_{-+'}+\CY_{+-'}+\CY_{++'}-\CY_{--'}\\
	&\tilde\CY_{+-'}=\CY_{+-'} \label{eq:AD_1_pops_bis}
\ea
in agreement with the above result.
% }

Now, having a line defect, we can consider its expectation value at some angle $\vartheta$: before the omnipop
\ba
	\langle L_{\wp,\zeta} \rangle \, = \, \CY_{++'} + \CY_{--'}. \label{eq:line_exp_val}
\ea
Since $\langle L_{\wp,\zeta} \rangle$ must not jump as $\vartheta$ varies, we can rewrite it at $\vartheta + \pi$ by inverting eq.s(\ref{eq:AD_1_framed_pops}), (\ref{eq:AD_1_pops_bis}) and plugging into (\ref{eq:line_exp_val}). This gives
\ba
	\langle L_{\wp,\zeta} \rangle \, = \, \tilde\CY_{++'} + \tilde\CY_{--'}.
\ea
We see no framed wall crossing. Comparing with \S 8.1.1 of \IV the explanation is the following one: as is shown in the reference, referring to fig.19 therein, framed wall crossing occurs when we compare between situations such as (A) and (B) (if e.g. in situation (A) $\vartheta=\vartheta_0$, situation (C) has $\vartheta=\vartheta_0+\pi$, and (B) is some particular intermediate situation), but no difference exists between situations (A) and (C). This agrees with eq.(8.10) of \IV.

\subsubsection{Large angular separation}
\begin{figure}[h!]
\begin{center}
\leavevmode
\includegraphics[width=0.25\textwidth]{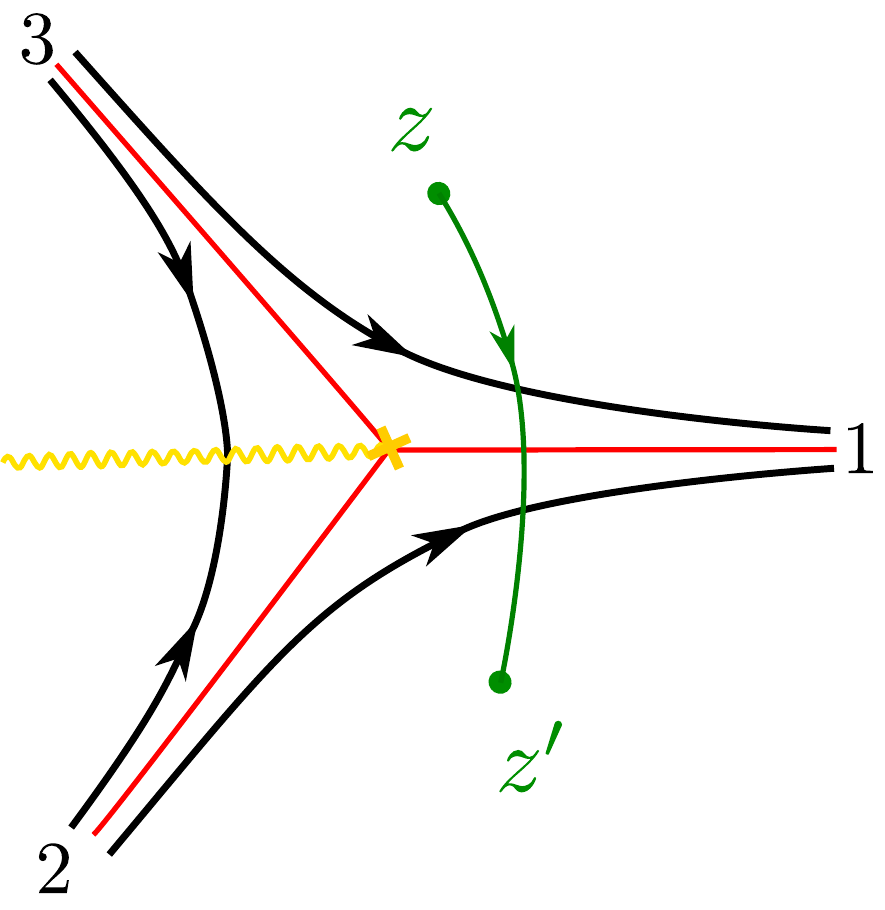}%
\hspace{0.15\textwidth}%
\includegraphics[width=0.25\textwidth]{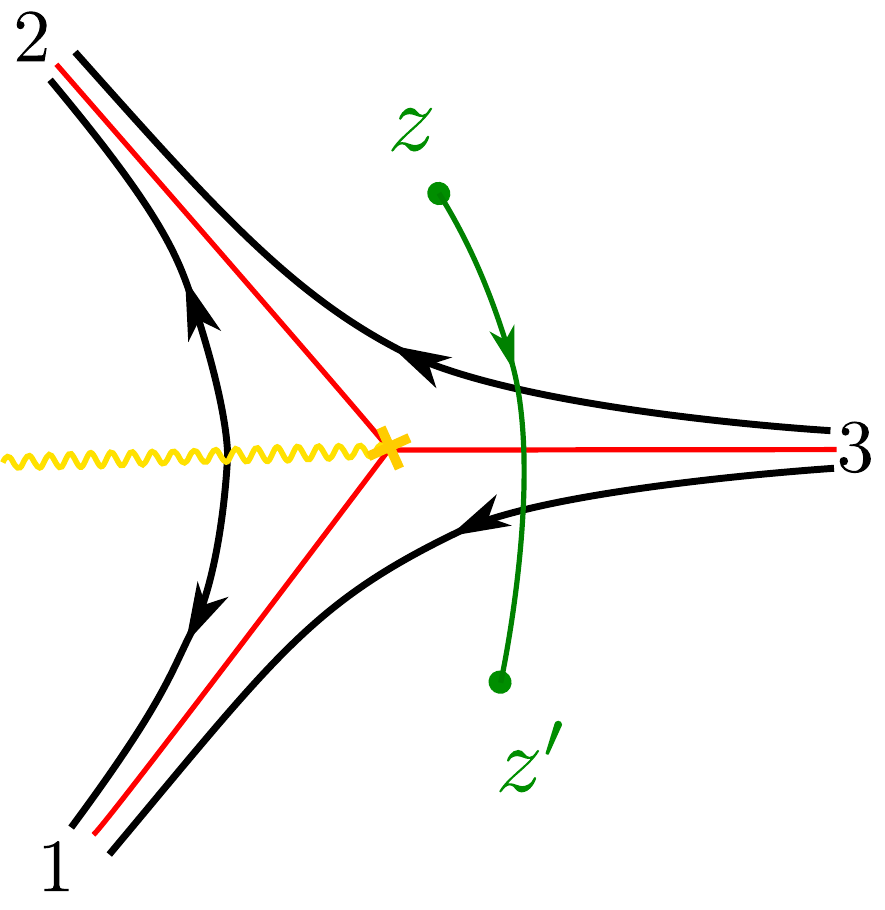}
\end{center}
\caption{The $+$ sheet, on the left at angle $\vartheta=0$, on the right at angle $\vartheta=-\pi$.}\label{fig:AD_1_framed_large}
\end{figure}

Let us now consider the case when $z,\,z'$ are separated by an angle between $2\pi/3$ and $\pi$, as shown in fig.\ref{fig:AD_1_framed_large}. 
In \IV the \emph{clockwise} transformation for $\vartheta\mapsto\vartheta-\pi$ is considered (see end of section 8.1.1 there), our methods are suitable for studying an omnipop, i.e. an increase of $\vartheta$ by $\pi$, nonetheless once we obtain the transformation for this, we can match with \IV by recalling that in the sector $[\vartheta,\vartheta-\pi]$ the spectrum consists of the corresponding \emph{antiparticles}.

Let us therefore study the regular omnipop first. This case matches with the analysis carried out in section \ref{sec:variation}: comparing figures \ref{fig:line_defect_neigboring_clockwise} and \ref{fig:AD_1_framed_large} entails the identification
\ba
    \text{sheet} 1 \, \leftrightarrow \, \text{sheet} +
\ea
therefore, we can employ eq.s (\ref{eq:CY_pop_framed_variation}): setting all $\mu^{2}\mapsto 0,\, \Sigma\mapsto 1,\, \Xi\mapsto 1$, they read
\ba
      \left\{\begin{array}{l}
      \tilde \CY_{11'} = \CY_{11'} + \CY_{22'} + \CY_{12'} - \CY_{21'} \\
      \tilde \CY_{22'} = \CY_{22'} \\
      \tilde \CY_{12'} = \CY_{12'}+\CY_{22'} \\
      \tilde \CY_{21'} = \CY_{21'} - \CY_{22'}
      \end{array} \right.\leftrightarrow %
      \left\{\begin{array}{l}
	\tilde\CY_{++'}=\CY_{++'}+\CY_{--'}+\CY_{+-'}-\CY_{-+'}\\
	\tilde\CY_{--'}=\CY_{--'}\\
	\tilde\CY_{+-'}=\CY_{+-'}+\CY_{--'}\\
	\tilde\CY_{-+'} = \CY_{-+'}-\CY_{--'}       
      \end{array}\right. \label{eq:N_1_large_pop}
\ea
these are generated by\footnote{Here we are using the following set of twisting functions: 
\ba
  \sigma(-+,++')=\sigma(+'-',++')=1=-\sigma(+'-',-+')=-\sigma(-+,+-')
\ea}
\ba
	{\bf S}(\vartheta+\pi,\vartheta)\,=\, {\cal S}_{-+} {\cal S}_{+'-'}
\ea
meaning that $\mu(-+)=+1=\mu(+'-')$.

We therefore conclude that the spectrum generator in the other half-plane will be
\ba
	{\bf S}(\vartheta,\vartheta-\pi)\,=\, {\cal S}_{+-} {\cal S}_{-'+'} \label{eq:SG-AD-large}
\ea
Now, to match with the example in \IV, we need to make the following consideration: inverting eq.s (\ref{eq:N_1_large_pop}) amounts to switching the tildes and all $ij\to ji$ as well as $i'j'\to j'i'$ namely, we get
\ba
	\left\{\begin{array}{l}
	\CY_{++'}=\tilde\CY_{++'}+\tilde\CY_{--'}-\tilde\CY_{+-'}+\tilde\CY_{-+'}\\
	\CY_{--'}=\tilde\CY_{--'}\\
	\CY_{+-'}=\tilde\CY_{+-'}-\tilde\CY_{--'}\\
	\CY_{-+'} = \tilde\CY_{-+'}+\tilde\CY_{--'}       
      \end{array}\right.
\ea
this would be the action of ${\bf S}(\vartheta,\vartheta+\pi)={\bf S}(\vartheta+\pi,\vartheta)^{-1}$, describing the \emph{clockwise} transformation from $\vartheta=\pi$ to $\vartheta=0$. What we found is that, in this example switching $ij\leftrightarrow ji$ in the $\CY$'s labels corresponds precisely to switching the \emph{clockwise}$\leftrightarrow$\emph{counterclockwise} jumps of $\vartheta$ by angles of $\pi$. We claim that this also holds for the spectrum generator in the other half plane, we prove this in the following subsection.

Now, according to remark (\ref{eq:change}) this switching of labels is just what we must employ to match our conventions with those of \IV. Therefore, we arrive at the conclusion that our ${\bf S}(\vartheta,\vartheta-\pi)$ is precisely the transformation they find in going \emph{clockwise} from $\vartheta=0$ to $\vartheta=-\pi$. Indeed they match, see eq.s (8.12) therein: ${\cal E}_C={\cal S}_{+-} {\cal S}_{-'+'}{\cal E}_A$.\\

The above derivation of the spectrum was straightforward, although the matching with \IV might appear a bit artificial. Let us prove that this is actually correct: we now give a full derivation of the spectrum generator when $\vartheta\mapsto\vartheta-\pi$ (i.e. following the direction taken by the reference).

We start with $\vartheta=0$, our building blocks are now
\ba
	s_+(z)=s_1(z)\,(s_3,s_2) \qquad & s_{+'}(z')=s_1(z')\,(s_3,s_2) \\
	s_-(z)=s_3(z)\,(s_2,s_1) \qquad & s_{-'}(z')=s_2(z')\,(s_1,s_3) \\
	\nu_+=-\nu_-= - = -\nu_{+'}&=\nu_{-'}
\ea
therefore, from (\ref{eq:line_morphisms}) we have
\ba
	\CY_{++'}^\vartheta =\frac{s_3(z)\otimes s_1(z')}{(s_1,s_3)} \qquad & \CY_{--'}^\vartheta =\frac{s_1(z)\otimes s_2(z')}{(s_2,s_1)} \\
	\CY_{+-'}^\vartheta=s_1(z)\otimes s_1(z')\,\frac{(s_3,s_2)}{(s_1,s_3)(s_2,s_1)} \qquad & \CY_{-+'}^\vartheta=\frac{s_3(z)\otimes s_2(z')}{(s_2,s_3)},
\ea
setting $\vartheta=-\pi$ gives the situation of fig.\ref{fig:AD_1_framed_large}.
\ba
	\tilde s_+(z)= s_2(z)\,( s_1, s_3) \qquad & \tilde s_{+'}(z')= s_1(z')\,( s_3, s_2) \\
	\tilde s_-(z)= s_3(z)\,( s_2, s_1) \qquad & \tilde s_{-'}(z')= s_3(z')\,( s_2, s_1) \\
	\tilde \nu_+=- \tilde\nu_-= + = - \tilde\nu_{+'}&=\tilde \nu_{-'}
\ea
where we now understand $\tilde\CY:=\CY^{-\pi}$. Therefore, rules (\ref{eq:line_morphisms}) now give
\ba
	\tilde \CY_{++'}= \frac{ s_3(z)\otimes  s_1(z')}{( s_1, s_3)}\qquad & %
	\tilde \CY_{--'}= \frac{ s_2(z)\otimes  s_3(z')}{( s_3, s_2)}\\
	\tilde \CY_{+-'}= \frac{ s_2(z)\otimes  s_1(z')}{( s_1, s_2)}\qquad & %
	\tilde \CY_{-+'}=   s_3(z)\otimes s_3(z') \frac{( s_2, s_1)}{( s_1, s_3)( s_3, s_2)} .
\ea
We can employ eq. (\ref{eq:section_identity}), and get directly
\ba
	&\tilde\CY_{++'}=\CY_{++'}\\
	&\tilde\CY_{--'}=\CY_{--'} + \CY_{++'} + \CY_{+-'} - \CY_{-+'}\\
	&\tilde\CY_{-+'}=\CY_{-+'}-\CY_{++'}\\
	&\tilde\CY_{+-'}=\CY_{+-'}+\CY_{++'} \label{eq:AD_1_pops_large}
\ea
These transformations correspond to going \emph{clockwise} from $\vartheta=0$ to $\vartheta=-\pi$, they are the inverse of the \emph{counterclockwise} $\vartheta=-\pi\mapsto\vartheta=0$ which reads
\ba
	&\CY_{++'}=\tilde\CY_{++'}\\
	&\CY_{--'}=\tilde\CY_{--'} + \tilde\CY_{++'} - \tilde\CY_{+-'} + \tilde\CY_{-+'}\\
	&\CY_{-+'}=\tilde\CY_{-+'}+\tilde\CY_{++'}\\
	&\CY_{+-'}=\tilde\CY_{+-'}-\tilde\CY_{++'}
\ea
these result precisely from the action of
\ba
	{\bf S}(\vartheta,\vartheta-\pi) = {\cal S}_{+-} {\cal S}_{-'+'}
\ea
in agreement with our previous derivation.\\
With a line defect at hand, we can examine the associated expectation value at some angle $\vartheta$: before the omnipop (cf. eq.(8.11) in \IV)\footnote{Note: as remarked in eq.(\ref{eq:change}) for us $\CY_{ij'}$ have $i$ and $j$ switched, as compared to \IV, so here our $\CY_{+-'}$ corresponds to their $\CY_{-+'}$, and vice versa}:
\ba
	\langle L_{\wp,\zeta} \rangle \, = \, \CY_{++'} + \CY_{+-'} + \CY_{--'}. \label{eq:line_exp_val_large}
\ea
Since $\langle L_{\wp,\zeta} \rangle$ must not jump as $\vartheta$ varies, we can rewrite it at $\vartheta = - \pi$ by inverting eq.s(\ref{eq:AD_1_pops_large}) and plugging into (\ref{eq:line_exp_val_large}). This gives
\ba
	\langle L_{\wp,\zeta} \rangle \, = \, \tilde\CY_{++'} + \tilde\CY_{-+'} + \tilde\CY_{--'}.
\ea
this corresponds to acting on $\langle L_{\wp,\zeta} \rangle$ with
\ba
      & \, (1-\CY_{-'+'})(1-\CY_{+-}) \langle L_{\wp,\zeta} \rangle (1+\CY_{+-})(1+\CY_{-'+'}) \\
      & \equiv \, (1-\CY_{+-}) \langle L_{\wp,\zeta} \rangle (1+\CY_{-'+'})
\ea
that means ${\bf S}(\vartheta,\vartheta-\pi)={\cal S}_{\gamma_{-'+'}}{\cal S}_{\gamma_{+-}}$, together with $\mu(\gamma_{-'+'})=1=\mu(\gamma_{+-})$.%
\footnote{%
Note on twisting functions: we are using 
\ba
  \sigma(+-,-+')= +1 \qquad & \sigma(+-,--')=-1 \\
  \sigma(--',-'+')=-1 \qquad & \sigma(+-',-'+')=+1
\ea
The first row is explicitly employed in \IV, see footnote at p.114; the first sign of the second row is implicitly used in eq. (8.10) in that reference; the last sign is derivable by means of the cocycle condition for twisting functions (2.24).%
}

\subsection{$N=2$ AD theory}
We now turn to the case of $N=2$ AD theory, label-matching is slightly more delicate here, hence we first give a derivation of ${\bf S}$ by carrying out a complete analysis of how $T_{\text{WKB}}$ evolves with $\vartheta$, later we obtain ${\bf S}$ relying only on the results of \ref{sec:soliton_pop}, this will be much faster and agree exactly with the former derivation, as well as with the analysis of \IV.

\subsubsection{Full derivation of the spectrum generator}
\begin{figure}[h!]
\begin{center}
\leavevmode
\includegraphics[width=0.40\textwidth]{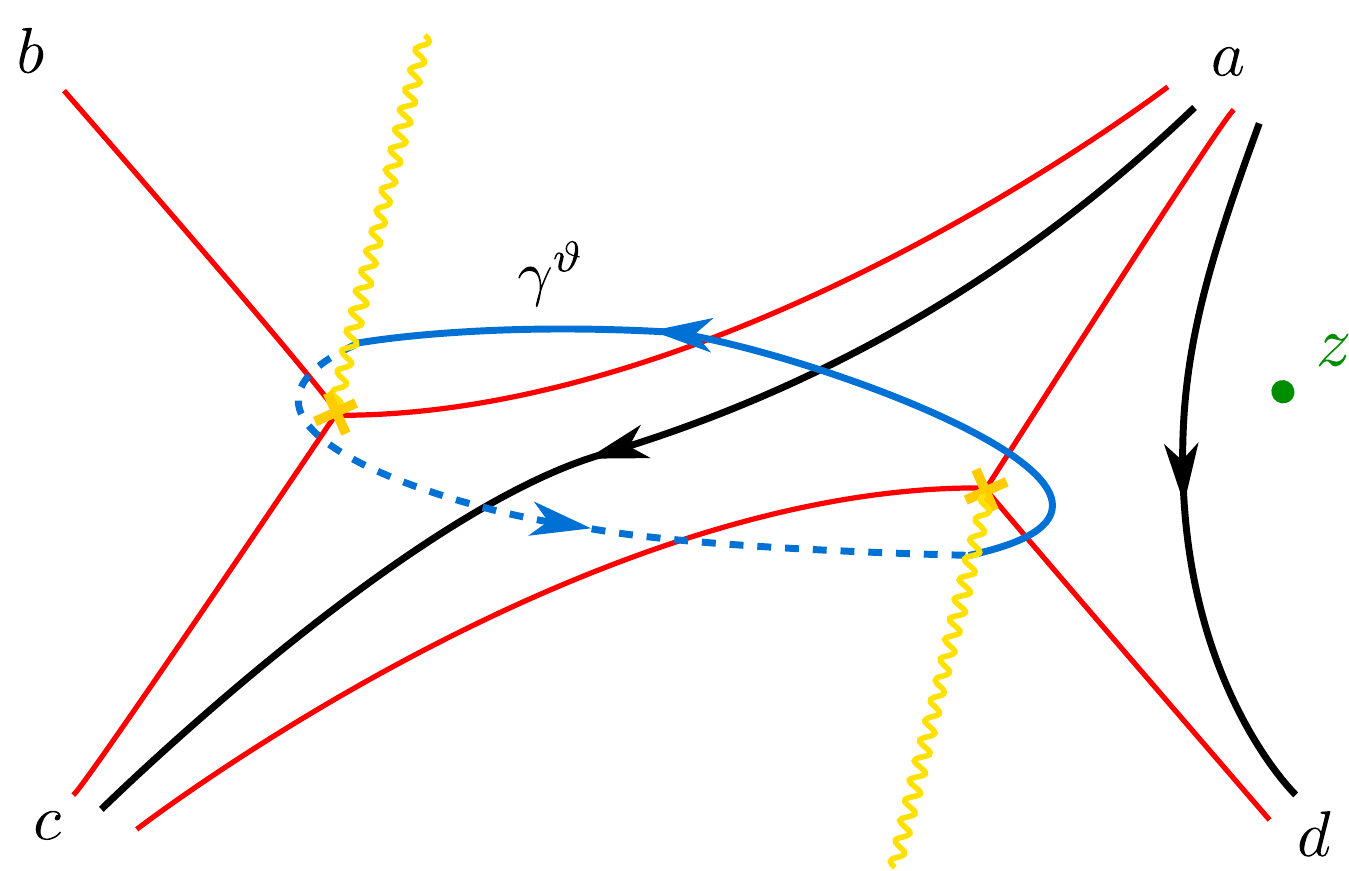}\hspace{0.12\textwidth} \includegraphics[width=0.40\textwidth]{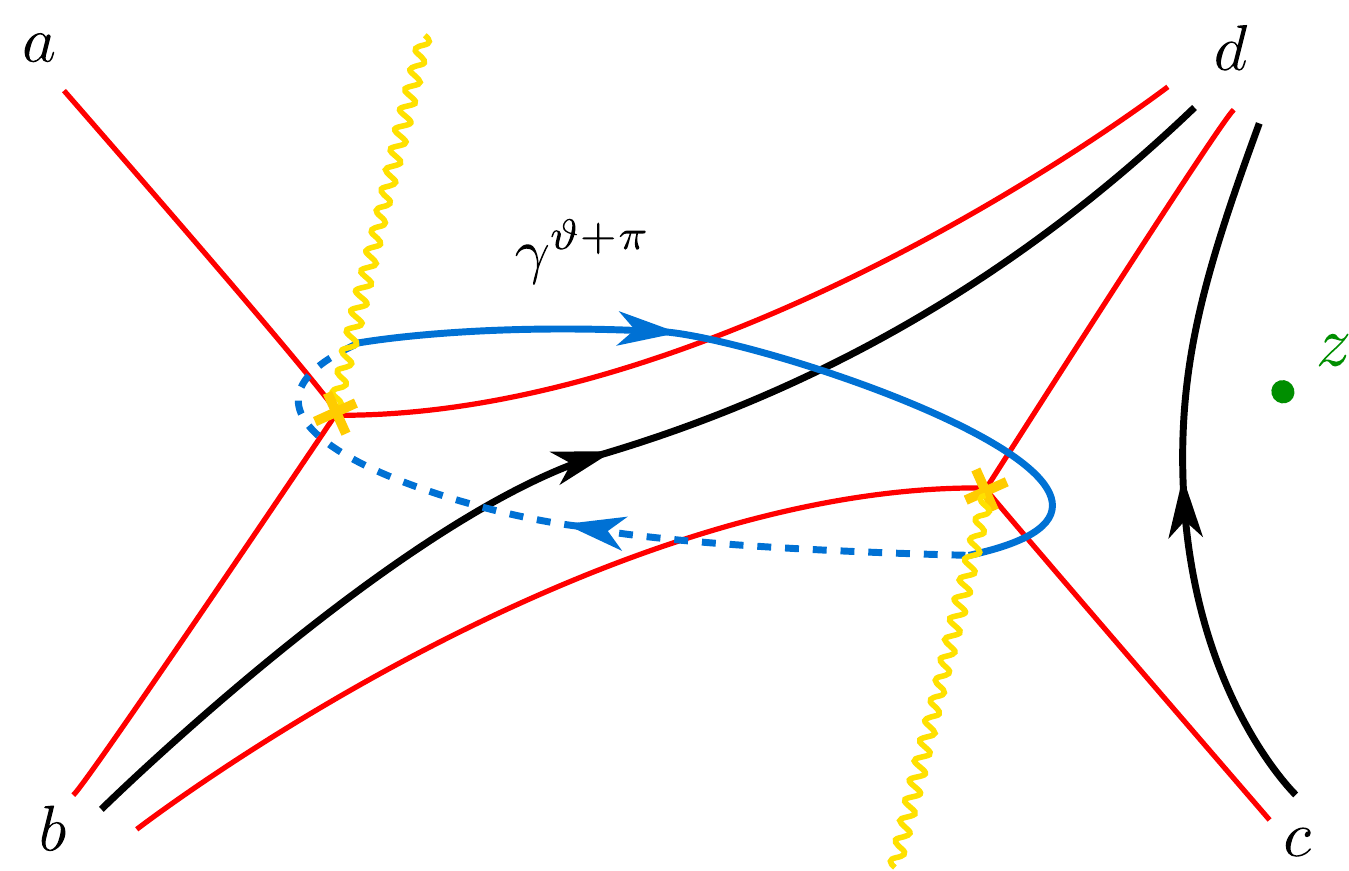}
\end{center}
\caption{The WKB triangulation on sheet 2 as seen from the \emph{north pole} of $\mathbb{CP}^1$: on the left at $\vartheta$ and on the right at $\vartheta+\pi$. The topology remains the same, while the flow of WKB lines switches direction, moreover there is a cyclic permutation of the vertices, since they all belong to the same irregular puncture at the south pole.}\label{fig:N_2_AD}
\end{figure}
\begin{figure}[h!]
\begin{center}
\leavevmode
\includegraphics[width=0.35\textwidth]{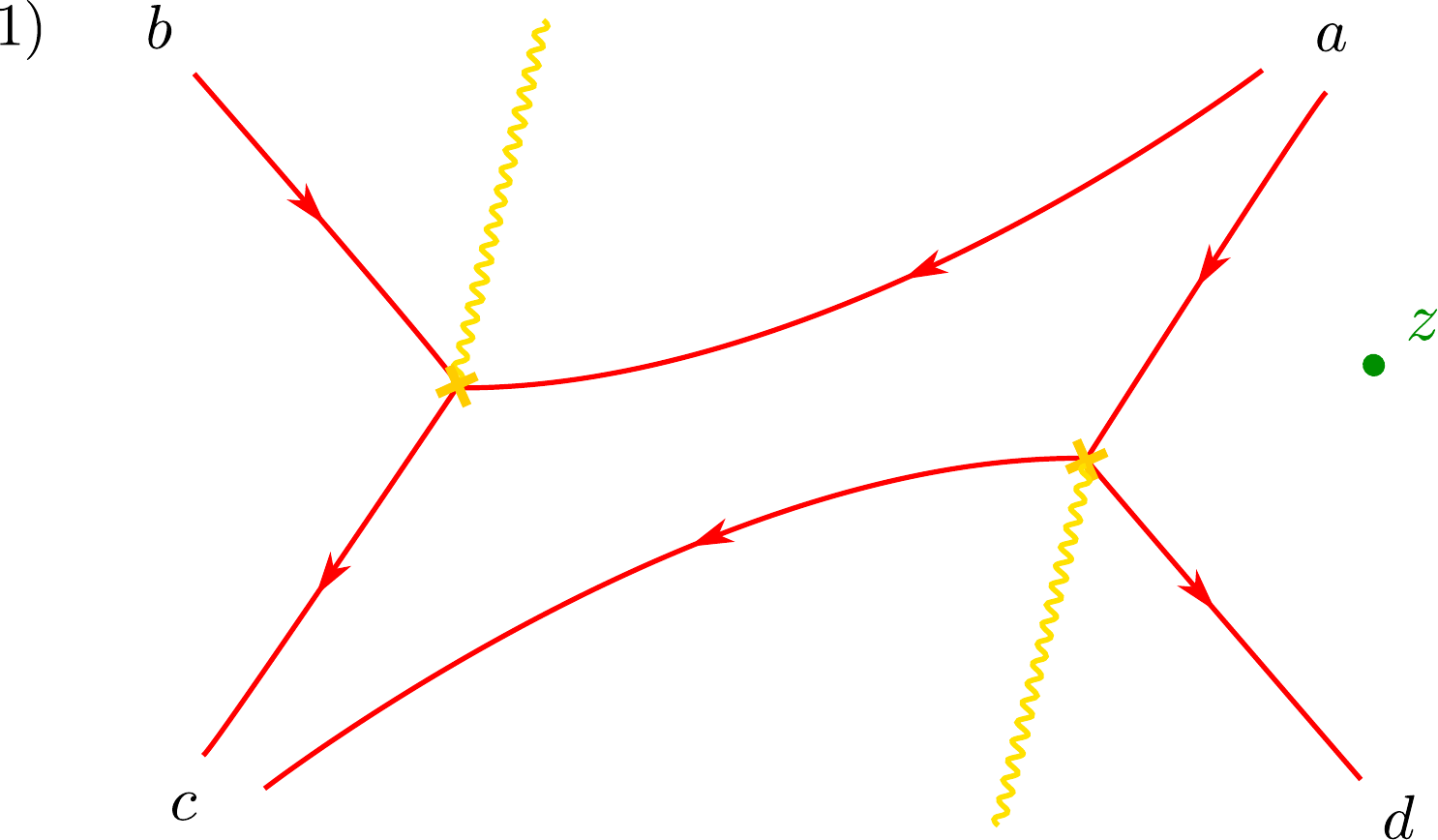} \hspace{0.15\textwidth} \includegraphics[width=0.35\textwidth]{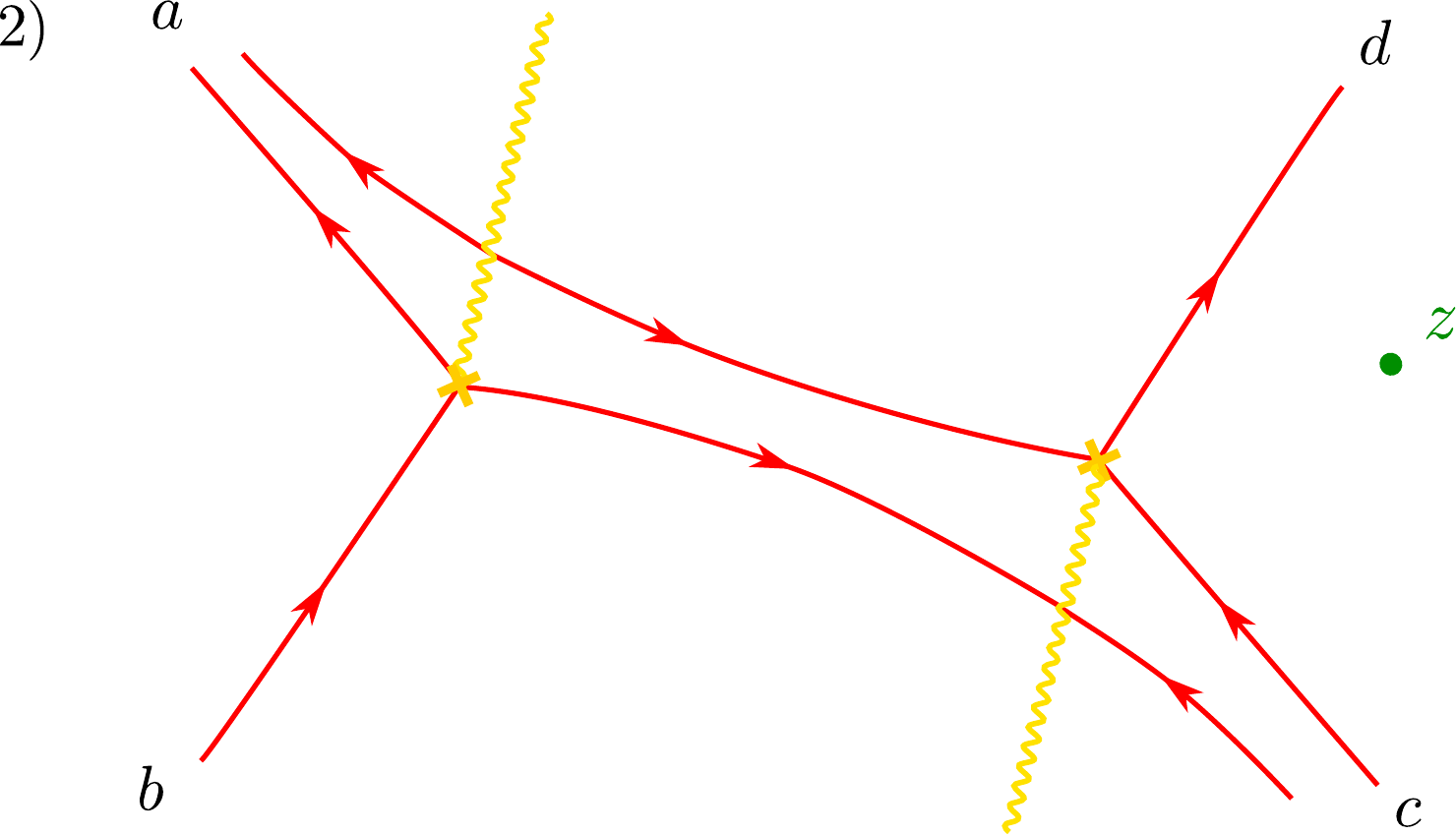} \\
\vspace{0.04\textwidth}
\includegraphics[width=0.35\textwidth]{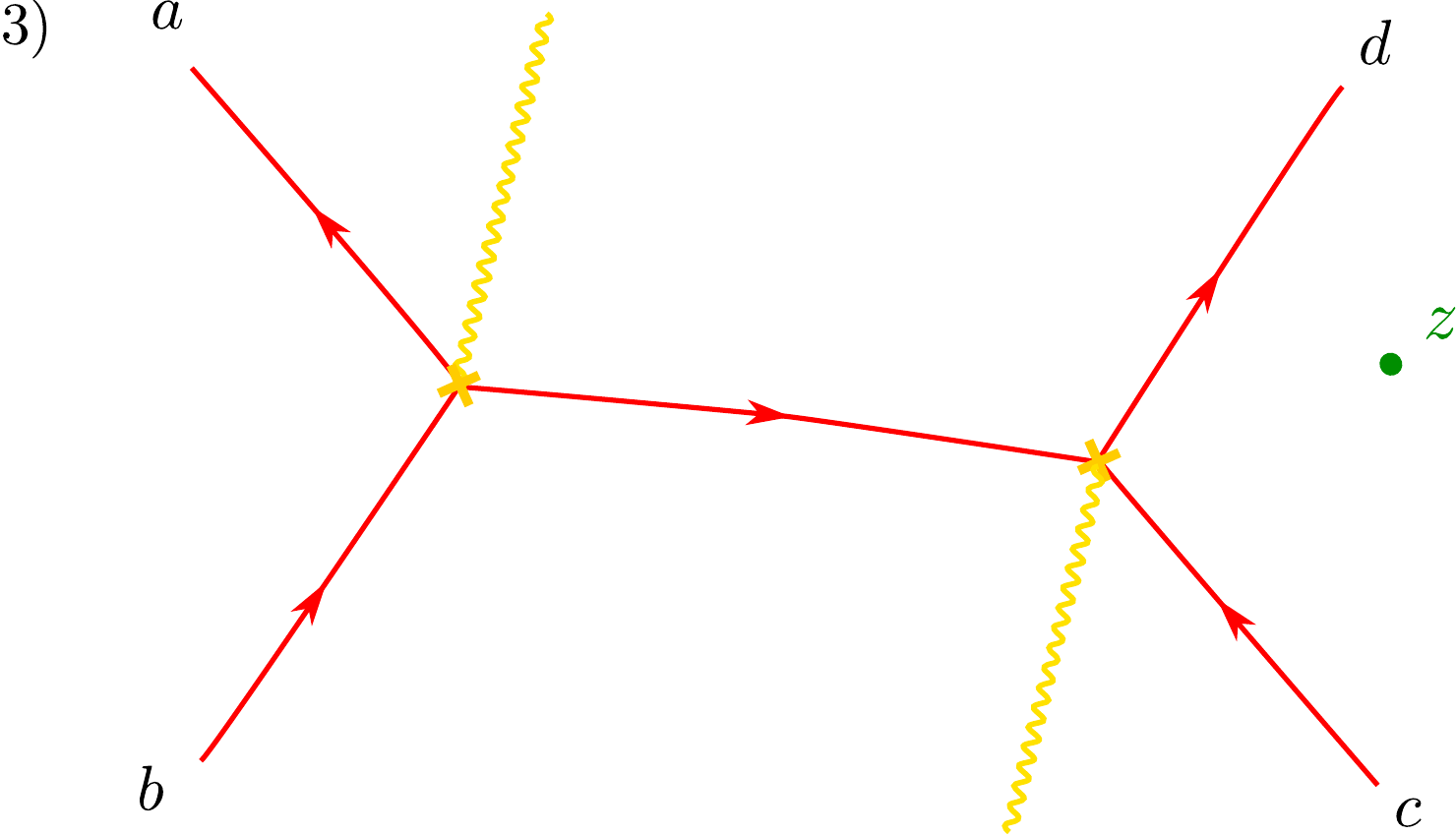} \hspace{0.15\textwidth} \includegraphics[width=0.35\textwidth]{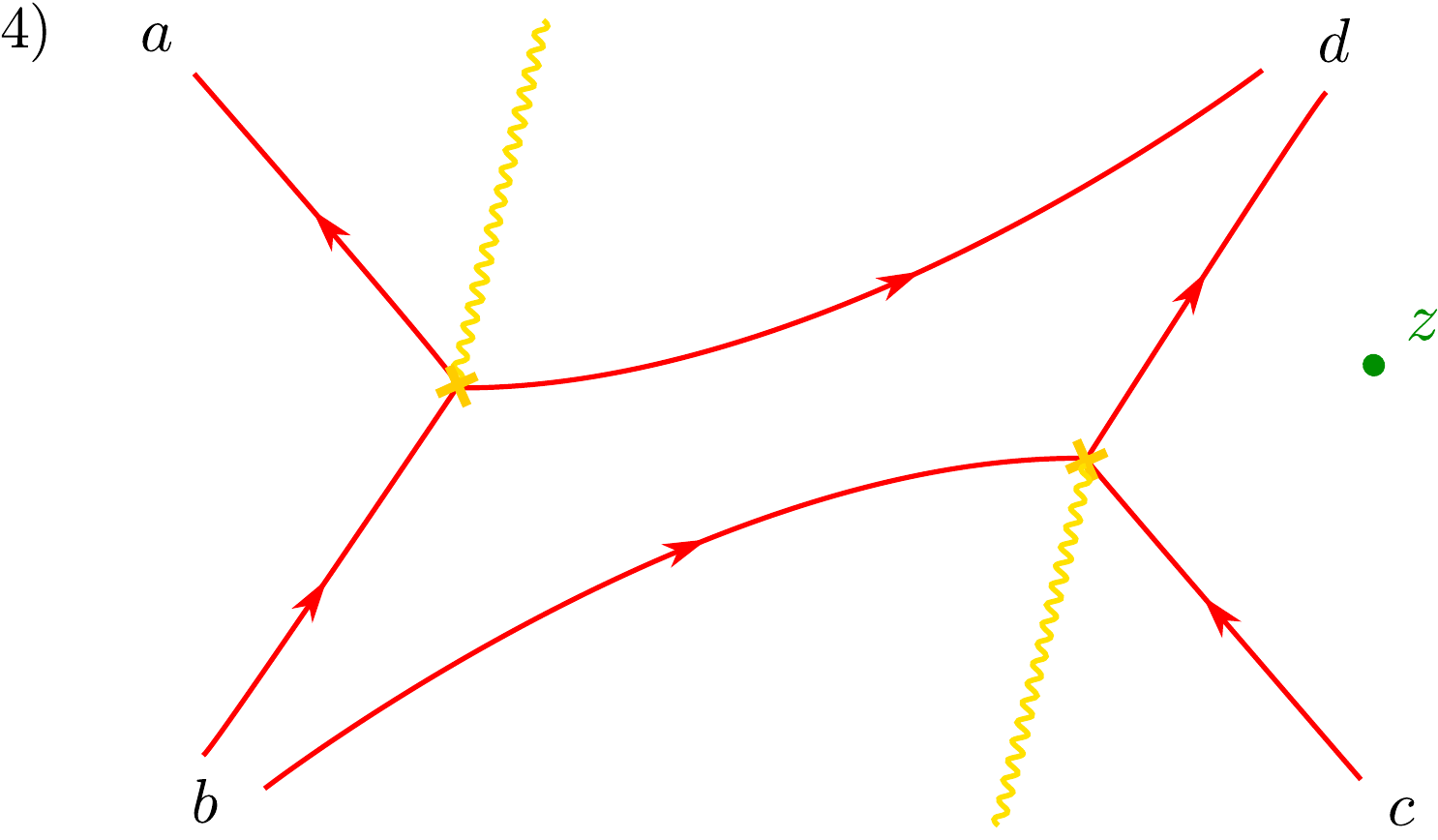}
\end{center}
\caption{Evolution of the WKB triangulation as $\vartheta\mapsto\vartheta+\pi$. Between stages 1 and 2 an ${\cal S}$ factor occurs, since the separating line at $d$ crosses $z$, at stage 3 we have a jump from the hypermultiplet as indicated by the flip, hence the spectrum will contain a ${\cal K}$ factor as well.}\label{fig:N_2_AD_evo}
\end{figure}

In this theory  the Seiberg-Witten differential is given by $P_2(z)=z^2+2m$. The only deformation is log-normalizable (see \II), hence the Coulomb branch is a single point. Let us adopt the conventions of section 8.1.2 in \IV, in particular we will choose the same conventions for the orientation of paths on $C$.
Figure \ref{fig:N_2_AD} shows how the WKB triangulation evolves in passing from $\vartheta$ to $\vartheta+\pi$ on sheet 2. This is justified as follows: on the one hand, the topology of the triangulation must be the same, with the orientation of WKB lines inverted; on the other hand one can follow how the triangulation evolves step by step, as shown in fig.\ref{fig:N_2_AD_evo}\footnote{Since we have an irregular singularity at infinity, the WKB rays will rotate as $\vartheta$ increases, for $N=4$ we have a rotation by $\pi/2$ for $\vartheta\mapsto\vartheta+\pi$ as the labels show} (also see fig. 45 in \II), which explains manifestly the nature of the spectrum: we must have an ${\cal S}$ factor due to the fact that a separatrix crosses $z$, and then a ${\cal K}$ factor because of the flip; this interpretation is confirmed by equation (8.14) in the reference, moreover by placing $z$, say, in the upper region, one sees that there will be 2 ${\cal S}$ factors  instead of 1, because two separatrices cross $z$, as confirmed by eq. (8.15) in the reference, and so on.
To begin our analysis of the spectrum in the region on the right, we first identify, as usual
\ba
	s_2=s_d(s_a,s_c) \quad s_1=s_a (s_c,s_d) \quad \nu_1=+=-\nu_2
\ea
After the omnipop, we'll have
\ba
	\tilde s_2= s_d(s_b,s_c) \quad \tilde s_1= s_c (s_d,s_b) \quad \tilde \nu_1=-=-\tilde \nu_2
\ea
Following rules (\ref{eq:fiber_endomorphisms}) we have immediately
\ba
	\CY_{12} = s_a \otimes s_a \, \frac{(s_c,s_d)}{(s_d,s_a)(s_a,s_c)} \qquad & \CY_{21} = s_d \otimes s_d \, \frac{(s_a,s_c)}{(s_d,s_a)(s_c,s_d)} \\
	\CY_{11} = \frac{s_d \otimes s_a}{(s_a,s_d)} \qquad & \CY_{22} = \frac{s_a \otimes s_d}{(s_d,s_a)}
\ea
together with
\ba
	\CY_\gamma = \frac{(s_a,s_b)(s_c,s_d)}{(s_b,s_c)(s_d,s_a)} = (\CY_{-\gamma})^{-1}.
\ea
After sending $\vartheta\mapsto\vartheta+\pi$ we have instead
\ba
	\tilde \CY_{12} &= - s_c \otimes s_c \frac{( s_d, s_b)}{(s_d, s_c)(s_b,s_c)} \\
	& = \frac{(s_a,s_c)(s_d,s_b)}{(s_b,s_c)(s_d,s_a)} \,\left[ \CY_{12} + \CY_{21} +\CY_{11} - \CY_{22}\right]
\ea
extrapolating $s_b$ in terms of $s_a, \,s_c$ from the analogue of (\ref{eq:section_identity}) we get
\ba
  (s_a,s_c)(s_d,s_b)=(s_b,s_c)(s_d,s_a)-(s_a,s_b)(s_c,s_d)
\ea
hence,
\ba
      \tilde \CY_{12} & = (1-\CY_\gamma) \,\left[ \CY_{12} + \CY_{21} +\CY_{22} - \CY_{11}\right] \\
      &= (1-\CY_\gamma) \, (1-\CY_{21})\,\CY_{12}\,(1+\CY_{21})
\ea
It's straightforward to follow the rules and obtain the other coordinates: by employing (\ref{eq:section_identity}) we find
\ba
      \tilde \CY_{21} &= \tilde \nu_2\frac{\tilde s_2  \otimes  \tilde s_2}{(\tilde s_1, \tilde s_2)} = (1-\CY_\gamma)^{-1}\,\CY_{21} \\
      \tilde \CY_{11} &= \frac{\tilde s_2  \otimes  \tilde s_1}{(\tilde s_1, \tilde s_2)} = \CY_{11}-\CY_{21} \\
      \tilde \CY_{22} &= \frac{\tilde s_1  \otimes  \tilde s_2}{(\tilde s_2, \tilde s_1)} = \CY_{22}+\CY_{21} 
\ea
recalling that $\langle\gamma_{ii},\gamma\rangle=0$, we recognize the spectrum generator
\ba
      {\bf S}={\cal S}_{21}\, {\cal K}_{\gamma} \label{eq:SG-AD-N2}
\ea
in agreement with eq.(8.14) of \IV.

\subsubsection{Adapting the general spectrum generator}
\begin{figure}[h!]
\begin{center}
\leavevmode
\includegraphics[width=0.45\textwidth]{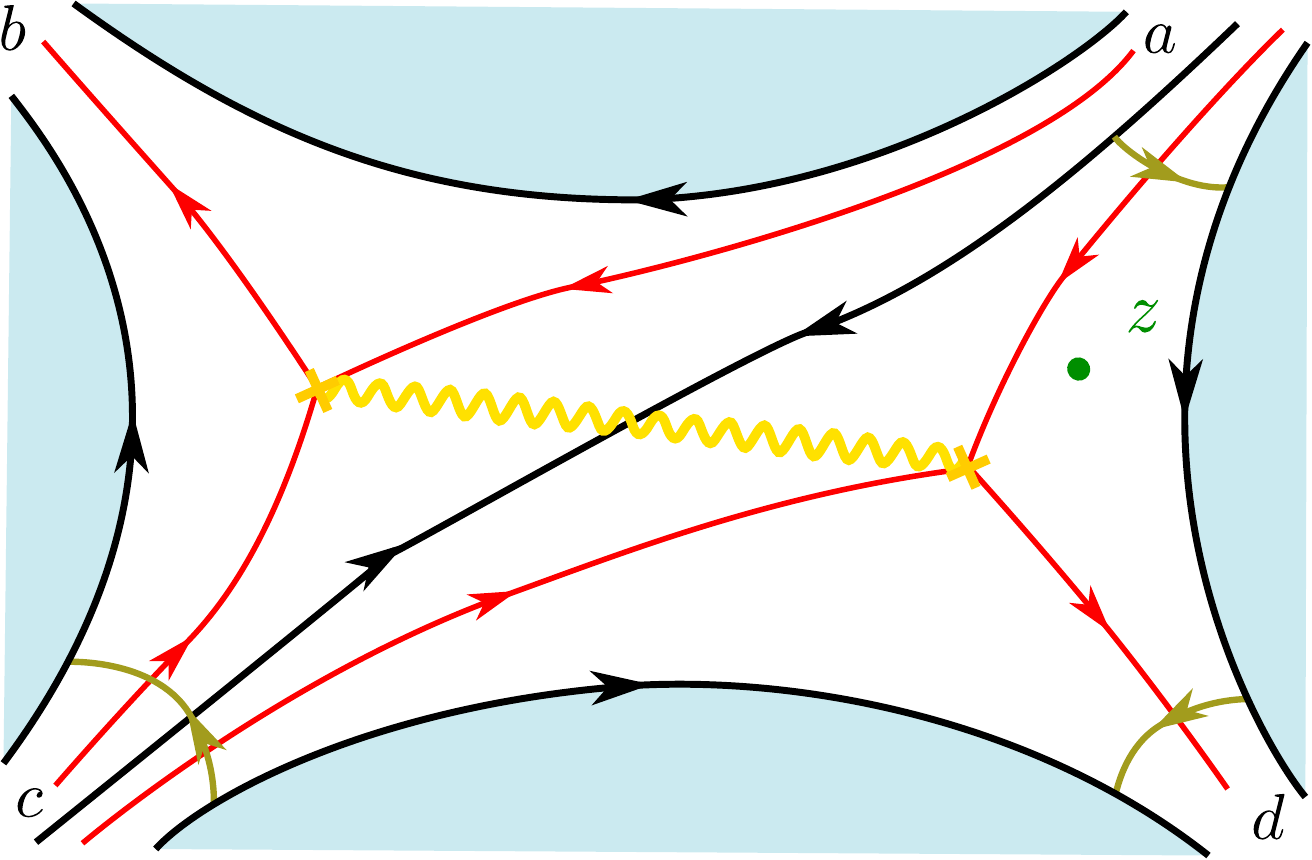}
\end{center}
\caption{The WKB triangulation at angle $\vartheta$, on sheet 2. Indicated are the paths for computing $\Sigma_a^{c \to d},\,\Sigma_{c}^{d \to b},\,\Sigma_d^{a \to c}$ around vertices $a,c,d$ respectively}\label{fig:N_2_AD_bis}
\end{figure}
We now apply the results of section \ref{sec:soliton_pop} to derive ${\bf S}$.
As usual this just requires matching labels correctly, this is slightly more delicate than in previous examples, therefore we will explain it in some detail.

In order to match the two situations, comparing figures \ref{fig:scan} and \ref{fig:N_2_AD_bis}, it is straightforward to identify
\ba
  \text{sheet } 2 \,\leftrightarrow \, \text{sheet } 2
\ea
in order to read the $\tilde \CY$'s from those in eq.s (\ref{eq:CY_pop_reg}), (\ref{eq:CY_pop_reg_bis}) we must take two steps
\begin{enumerate}
 \item replace labels according to
\ba
	a \to d \qquad b \to a \qquad c \to c 
\ea
 \item perform the replacements suitable to irregular punctures, as explained in section \ref{sec:irregular_general}. For this example we have simply (by definition, or by combining (\ref{eq:irr_pop}) with (\ref{eq:section_identity}))
  $$ \tilde s_a = s_d, \quad\tilde s_b = s_a, \quad\tilde s_c = s_b, \quad\tilde s_d = s_c$$
\end{enumerate}
Labels only occur within $\Xi$'s and $\Sigma$'s, therefore we only need to replace
\ba
      &\Sigma_{a}^{b \to b}\,\stackrel{(1)}{\mapsto} \, \Sigma_{d}^{a \to a}\,\stackrel{(2)}{\mapsto} \, \Sigma_{d}^{a \to c} \,=\, 1\\
      &\Sigma_{b}^{c \to c}\,\stackrel{(1)}{\mapsto} \, \Sigma_{a}^{c \to c}\,\stackrel{(2)}{\mapsto} \, \Sigma_{a}^{c \to d} \,=\, 1\\
      &\Sigma_{c}^{a \to a}\,\stackrel{(1)}{\mapsto} \, \Sigma_{c}^{d \to d}\,\stackrel{(2)}{\mapsto} \, \Sigma_{c}^{d \to b} \,=\, 1+\CX_\gamma=1-\CY_\gamma
\ea
Where, in the last column we used the definition of $\Sigma$ (11.9) of \II. Notice that, since $a,b,c,d$ all belong to an irregular puncture, all $\mu^{2}$ must be set to zero, as explained in section \ref{sec:irregular_single}. Employing the definition of the $\Xi$'s (\ref{eq:Xi_def}), we can perform the following replacement for the $\Xi$'s:
\ba
      & \Xi(\Sigma_a^{b \to b},\Sigma_b^{c \to c};\mu_a,\mu_b) \,\mapsto \, 1 \\
      & \Xi(\Sigma_b^{c \to c},\Sigma_c^{a \to a};\mu_b,\mu_c) \,\mapsto \, 1-\CY_\gamma \\
      & \Xi(\Sigma_c^{a \to a},\Sigma_a^{b \to b};\mu_c,\mu_a) \,\mapsto \, 1
\ea
Given these, we can now directly read off the omnipop transformation from (\ref{eq:CY_pop_reg}), (\ref{eq:CY_pop_reg_bis})
\ba
      \tilde \CY_{12} & = (1-\CY_\gamma) \,\left[ \CY_{12} + \CY_{21} +\CY_{22} - \CY_{11}\right] \\
      \tilde \CY_{21} &= (1-\CY_\gamma)^{-1}\,\CY_{21} \\
      \tilde \CY_{11} &= \CY_{11}-\CY_{21} \\
      \tilde \CY_{22} &= \CY_{22}+\CY_{21} 
\ea
in agreement with our full derivation.\\

\noindent{\bf Wall crossing formulae:} by repeating the above reasoning, one can determine the spectra for various deformations of the surface defect. According to our derivation, there will be a jump whenever the labeling changes: this occurs precisely when we move from one cell to another (the vertex labels change), or when we cross a branch cut (the sheet labels change). Therefore we expect precisely 6 different regions with their own spectra. This is exactly what \IV found.

\subsection{The $\CP^1$ sigma model}
The next example we consider is that of the $\CP^1$ sigma model presented in \IV. As pointed out there, the 4d dynamics is trivial (no wall crossing), while we do have some 2d phenomena.

\begin{figure}[h!]
\begin{center}
\leavevmode
\includegraphics[width=0.75\textwidth]{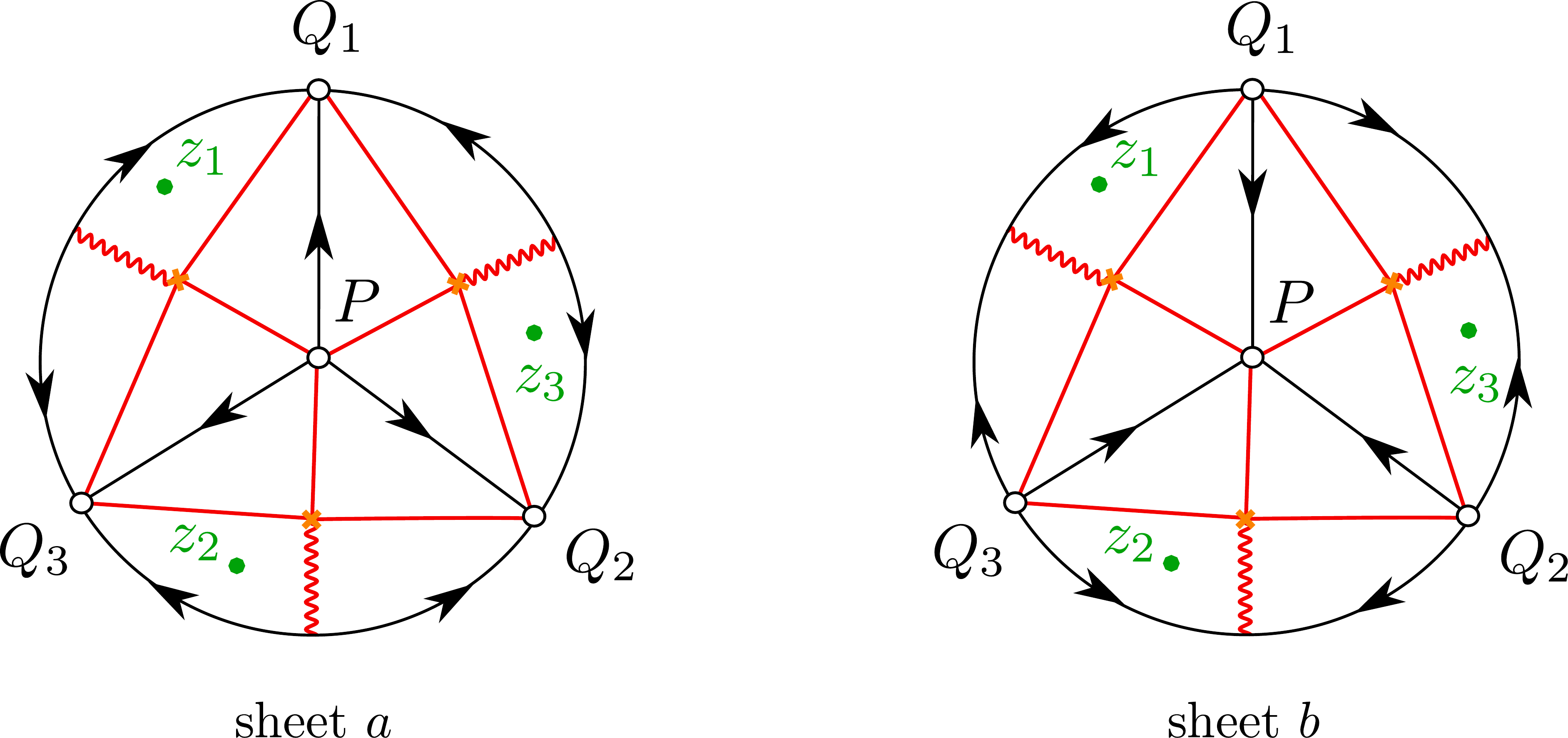}
\end{center}
\caption{The six-fold cover at angle $\vartheta$, the surface defect sitting at strong coupling. For convenience points at infinity have been mapped at finite distance.}\label{fig:CP1-sigma}
\end{figure}

The Seiberg-Witten curve is described by
\ba
    \lambda^2=\left(  \frac{\Lambda^2}{z}  + \frac{m^2}{z^2}  \right)\,\d z^2
\ea
which encodes a turning point at $z=-m^2/\Lambda^2$, as well as a regular singularity in $z=0$ and an irregular singularity at infinity with one Stokes sector. The triangulation consists therefore of one degenerate triangle on each of the two sheets. In order to build the coordinates, we must locally pass to a three-fold covering on each of the two sheets, as pointed out in \II.

After doing so, we end up with the six-fold cover illustrated in fig.\ref{fig:CP1-sigma}. Let us start with the so-called strong coupling regime, and choose to work within the triangle denoted by vertices $Q_1,P,Q_3$. We will understand $z$ to stand for $z_1$. Let $M$ be the monodromy matrix for parallel transport around $P$, then $M \cdot s_3=s_1=M^{-1} \cdot s_2$, and we denote $\CY_\gamma=\CY_{\gamma_{i,P}},\,\,\forall i=1,2,3$. The vertices $Q_i$ belong to an irregular vertex, while $P$ is regular, therefore we must employ the rules of \S \ref{sec:irregular_general} in order to match the spectrum generator correctly. Taking a look at fig. \ref{fig:scan}, we have the following identifications:
\ba
      & Q_1 \to {\cal P}_a  \qquad Q_3 \to {\cal P}_b  \qquad  P \to {\cal P}_c &\\
      & \qquad \qquad \text{sheet }a\,\,\to\,\, \text{sheet }2 &
\ea
since $Q_1,Q_3 $ are irregular we set $\mu_a,\mu_b\to0$. Notice that, since we have an irregular singularity at infinity 
\ba
    \tilde s_1 = s_2 \qquad \tilde s_2 = s_3 \qquad \tilde s_3 = s_1 \qquad
\ea
therefore we evaluate the modified $\Sigma$'s according to our prescriptions\footnote{we use eq. (11.9) of \II together with the fact that $\CY_\gamma=-{\cal X}_\gamma$, as pointed out in appendix F of \IV.}
\ba
      &\Sigma_a^{b\to \tilde a} = \Sigma_{1}^{3\to 2} \,= 1-\CY_{\gamma_{1P}} \,= 1-\CY_\gamma \\
      &\Sigma_b^{c\to \tilde b} = \Sigma_{3}^{P\to 1} \,= 1 \\
      &\Sigma_c^{a \to a} = \Sigma_{P}^{1\to 1} \,= 1-\CY_{\gamma_{3P}}+\CY_{\gamma_{3P}}\CY_{\gamma_{2P}} \,= 1-\CY_\gamma +\CY_\gamma^2
\ea
entailing the explicit expressions
\ba
      & \Xi(\Sigma_a^{b \to \tilde a},\Sigma_b^{c \to \tilde b};\mu_a,\mu_b)\,=\,1 \\
      & \Xi(\Sigma_b^{c \to \tilde b},\Sigma_c^{a \to a};\mu_b,\mu_c)\,=\, 1-\CY_\gamma+\CY_\gamma^2 \\
      & \Xi(\Sigma_c^{a \to a},\Sigma_a^{b \to \tilde a};\mu_c,\mu_a)\,=\, 1-\CY_\gamma+\CY_\gamma^2
\ea
We are now ready to read the omnipop transformation from eqs. (\ref{eq:CY_pop_reg}), (\ref{eq:CY_pop_reg_bis}): they read
\ba
	\tilde \CY_{aa}&=\CY_{aa} + (1-\CY_{\gamma}) \CY_{ab} \\
	\tilde \CY_{bb}&=\CY_{bb} - (1-\CY_{\gamma}) \CY_{ab} \\
	\tilde \CY_{ab}&=\CY_{ab} \\
	\tilde \CY_{ba}&=\CY_{ba} + (1-\CY_{\gamma})^2 \CY_{ab} + (1-\CY_{\gamma}) (\CY_{aa}-\CY_{bb})
\ea
by direct inspection\footnote{we use $\sigma{(\gamma,\gamma_{\alpha\beta})=1},\,\,\forall\alpha,\beta\in\{a,b\}$}, these correspond to the spectrum generator\footnote{we employ the following twisting functions: 
\ba
    \sigma(aa,ab)=\sigma(bb,ab)=\sigma(aa+\gamma,ab+\gamma)=\,+ \\
    \sigma(aa,ab+\gamma)=\sigma(bb,ab+\gamma)=\sigma(ba,ab)=\sigma(ba,ab+\gamma)=\,-
\ea} 
\ba
	{\bf S} \,&=\, {\cal S}_{\gamma_{ab}+\gamma}\,{\cal S}_{\gamma_{ab}} \label{eq:SG-CP1}
\ea
we find two 2d solitons whose charges differ by the flavor charge $\gamma$. This is consistent with what is found in \S 8.2 of \IV, where they find a soliton with charge $\tilde\gamma_{12}$ and one with charge $\gamma_{12}= \gamma_f-\tilde \gamma_{21}=\gamma_f+\tilde \gamma_{12}$.\\
\begin{figure}[h!]
\begin{center}
\leavevmode
\includegraphics[width=0.75\textwidth]{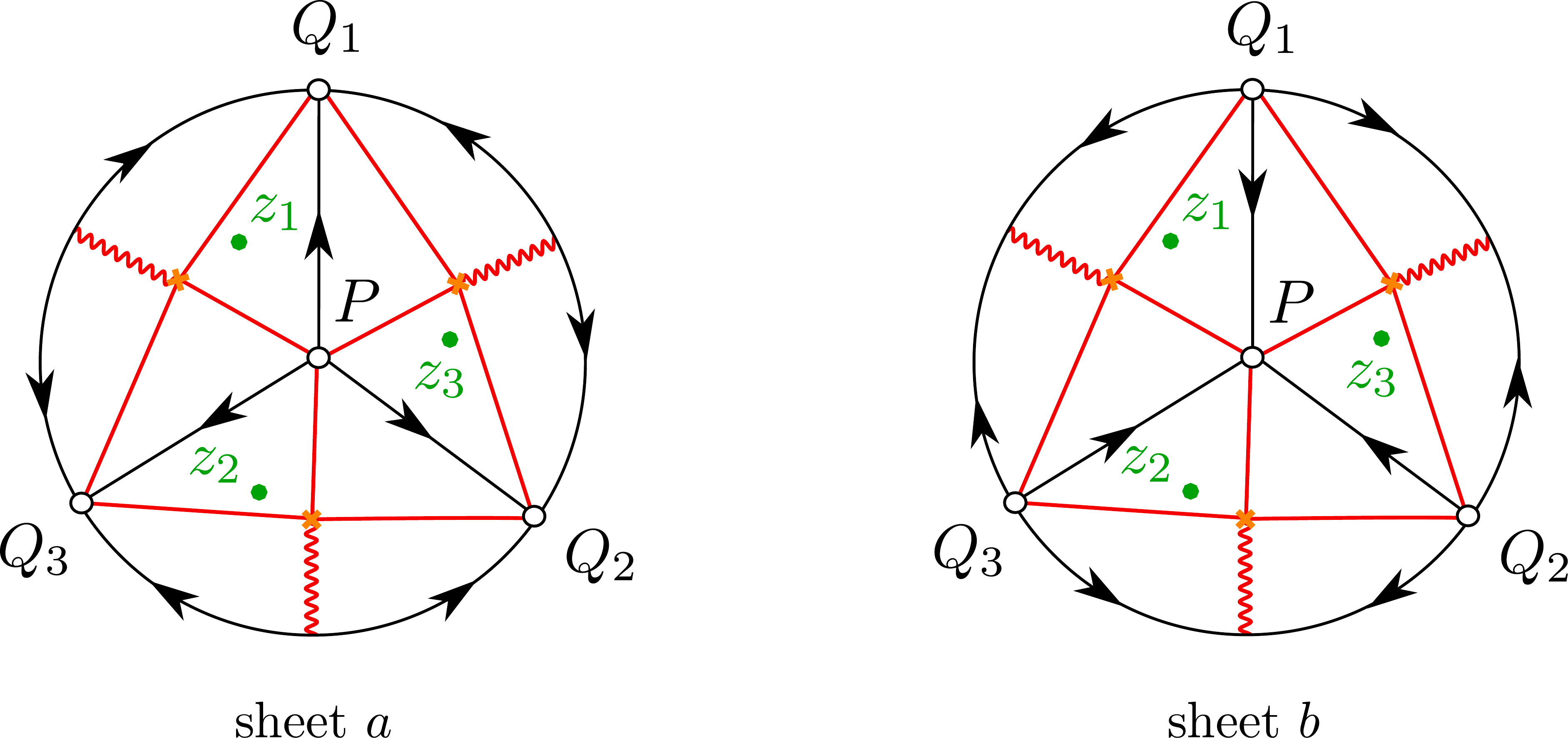}
\end{center}
\caption{The six-fold cover at angle $\vartheta$ with the surface defect in the weak coupling region}\label{fig:CP1-sigma-bis}
\end{figure}
Let us now investigate the weak coupling regime. Suppose that $z$ lies in the sector indicated in fig. \ref{fig:CP1-sigma-bis}. Then, looking at fig.\ref{fig:scan} we identify
\ba
      & Q_1 \to {\cal P}_b  \qquad Q_3 \to {\cal P}_c  \qquad  P \to {\cal P}_a &\\
      & \qquad \qquad \text{sheet }a\,\,\to\,\, \text{sheet }1 &
\ea
since $Q_1,Q_3 $ are irregular we set $\mu_b,\mu_c\to0$, while $\mu_a^{2}=\mu^{2}=-\CY_\gamma$\footnote{cf. eq.(7.6) in \II}. We evaluate the modified $\Sigma$'s according to our prescriptions
\ba
      &\Sigma_a^{b\to b} = \Sigma_{P}^{1\to 1} \,= 1-\CY_\gamma + \CY_\gamma^2 \\
      &\Sigma_b^{c\to \tilde b} = \Sigma_{1}^{3\to 2} \,= 1-\CY_\gamma \\
      &\Sigma_c^{a \to \tilde c} = \Sigma_{3}^{P\to 1} \,= 1
\ea
entailing the explicit expressions
\ba
      & \Xi(\Sigma_a^{b \to b},\Sigma_b^{c \to \tilde b};\mu_a,\mu_b)\,=\,1+\CY_\gamma^2-\CY_\gamma^3 \\
      & \Xi(\Sigma_b^{c \to \tilde b},\Sigma_c^{a \to \tilde c};\mu_b,\mu_c)\,=\, 1 \\
      & \Xi(\Sigma_c^{a \to \tilde c},\Sigma_a^{b \to b};\mu_c,\mu_a)\,=\, 1-\CY_\gamma+\CY_\gamma^2
\ea
therefore, from eqs. (\ref{eq:CY_pop_reg}), (\ref{eq:CY_pop_reg_bis}) we read off the omnipop transformation
\ba
	\tilde \CY_{aa}&=(1+\CY_{\gamma}^2 -\CY_{\gamma}^3)^{-1} \\ 
	& \times \left[ (1+\CY_{\gamma})\CY_{aa}-(\CY_{\gamma}-\CY^2_{\gamma}+\CY_{\gamma}^3)\CY_{bb} - (\CY_{\gamma}+\CY_{\gamma}^2)\CY_{ab} -(1-\CY_{\gamma}+\CY_{\gamma}^2) \CY_{ba} \right] \\
	\tilde \CY_{bb}&=(1+\CY_{\gamma}^2 -\CY_{\gamma}^3)^{-1}  \\ 
	& \times  \left[ (1+\CY_{\gamma})\CY_{bb}-(\CY_{\gamma}-\CY^2_{\gamma}+\CY_{\gamma}^3)\CY_{aa} - (\CY_{\gamma}+\CY_{\gamma}^2)\CY_{ab} +(1-\CY_{\gamma}+\CY_{\gamma}^2) \CY_{ba} \right] \\
	\tilde \CY_{ab}&= \left[(1+\CY_{\gamma}^2 -\CY_{\gamma}^3)(1-\CY_{\gamma} +\CY_{\gamma}^2)\right]^{-1}  \\ 
	& \times  \left[ (1+\CY_{\gamma})^2\CY_{ab} +(1-\CY_{\gamma}+\CY_{\gamma}^2)^2 \CY_{ba}+ (1+\CY_{\gamma}^2 - \CY_{\gamma}^3)(1+\CY_{\gamma})(\CY_{bb} - \CY_{aa}) \right]\\
	\tilde \CY_{ba}&=\left[(1+\CY_{\gamma}^2 -\CY_{\gamma}^3)(1-\CY_{\gamma} +\CY_{\gamma}^2)\right]^{-1}  \\ 
	& \times  \left[ \CY_{ba}+\CY_{\gamma}^2\CY_{ab} + \CY_{\gamma}(\CY_{bb} - \CY_{aa}) \right]
\ea
It is not easy to read the spectrum generator off these transformations. As pointed out in \IV, there should be two solitons with charges $\gamma+\gamma_{aa},\,\gamma+\gamma_{bb}$, as well as two towers of states with charges $\gamma_{ab}+n\gamma,\, \gamma_{ba}+n\gamma$.

\section{Conclusions}
We explicitly derived the expressions for the $\tilde\CY_a$ as functions of the $\CY_a$ both for 2d-4d BPS solitonic charges, and for framed 2d-4d BPS solitons. The cases we analyzed in \S\ref{sec:formal_work} serve as building blocks for analyzing more complicated situations. Our results have been confirmed by correctly recovering the well-known spectra of certain variations of $N=1,2$ AD theories and of the $\mathbb{CP}^1$ sigma model. As a byproduct we have shown how our formulae can be successfully employed to obtain, in a fairly straightforward way, the action of the 2d-4d spectrum generator for several 2d-4d systems of the $A_1$ type. We must recall, however, that spectrum generating functions only implicitly encode the BPS spectrum, which is ultimately obtained by recovering a factorization of the spectrum generator $\mathbb{S}$, this is a nontrivial task, and is currently being investigated. An interesting open problem would be to generalize our derivation to systems of higher rank, presumably employing the newly developed techniques of \cite{GMN5}.

\section*{Acknowledgements}
\noindent I wish to thank Greg Moore and Andy Royston for helpful explanations and discussions. \\ The work of P.L. is supported by the DOE under grant DE-FG$02$-$96$ER$40959$.

\nocite{*}
%\cleardoublepage
\addcontentsline{toc}{section}{Bibliography}
\bibliographystyle{bibliostyle}
\bibliography{Bibliography} 
\end{document}